\documentclass[12pt,12pt]{article}
\usepackage{graphicx}
\usepackage{epsfig}
\usepackage{amsmath,amssymb}
\usepackage{slashed} 
\font\mathbf cmbxti10 at 12pt

\catcode`\@=11
\newdimen\ex@
\font\dozeb=cmmib10 scaled \magstep1
\font\dozesyb=cmbsy10 scaled \magstep1
\font\dezb=cmmib10
\def\bm{\fam9}
\def\beq{\begin{equation}}
\def\eeq{\end{equation}}
\def\beqa{\begin{eqnarray}}
\def\eeqa{\end{eqnarray}}
\newcommand\BA{\begin{array}}
\newcommand\EA{\end{array}}

\def\E{{\bm E}}
\def\K{{\bm K}}

\begin{document}
\def\thefootnote{\fnsymbol{footnote}}

\title{\bf $\!\!$Mean-field theory based on the $\mathfrak{Jacobi~hsp}$\\
:= semi-direct sum  
 $\mathfrak{h}_N \!\rtimes\! \mathfrak{sp}(2N,\mathbb{R})_\mathbb{C}$
algebra \\of boson operators}
\vskip1cm
\author
{Seiya NISHIYAMA\footnotemark[1]
and Jo\~ao da PROVID\^{E}NCIA\footnotemark[2] \\
\\[-0.1cm]
Centro de F\'\i sica,
Departamento de F\'\i sica,\\ 
Universidade de Coimbra\\
P-3004-516 Coimbra, Portugal
}

\maketitle

\footnotetext[1]
{Corresponding author.
~E-mail address: seikoceu@khe.biglobe.ne.jp}
\footnotetext[2]
{E-mail address: providencia@teor.fis.uc.pt}
  
\vspace{-1cm}

\begin{abstract}
$\!\!\!\!\!\!\!\!\!$In this paper,
we give an expression 
for canonical transformation group
with Grassmann variables,
basing on the 
$\mathfrak{Jacobi~hsp}$ 
\!:=\! 
semi-direct sum
$
\mathfrak{h}_{N} 
\!\rtimes\! 
\mathfrak{sp}(2N,\mathbb{R})_\mathbb{C}
$
algebra of  boson operators.
We assume a mean-field Hamiltonian (MFH)
linear in the $\mathfrak{Jacobi}$ generators. 
We diagonalize the boson MFH.
We show a new aspect of eigenvalues of the MFH.
An excitation energy
arisen from additional self-consistent field (SCF) parameters
has never been seen in the traditional boson MFT.
We derive this excitation energy.
We extend the Killing potential 
in the 
$\!\frac{Sp(2N,\mathbb{R}\!)_\mathbb{C}}{U(N\!)}\!$ 
coset space  
to the one 
in the 
$\!\frac{Sp(2N \!+\! 2,\mathbb{R})_\mathbb{C}}{U(N \!+\! 1)}\!$ 
coset space
and make clear the geometrical structure of K\"{a}hler manifold, 
a non-compact symmetric space 
$\frac{Sp(2N \!\!+\!\! 2,\mathbb{R})_\mathbb{C}}{U(N \!+\!1)}$.
$\!\!$
The $\mathfrak{Jacobi~hsp}$ transformation group is embedded
into an $Sp(2N \!\!+\!\! 2,\mathbb{R})_\mathbb{C}$ group  
and  
an $\frac{Sp(2N \!+\! 2,\mathbb{R})_\mathbb{C}}{U(N \!+\! 1)}$ 
coset variable is introduced.
Under such mathematical manipulations,
extended bosonization 
of $Sp(2N \!\!+\!\! 2,\mathbb{R})_\mathbb{C}$ Lie operators,
vacuum function and differential forms
for extended boson are presented 
by using integral representation of boson state on
the $\frac{Sp(2N \!+\! 2,\mathbb{R})_\mathbb{C}}{U(N \!+\! 1)}$ coset variables.
\end{abstract}
\vskip0.3cm












\newpage

\setcounter{equation}{0}
\renewcommand{\theequation}{\arabic{section}.\arabic{equation}}

\section{INTRODUCTION}

\vspace{-0.2cm}

~~~~
In nuclear and condensed matter physics,
the time dependent Hartree-Bogoliubov (TDHB) theory
\cite{BCS.57,Bog.59}
has been regarded as a standard tool 
in many-body theoretical descriptions of 
superconducting fermion systems$\!$
\cite{RS.80,BR.86}.
$\!$An HB wavefunction (WF) for such systems represents 
bose condensate states of fermion pairs. 
Standing on the Lie-algebraic viewpoint,
the  pair operators of fermion with $N$ single-particle states 
form the $SO(2N)$ Lie algebra
which contains the $U(N)$ Lie algebra as a sub algebra. 
The $SO(2N)$ and $U(N)$ denote a special orthogonal group 
of $2N$-dimension, say $g$ and a unitary group of $N$-dimension.
We can give an integral representation of a state vector on the group $g$,
exact {\it coherent state} representation (CS rep) of a fermion system
\cite{Perelomov.86,Perelomov.87}.
It makes possible global approach to such problems, e.g. see Ozaki
\cite{Ozaki1.85,Ozaki2.85}. 
The canonical transformation (TR) of the fermion operators generated by 
the Lie operators in the $SO(2N)$ Lie algebra arises 
the famous generalized Bogoliubov TR for fermions. 

For providing a general microscopic means
for a unified self-consistent field (SCF) description 
for Bose and Fermi type collective excitations in fermion systems,  
Fukutome, Yamamura and one of the present authors (Nishiyama)
proposed a new fermion many-body theory
basing on the $SO(2N \!\!+\!\! 1)$ Lie algebra of fermion operators 
\cite{FYN.77,FukNishi.84}.
The fermion  
creation-annihilation and pair operators
form the 
Lie algebra of $SO(2N \!\!+\!\! 1)$ group.
A rep of an $SO(2N \!\!+\!\! 1)$ group is derived 
by group theoretical extension of
the $SO(2N)$ fermion Bogoliubov TR 
to a new canonical TR group.  
The fermion Lie operators, when operated onto 
the integral rep of 
the $SO(2N \!\!+\!\! 1)$ WF, are mapped into
the regular rep of the $SO(2N \!\!+\!\! 1)$ group and 
are given by the Schwinger-type boson rep 
\cite{Sch.65,YN.76}.
The boson images of all the Lie operators
are also expressed by closed first order differential forms. 
Along the above way,
using Grassmann variables,
we give a new mean-field theory (MFT),
based on the 
$\mathfrak{Jacobi~hsp}$ 
\!:=\! 
semi-direct sum
$
\mathfrak{h}_{\!N} 
\!\rtimes\! 
\mathfrak{sp}(2N\!,\!\mathbb{R})_\mathbb{C}
$\!
algebra 
($\mathfrak{Jacobi~hsp}$ algebra) of boson operators, 
which is studied intensively by Berceanu
\cite{Berceanu.12,Berceanu.6}.
We take an Hamiltonian
consisting of the generalized HB (GHB) 
MF Hamiltonian (MFH)
and assume a linear MFH
expressed in terms of the generators of 
the $\mathfrak{Jacobi~hsp}$  algebra.
We diagonalize the boson MFH.
A new aspect of eigenvalues of the MFH is shown.
An excitation energy
arisen from additional SCF parameters
has never been seen in the traditional boson MFT.
We derive this excitation energy.
We further extend the Killing potential 
in the 
$\frac{Sp(2N\!,\mathbb{R})_\mathbb{C}}{U(N)}$ 
coset space  
to the Killing potential 
in the 
$\frac{Sp(2N \!+\! 2\!,\mathbb{R})_\mathbb{C}}{U(N \!+\! 1)}$ 
coset space.
The extended Killing potential is equivalent with 
the generalized density matrix (GDM). 
Embedding the $\mathfrak{Jacobi~hsp}$ group 
into an $Sp(2N \!\!+\!\! 2,\mathbb{R})_\mathbb{C}$ group and 
using 
$\frac{Sp(2N \!+\! 2\!,\mathbb{R})_\mathbb{C}}{U(N \!+\! 1)}$ 
coset variables,
we give an extended boson rep 
on a non-compact symmetric space (NCSS) 
$\frac{Sp(2N \!+\! 2\!,\mathbb{R})_\mathbb{C}}{U(N \!+\! 1)}$.
 
In \S 2,
we give an expression 
for canonical TR group
with Grassmann variables
based on the $\mathfrak{Jacobi~hsp}$.
In \S 3,
we give an inverse of the TR.
In \S 4,
we give a boson MFH and its diagonalization
and show a new aspect of eigenvalues of the MFH.
In \S 5,
we show that the additional SCF parameters
are restricted by a new SCF condition.
We study the MFH linear in the $\mathfrak{Jacobi}$ generator.
In \S 6,
we investigate a coset space
$\frac{Sp(2N \!+\! 2\!,\mathbb{R})_\mathbb{C}}{U(N \!+\! 1)}$
and make clear geometrical structure of K\"{a}hler manifold, 
NCSS 
$\frac{Sp(2N \!+\! 2\!,\mathbb{R})_\mathbb{C}}{U(N \!+\! 1)}$.
In \S 7,
a Killing potential in the coset space
$\frac{Sp(2N \!+\! 2\!,\mathbb{R})_\mathbb{C}}{U(N \!+\! 1)}$ 
is given and 
its equivalence with the GDM is proved.
Finally, in \S 8 
we give discussions and summary. 
In Appendices,
embedding of the $\mathfrak{Jacobi~hsp}$ group into 
an $Sp(2N \!\!+\!\! 2,\mathbb{R})_\mathbb{C}$ group is made 
and introduction of 
$\frac{Sp(2N \!+\! 2\!,\mathbb{R})_\mathbb{C}}{U(N \!+\! 1)}$ 
coset variables is made.
We give
an extended bosonization procedure 
of the $\mathfrak{sp}(2N \!\!+\!\! 2,\mathbb{R})_\mathbb{C}$ Lie operators,
vacuum function and differential forms for extended boson expressed 
in terms of the
$\frac{Sp(2N \!+\! 2\!,\mathbb{R})_\mathbb{C}}{U(N \!+\! 1)}$ 
coset variables.


\newpage

\setcounter{equation}{0}
\renewcommand{\theequation}{\arabic{section}.\arabic{equation}}

\section{$\mathfrak{Jacobi~hsp}$ ALGEBRA OF BOSON OPERATORS AND
EXTENSION OF BOSON BOGOLIUBOV TRANSFORMATION}

\vspace{-0.5cm}

\def\bra#1{{<\!#1\,|}} 
\def\ket#1{{|\,#1\!>}}

~~We consider a  boson system with 
$N$ single-particle states.
Let $a_i$ and $a^{\dag}_i (i \!=\! 1, \!\cdots\!, N)$ be 
the annihilation-creation operators satisfying 
the canonical commutation relation for the boson\\[-22pt]
\begin{eqnarray}
[a_i,~a^{\dag }_j ] = \mathbb{I}\delta _{ij},~~
[a_i,~ a_j ] = 0,~~
[a^{\dag }_i,~a^{\dag }_j ] = 0 ,~
\mathfrak{h}
=
\{a_i,~a^{\dag }_i,~\mathbb{I}\} ,
\label{commutationrelation}
\end{eqnarray}\\[-22pt]
The $\mathfrak{h}$ spans
the $\mathfrak{Heizenberg}~\mbox{algebra}$.
$\!\!$The Roman indices $i,j, \cdots\!$ denote 
the given $\!N\!$ single-particle states.
Following Gilmore
\cite{Gilmore.74},
Zhang and Feng
\cite{ZhangFengGilmore.90},
the {\em two-photon} algebra is spanned by the following operators:\\[-26pt] 
\beqa
\left.
\BA{l}
a^{\dag }_i a^{\dag }_j
\equiv
E^{i j} ,~~
a^{\dag }_i a_j + \frac{1}{2} \mathbb{I} \delta _{ij}
\equiv
E^{i }_{~j },~~
a_i a_j
\equiv
E_{i j},~~
a^{\dag }_i,~a_j,~~\mathbb{I} , \\
\\[-8pt]
E^{i }_{~j }
=
E^{\dag j }_{~i },~
E^{i j}
=
E^{j i},~
E_{i j}
=
E_{j i},~
E^{\dag i j}
=
E_{j i} ,
\EA
\right\}
\label{twophotonalgebra}
\eeqa\\[-14pt]
and introduce another notation of the standard Cartan basis
$
\{\!{\bf E}_{e_i + e_j},{\bf E}_{- e_i - e_j},{\bf E}_{e_i - e_j},H_i~
\mbox{if}~i \!=\! j \!\}
$
of the {\em two-photon} algebra.
They pointed out that the operators 
(\ref{twophotonalgebra})
can be regarded as a graded algebra 
with grading $d$ and made the following identifications: \\[-26pt]
\begin{eqnarray}
\left.
\begin{array}{rl}
d
=
2
=
&\!\!\!\!
(2,2,2):~
a^{\dag }_i a^{\dag }_j~({\bf E}_{e_i + e_j}),~~
a^{\dag }_i a_j + {\displaystyle \frac{1}{2}} \mathbb{I} \delta_{ij}
~({\bf E}_{e_i - e_j},~H_i~\mbox{if}~i = j),~~
a_i a_j~({\bf E}_{- e_i - e_j}),\\
\\[-12pt]
d
=
1
=
&\!\!\!\!
(1,1):~
a^{\dag }_i~({\bf E}_{e_i - e_{{}_{N + 1}}}),~~
a_i~({\bf E}_{- e_i - e_{{}_{N+1}}}),\\
\\[-10pt]
d
=
0
=
&\!\!\!\!
(0):~
\mathbb{I}~({\bf E}_{-2e_{{}_{N+1}}}) ,
\end{array}
\right\}
\label{operatorset} 
\end{eqnarray}\\[-14pt]
whose algebra has dimension $(2N\!+\!1) \!\times\! (N\!+\!1)$
and whose generators can be identified with 
a subset of generators of the $\mathfrak{Jacobi~hsp}$ group.
Due to the commutation relation
(\ref{commutationrelation}),
the commutation relations for the boson operators
(\ref{operatorset})
in the $\mathfrak{Jacobi~hsp}$ algebra are \\[-26pt]
\beqa
\!\!\!\![E^{i }_{~j },~E^{k }_{~l }]
=
\delta_{k j }E^{i }_{~l } 
- 
\delta_{i l}E^{k }_{~j },~~
(\mathfrak{u}(N)~\mbox{algebra})
\label{commurel1}
\eeqa
\vspace{-1.0cm}
\beqa
\left.
\BA{ll}
&[E^{i }_{~j},~E_{k l }]
=
-
\delta_{i l }E_{j k } 
- 
\delta_{i k }E_{j l },~~
[E^{i }_{~j},~E^{l k }]
=
\delta_{j l }E_{ k i} 
+ 
\delta_{j k }E_{l j }, \\
\\[-8pt]
&[E^{i j },~E_{k l }]
=
-
\delta_{i l }E^{j }_{~k } 
- 
\delta_{j k }E^{i }_{~l}
-
\delta_{i k }E^{j }_{~l } 
-
\delta_{j l }E^{i }_{~k }, \\
\\[-8pt]
&[E_{i j },~E_{k l }]
=
0,~~
[E^{i j },~E^{k l }]
=
0,
\EA
\right\}
\label{commurel2}
\eeqa
\vspace{-0.7cm}
\beqa
\left.
\BA{ll}
&[a_{i },~E^{j }_{~k }]
=
\delta_{i j }a_{k },~~
[E^{k }_{~j },~a^{\dag }_{i}]
=
\delta_{i j }a^{\dag }_{k },\\
\\[-8pt]
&[a_{i },~E_{j k }]
=
0 ,~~
[E^{ k j },~a^{\dag }_{i}]
=
0 ,\\
\\[-8pt]
&[a_{i },~E^{j k }]
=
\delta_{i j }a^{\dag }_{k }
+
\delta_{i k}a^{\dag }_{j },~~
[E_{k j },~ a^{\dag }_{i}]
=
\delta_{i j }a_{k }
+
\delta_{i k}a_{j } .
\EA
\right\}
\label{commurel3}
\eeqa\\[-10pt]
The $\mathfrak{Jacobi~hsp}$ algebra of the boson operators 
(\ref{operatorset})
contains
$\mathfrak{u}(N) ( \!=\! \{E^{i }_{~j }\})$ and 
$\mathfrak{sp}(2N\!,\mathbb{R})_\mathbb{C}
( \!=\! \{E^{i }_{~j }, E^{i j }, E_{i j}\} )$ 
Lie algebras of pair operators 
as sub algebras.
The smallest faithful matrix representation of
an extended {\em two-photon} algebra is a 
$(2N \!+\! 1) \!\times\! (2N \!+\! 1)$
matrix representation of the algebra.
All the above algebras often called the 
$\mathfrak{Jacobi}$ algebra
have been investigated intensively by Eichler-Zagier
\cite{EichlerZagier.85}
and Berndt-Schmidt
\cite{BerndtSchmidt.98}.
The $\mathfrak{Jacobi}$ algebra
deals with the complex Heisenberg Lie algebra and
with the Lie algebra
$
\mathfrak{sp}(2N\!,\mathbb{R})_\mathbb{C}
\approx
\mathfrak{sp}(2N\!,\mathbb{C})
\cap
\mathfrak{u}(N\!,N)
$.
Rowe and Kramer have also studied the 
$\mathfrak{sp}(2N,\mathbb{R})_\mathbb{C}$ algebra
\cite{Rowe.84,Kramer1.82,Kramer2.82}.
In this paper,
we introduce the $2N \!\!+\!\! 1$-dimensional row vector
$\!\left[(a_i), (a^{\dag }_i),\!
\frac{1}{\sqrt{2}} \mathbb{I} \right]$
and $N$-dimensional unit matrix $1_{\!N}$.

$\!\!\!\!$In the above graded algebra,
we furthermore notice the role of the $(- \! 1)^n\!$
where $n$ is the boson number operator
$n = a^{\dag }_i a_i$
(Einstein summation convention over repeated indices). 
As was already pointed by the Ozaki's original idea
\cite{Ozaki.08},
the operator
$(\!- 1\!)^n$ anti-commutes with 
$a_i$ and $a^{\dag }_i$,\\[-20pt]
\begin{eqnarray}
\{ a_i,~(- 1)^n \} 
= 
\{ a^{\dag }_i,~(- 1)^n \} = 0 .
\label{anticommutation}
\end{eqnarray}
This is proved by using the relations
$a_i f(n_i) \!=\! f(n_i \!+\! 1)a_i$
and
$a^\dag_i f(n_i) \!=\! f(n_i \!-\! 1)a^\dag_i,~n_i  \!=\! a^\dag_i a_i$
$(\mbox{not summed for}~i)$
for any function
$\!f(n_i)\!$
\cite{HwaNuyts.66}.
The operator
$\!(- 1)^n\!$
makes a crucial role to generate
the generalized boson Bogoliubov transformation (TR)
quite parallel to
the generalized fermion Bogoliubov TR,
$SO(2N \!+\! 1)$ canonical TR
by Fukutome
\cite{Fuk.81}.

In order to realize 
a matrix representation,
first we introduce a Berezin-type operator
$
\Theta 
\!\!\equiv\!\!
a^{\dag }_i \theta ^B _i \!-\!  a_i \overline{\theta} ^{B}_i 
\!\!$
\cite{Berezin.66},
a free-boson vacuum $\ket {\!0\!}$
as
$a_i \ket {\!0\!} \!\!=\!\! 0$
and a TR $U(G)$ 
defined by\\[-20pt]  
\begin{eqnarray}
\!\!\!\!
U(G)
\!\equiv\!
e^{\Theta }\!e^{\Lambda }\!e^{\Gamma }
\!\!=\!\!
U(\!G_X\!)U(\!F_\lambda\!)U(\!F_u\!) ,
\Gamma
\!\equiv\!
a^{\dag }_i \gamma ^B _{ij} a_j 
\left( \!
\gamma ^{B\dag } 
\!\!=\!\!
- 
\gamma ^B \!
\right) \! , 
\Lambda
\!\equiv\!\!
{\displaystyle \frac{1}{2}} \!
\left( \!
a^{\dag }_i \lambda ^B _{ij} a^{\dag }_j 
\!\!-\!\! 
a_i \overline{\lambda}^{B} _{ij} a_j \!
\right) \!
\left( \!
\lambda ^{B\mbox{\scriptsize T}} 
\!\!=\!\! 
\lambda ^B \!
\right) \! .
\label{UGop}
\end{eqnarray}\\[-20pt]
The $U(G)$ acts on a boson state vector $\ket {\!\Psi\!}$ 
corresponding to a function 
$\Psi (G)$ in $G \!\in\! \mathfrak{Jacobi~hsp}$\\[-2pt]
as   
$
{\displaystyle
\ket {\!\Psi\!}
\!\!=\!\!\!\!
\int \!\!\! U \! (G) \ket {\!0\!} \bra {\!0\!} U^\dag \! (G) \ket {\!\Psi\!}dG
\!\!=\!\!\!\!
\int \!\!\! U \! (G) \ket {\!0\!} \Psi (G) dG
}
\!$,
shown in Appendixes A-D.
The symbols $\dag$,$\mbox{\scriptsize T}$ and overline
denote hermitian-,transpose- and
complex-conjugation of matrices and vectors.

Using the operator identity 
$
e^X\!Ae^{-X}
\!\!=\!\!
A
\!+\!
[X,A]
\!+\!
\frac{1}{2!}[X,[X,A]]
\!+\!
\cdots 
$,
we obtain\\[-20pt] 
\begin{eqnarray}
\begin{array}{c}
e^{\Gamma }a_ie^{-\Gamma }
\!=\!
a_j \overline{u}^{B}_{ji},~
e^{\Gamma }a^{\dag }_ie^{-\Gamma }
\!=\!
a^{\dag }_j u^B _{ji}, ~~
u^B \equiv \exp (\gamma ^B),~~
u^{B\dag }u^B \!=\! u^Bu^{B\dag } \!=\! 1_N ,
\end{array}
\label{UNmatrix}
\end{eqnarray}\\[-20pt]
where $u^B$ is a $U(N)$ matrix.
The $U(N)$ matrix $u^B$ is identified
with $u^B$ induced by $U(F_u)$.
The TR
(\ref{UNmatrix})
is the well known Thouless TR
\cite{Thouless.60}
for boson.
We also obtain\\[-20pt]
\begin{eqnarray}
\begin{array}{c}
e^{\Lambda}a_ie^{-\Lambda}
\!=\!
a_j [Ch(\lambda ^B)]_{ji} 
- 
a^{\dag }_j [Sh(\lambda ^B)]_{ji}, ~
e^{\Lambda}a^{\dag }_ie^{-\Lambda}
\!=\!
a^{\dag }_j [\overline{Ch}(\lambda ^B)]_{ji} 
- 
a_j [\overline{Sh}(\lambda ^B)]_{ji},
\end{array}
\label{Sp2NMatrix} 
\end{eqnarray}\\[-20pt]
where the
$Ch(\lambda ^B)$ and $Sh(\lambda ^B)$ are 
$N \!\!\times\!\! N$ matrices
defined in terms of a symmetric matrix $\lambda ^B$
as\\[-20pt]
\begin{eqnarray}
\left.
\begin{array}{rl}
Ch(\lambda ^B)
\!\equiv\!
&\!\!
\sum_{n=0}^{\infty }
{\displaystyle \frac{1}{(2n)!}}
(\bar{\lambda} ^{B}\lambda ^B)^n, ~~~~~~~~~~~
Ch^{\dag }(\lambda ^B)
=
Ch(\lambda ^B), \\
\\[-14pt]
Sh(\lambda ^B)
\!\equiv\!
&\!\!
\sum_{n=0}^{\infty }
{\displaystyle \frac{1}{(2n+1)!}}
\lambda ^B (\bar{\lambda} ^{B}\lambda ^B)^n ,~~
Sh^{\dag }(\lambda ^B)
=
\overline{Sh}(\lambda ^B).
\end{array}
\right\}
\label{coshsinh} 
\end{eqnarray}\\[-16pt]
The matrices $Ch(\lambda ^B)$ and $Sh(\lambda ^B)$ 
have the 
properties analogous to the hyperbolic functions\\[-16pt]
\begin{eqnarray}
\!\!\!\!\!\!\!\!\!\!\!\!
\left.
\begin{array}{rl}
&
Ch^{\dag } (\lambda ^B) Ch(\lambda ^B)
\!-\!
Sh^{\dag } (\lambda ^B) Sh(\lambda ^B)
\!=\!
1_{\!N}, 
Ch(\lambda ^B) Ch^{\dag } (\lambda ^B)
\!-\!
\overline{Sh}(\lambda ^B) Sh^{\mbox{\scriptsize T}} (\lambda ^B)
\!=\!
1_{\!N},\\ 
\\[-6pt]
&
Sh^{\mbox{\scriptsize T}} (\lambda ^B) Ch(\lambda ^B)
\!-\!
Ch^{\mbox{\scriptsize T}} (\lambda ^B) Sh(\lambda ^B)
\!=\!
0, 
Ch(\lambda ^B) Sh^{\dag } (\lambda ^B)
\!-\!
\overline{Sh}(\lambda ^B) Ch^{\mbox{\scriptsize T}} (\lambda ^B)
\!=\!
0. 
\end{array}
\right\} 
\label{CS condition}
\end{eqnarray}\\[-10pt] 
Let $[(b_i), (b^{\dag }_i)]$
be the $N$-dimensional row vector.$\!$
We define the $\!N \!\times\! N\!$ matrices
$a  \!\!=\!\! (a_{ij}\!)$
and
$b  \!\!=\!\! (b_{ij}\!)$
and the $2N \!\times\! 2N$ matrix $F$
identified with the $F$
in 
$U(F) \!=\! U(F_\lambda)U(F_u)~
(\ref{UGop}),
(F \!=\! F_\lambda F_u)$, 
by\\[-18pt] 
\begin{eqnarray}
\!\!\!\!\!\!\!\!
\left.
\begin{array}{rl}
&\!\!
[b,~b^{\dag }]
=
[a,~a^{\dag }]F ,~~
a^B
=
Ch(\lambda ^B) \overline{u}^{B}, ~~
b^B
=
-Sh(\lambda ^B)\bar{u}^{B}, \\
\\[-6pt]
&\!\!
F
\!\!=\!\!
\left[ \!\!
\begin{array}{cc}
Ch(\lambda ^B)  & -\overline{Sh}(\lambda ^B)  \\
\\[-6pt]
-Sh(\lambda ^B)  & \overline{Ch}(\lambda ^B)    
\end{array} \!\!
\right] \!
\left[ \!\!
\begin{array}{cc}
\overline{u}  & 0  \\
\\[-4pt]
0    & u    
\end{array} \!\!
\right] 
\!\!=\!\!
\left[ \!\!
\begin{array}{cc}
a^B  &\!\! \overline{b}^{B}  \\
\\[-8pt]
b^B  &\!\! \overline{a}^{B}    
\end{array} \!\!
\right] , \!\!\!\!
\begin{array}{cc}
&\!\!
a^{B\dag }a^B \!\!-\!\! b^{B\dag }b^B
\!\!=\!\!
1_{\!N}, ~
a^Ba^{B\dag } \!\!-\!\! \overline{b}^{B}b^{B\mbox{\scriptsize T}}
\!\!=\!\!
1_{\!N}, \\ 
\\[-8pt]
&\!\!\!\!\!
b^{B\mbox{\scriptsize T}}a^B \!\!-\!\! a^{B\mbox{\scriptsize T}}b^B
\!\!=\!\!
0, ~
a^Bb^{B\dag } \!\!-\!\! \overline{b}^{B}a^{B\mbox{\scriptsize T}}
\!\!=\!\!
0. 
\end{array}
\end{array} \!\!\!\!\!\!
\right\} 
\label{matrixF}
\end{eqnarray}\\[-10pt]
The $U(F)$ induces a linear TR of
the complex bases 
$a_i$
and
$a^{\dag }_i$, i.e.,
the generalized boson Bogoliubov TR
given by an $Sp(\!2N\!,\mathbb{R})_\mathbb{C}$ matrix $F$
which is not unitary but satisfies\\[-18pt]
\begin{eqnarray}
F^{\dag }
\widetilde{I}_{2N}
F
=
F
\widetilde{I}_{2N}
F^{\dag }
=
\widetilde{I}_{2N} ,~~
\widetilde{I}_{2N}
\equiv
\left[ \!\!\!
\begin{array}{cc}
1_{\!N}  &\!\! 0  \\
\\[-8pt]
0    &\!\! -1_{\!N}     
\end{array} \!\!\!
\right] ,~~
\det F = 1,
\label{Fmat}
\end{eqnarray}\\[-12pt]
and can be transformed to
a real $Sp(2N, \mathbb{R})_\mathbb{C}$ matrix
by the transformation
$
S
\!=\!
V F V^{-1},
V
\!\equiv\!
\left[ \!\!\!\!
\begin{array}{cc} 
 \frac{i}{\sqrt{2}} 1_{\!N}  &\!\! 
 \frac{i}{\sqrt{2}} 1_{\!N}  \\
\\[-12pt]
-\frac{1}{\sqrt{2}} 1_{\!N}  &\!\! 
\frac{1}{\sqrt{2}} 1_{\!N} 
\end{array} \!\!\!\!
\right] \! .   
$
The $\Theta$ in
(\ref{UGop})
has
$\theta^B_i$ and $\overline{\theta }^B_i$
(anti-commuting Grassmann variables)
\cite{Casalbuoni1.76,FSS.2000}. 
This $\Theta$ is essentially different from a  $\Theta^\prime$ 
suggested by Ozaki
who uses no such variables
\cite{Ozaki.08}.
Instead, we should require that the variables
$\theta^B_i$
and
$\!\overline{\theta}^B_i\!$
even more anti-commute with
$a_i$ and $a^\dag_i\!$,\\[-8pt]
\beq
\{\theta^B_i, \theta^B_j\}
\!=\!
0 ,~
\{\theta^B_i, \overline{\theta}^B_j\}
\!=\!
0 ;~
\{\theta^B_i, a_j\}
\!=\!
0 ,~
\{\theta^B_i, a^\dag_j\}
\!=\!
0 ;~
\{\overline{\theta}^B_i, a_j\}
\!=\!
0 ,~
\{\overline{\theta}^B_i, a^\dag_j\}
\!=\!
0 .
\label{Grassmann}
\eeq\\[-12pt]
A possible quantization of fermion TDHB theory
in terms of Grassmann variables
based on the $SO(2N\!+\!1)$ algebra 
has also been made 
by Yamamura
\cite{Yamamura.80}.
Due to the relations 
(\ref{Grassmann})
which lead to
$
\Theta ^{2}
\!=\!
\overline{\theta }^B_i \! \theta^B_i
\!\equiv\!
\overline{\theta }^B \! \theta^B
$
and using the formal expression for
square root of
$\overline{\theta }^B \! \theta^B$,
we have\\[-14pt]
\beqa
\left.
\BA{ll}
e^\Theta
\!=\!
Z^B 
\!+\!
a^\dag_i X^B_i
\!-\!
 a_i \overline{X}^B_i
\!\equiv\!
Z^B 
\!+\!
\Theta_X
\left( \!
\Theta^\dag_X
\!=\!
-\Theta_X \!
\right) \! ,~
(Z^B)^2 \!-\! \Theta_X^2
\!=\!
(Z^B)^2  \!-\! \overline{X}^B_i \! X^B_i \!=\! 1 ,\\
\\[-14pt]
Z^B
\!=\!
\cosh |\theta| ,~~
X^B_i
\!=\!
{\displaystyle \frac{\theta^B_i }{|\theta|}}\!  \sinh |\theta| ,~~
|\theta|
 =
\sqrt{\overline{\theta }^B \! \theta^B} .
\EA \!\!
\right\}
\label{theta}
\eeqa\\[-12pt]
The even-Grassmann variable
$\overline{\theta }^B \! \theta^B$
corresponds to the density of
the probability of finding 
\underline{the violation of boson number conservation by odd numbers}
\cite{FYN.77,FukNishi.84}
and relates to the occupation number.
See Friedrichs
\cite{Friedrichs.53}
and
Berezin
\cite{Berezin.66}.
Then, the use of the $\overline{\theta }^B \theta^B$ 
meaning such the density and the occupation number
is recognized to be appropriate.
See 
Baez $et~al$.$\!$
\cite{Baez.92}.
The $\dagger$ operation on both the
$a^\dag_i \theta^B_i$ and $a_i \overline{\theta }^B_i$ 
is made as the usual $\dagger$ operation on both the
$a_i \overline{\theta }^B_i$ and $a^\dag_i \theta^B_i$
in the case of the boson.
The appearance of the hyperbolic functions
$\cosh  |\theta|$ and $\sinh  |\theta|$
is due to the introduction of
the anti-commuting Grassmann variables
$\theta^B_i$ and $\overline{\theta }^B_i$
for which we demand that they commute with $(-1)^n$.
From
(\ref{commutationrelation}),  (\ref{anticommutation}),
(\ref{Grassmann})
and
(\ref{theta}),
thus we obtain\\[-16pt]
\begin{eqnarray}
\begin{array}{rl}
&
e^{\Theta }
{\displaystyle \frac{1}{\sqrt{2}}}
(-1)^n e^{-\Theta }
\!=\!
{\displaystyle \frac{1}{\sqrt{2}}}
(Z^B \!+\! \Theta_X) (Z^B \!+\! \Theta_X) \! (-1)^n \\
\\[-10pt]
&
\!=\!
\left\{ \!
a_j \!
\left( \! -\sqrt{2}Z^B  \overline{X}^{B} _j \! \right) 
\!+\!
a^{\dag }_j \!
\left(\sqrt{2}Z^B X^B _j\right)
\!+\!
{\displaystyle \frac{1}{\sqrt{2}}} 
\left( \! 2 \! \left(Z^{B}\right)^2 \!\!-\!\! 1\! \right) \!
\right\} \!
(-1)^n  ,
\end{array}
\label{trans(-1)n} 
\end{eqnarray}\\[-10pt]
which shows
mixing of
$a_i (-1)^n$
and
$a^{\dag }_i (-1)^n$.
Then, we need to know transformation rules for both the
$e^{\Theta } a_i (-1)^n e^{-\Theta }$
and
$e^{\Theta } a^{\dag }_i  (-1)^n e^{-\Theta }$.
For the purpose of realizing the rules,
we notice the commutation relations
$ 
[a_i, Z^B \!+\! \Theta_X ]
\!=\!
- X^B_i
\!+\!
2 a_i \Theta_X 
$
and
$ 
[a^\dag_i, Z^B \!+\! \Theta_X ]
\!=\!
-\overline{X}^B_i
\!+\!
2 a^\dag_i \Theta_X. 
$
They play essential roles.
First
the former transformation is exactly calculated
as follows:\\[-14pt]
\begin{eqnarray}
\begin{array}{rl}
&
e^{\Theta } a_i (-1)^n e^{-\Theta }
\!=\!
a_i (-1)^n
+
X^B_i
(Z^B \!+\! \Theta_X)
(-1)^n \\
\\[-8pt]
\!=\!
&
\left\{ \!
a_j \!
\left( \!
\delta_{ji} 
\!-\!
\overline{X}^{B} _j X^B _i
\right) 
\!+\!
a^{\dag }_j \!
\left( X^B _j X^B _i\right)
\!+\!
{\displaystyle \frac{1}{\sqrt{2}}} 
\left( \sqrt{2} Z^B X^B _i\right) \!
\right\} \!
(-1)^n  .
\end{array}
\label{transa(-1)n} 
\end{eqnarray}\\[-10pt]
Similarly, we compute
the transformation rule for the latter one as\\[-16pt]
\begin{eqnarray}
\begin{array}{rl}
&
e^{\Theta } a^{\dag }_i (-1)^n e^{-\Theta }
\!=\!
a^\dag_i (-1)^n 
\!+\!
\overline{X}^B_i 
(Z^B \!+\! \Theta_X)
(-1)^n \\
\\[-6pt]
\!=\!
&
\left\{ \!
a_j \!
\left( -\overline{X}^B _j \overline{X}^B _i\right)
\!+\!
a^{\dag }_j \!
\left( \!
\delta_{ji} 
\!+\!
X^B _j \overline{X}^B _i
\right) 
\!+\!
{\displaystyle \frac{1}{\sqrt{2}}} 
\left( \sqrt{2} Z^B \overline{X}^B _i\right) \!
\right\} \!
(-1)^n .
\end{array}
\label{transadag(-1)n} 
\end{eqnarray}\\[-20pt]

Expressing 
(\ref{trans(-1)n}),
(\ref{transa(-1)n})
and
(\ref{transadag(-1)n})
in a lump,
we can obtain\\[-16pt] 
\beqa
\BA{c}
e^\Theta \!
\left[ a_i, ~a^\dag _i, 
~{\displaystyle \frac{1}{\sqrt{2}}}\mathbb{I}\right] \! (-1)^n e^{-\Theta }
\!=\!
\left[ a_j, ~a^\dag _j, 
~{\displaystyle \frac{1}{\sqrt{2}}}\mathbb{I}\right] \! (-1)^n G_{Xji} ,
\EA
\label{chiraloptrans}
\eeqa\\[-10pt] 
where
$G_{Xji}$
is defined as\\[-16pt] 
\beqa
\BA{c}
G_{Xji} 
\!\stackrel{\mathrm{def}}{=}\!
\left[ \!\!
\BA{ccc} 
\delta_{ji}  
\!-\! 
\overline{X}^B_j X^B_i &
-\overline{X}^B_j \overline{X}^B_i & 
-\sqrt{2}Z^B \overline{X}^B_j \\
\\[-8pt]
X^B_j X^B_i &
\delta_{ji}  
\!+\!
X^B_j \overline{X}^B_i & 
\sqrt{2}Z^BX^B_j  \\
\\[-8pt]
\sqrt{2}Z^BX^B_i & \sqrt{2}Z^B \overline{X}^B_i &
2 \! \left(Z^B\right)^2 \!-\! 1 
\EA \!
\right] \! .
\EA
\label{chiraloptrans2}
\eeqa\\[-16pt]
Let $G$ be a $(2N+1) \!\times\! (2N+1)$ matrix defined by\\[-16pt]
\begin{eqnarray}
\!\!\!\!
\begin{array}{rl}
G
&\!\equiv\!
G_X \!
\left[ \!\!
\begin{array}{ccc}
a^B& \overline{b}^{B}& 0 \\ \\[-6pt]
b^B& \overline{a}^{B}& 0 \\ \\[-6pt]
0  & 0     & 1  
\end{array} \!\!
\right] 
\!=\!
\left[ \!\!
\begin{array}{ccc}
a^B \!-\! \overline{X}^{B} Y^B & 
\overline{b}^{B} \!-\! \overline{X}^{B} \overline{Y}^{B}&
-\sqrt{2}Z^B \overline{X}^{B}\\ \\[-6pt]
b^B \!+\! X^B Y^B    & 
\overline{a}^{B} \!+\! X^B \overline{Y}^{B}   & 
\sqrt{2}Z^B X^B \\ \\[-6pt]
\sqrt{2}Z^B Y^B & \sqrt{2}Z^B \overline{Y}^{B} &
2 \! \left(Z^B\right)^2 \!-\! 1 
\end{array} \!
\right] 
\end{array}
\label{(2N+1)Gmatrix}
\end{eqnarray}\\[-10pt]
where $a^B\!$ and $b^B$
appear in $\!F\!$
(\ref{matrixF}) and $Y^{\!B}$
is a new row vector
with the essential difference having
a linear combination with minus sign 
in the fermion row vector $\!Y^{\!F}\!$
used in the $SO(2N\!\!+\!1)$ canonical transformation
\cite{FYN.77},
which is given by\\[-16pt]
\begin{eqnarray}
Y^B_i
\!=\!
X^B _j a^B_{ji} \!+\! \overline{X}^{B}_j b^B_{ji},~~
\left(Z^{B}\right)^2
\!-\!  
\overline{Y}^B _i Y^B _i
\!=\! 1.
\label{Yvector}
\end{eqnarray}\\[-20pt]
Introducing new matrices $A^B$ and $B^B$ and
a column vector $x^B$ and a row vector $y^B$ and
a scalar $z^B$,
the matrix $G$ is rewritten with the use of
the Grassmann variables $X^B$ and $Y^B$ as\\[-16pt]
\begin{eqnarray}
\!\!\!\!
\begin{array}{l}
G
\!=\!
\left[ \!
\begin{array}{ccc}
A^B                  & \overline{B}^{B}                  &
{\displaystyle -\frac{\overline{x}^{B}}{\sqrt{2}}} \\
B^B                  & \overline{A}^{B}                   & 
{\displaystyle \frac{x^B}{\sqrt{2}}} \\
{\displaystyle \frac{y^B}{\sqrt{2}}} & 
{\displaystyle \frac{\overline{y}^{B}}{\sqrt{2}}} & z^B  
\end{array} \!
\right] \! , 
\left.
\BA{l}
A^B 
\equiv 
a^B - \overline{X}^{B} Y^B
=
a^B
-
{\displaystyle \frac{\overline{x}^{B} y^B}{2(1+z^B)}} ,~\\
\\[-10pt]
B^B
\equiv 
b^B + X^B Y^B
=
b^B
+
{\displaystyle \frac{x^B y^B}{2(1+z^B)}} ,~\\
\\[-6pt]
x^B \!\equiv\! 2Z^B X^B,~\!y^B \!\equiv\! 2Z^B Y^B \! ,~\!
z^B \!\equiv\! 2 \! \left(Z^B\right)^2 \!-\! 1,
\end{array} \!
\right\}
\end{array}
\label{new(2N+1)Gmatrix}
\end{eqnarray}\\[-6pt]
which is formally almost the same form as
the $SO(2N\!+\!1)$ matrix
\cite{FYN.77}
and a form suggested by Ozaki
\cite{Ozaki.08}. 
Here we have used
the complex conjugation rule for products of
$\!\overline{X}^{B} Y^B\!$ and $\!X^B Y^B\!$
for Grassmann variables $\!X^B\!$ and $\!Y^B\!$.
This kind of rule is provided explicitly by Berezin
\cite{Berezin.66}:\\[-18pt] 
\beqa
\BA{l}
\overline{\overline{X}^{B} \! Y^B}
=
\overline{Y}^{B} \! X^B
=
- X^B \overline{Y}^{B} \! , ~~
\overline{X^B \! Y^B}
=
\overline{Y}^{B} \overline{X}^{B}
=
- \overline{X}^{B} \overline{Y}^{B} \! .
\end{array} 
\label{(2N+1)Amplitudes}
\end{eqnarray}\\[-20pt]
By the transformation
$U(G)=e^{\Theta }e^{\Lambda }e^{\Gamma }$
for bosons $[a, a^\dag,\frac{1}{\sqrt{2}} \mathbb{I}]$,
thus we obtain\\[-16pt] 
\begin{eqnarray}
\begin{array}{c}
U(G) \!
\left[a_i,~a^{\dag }_i,{\displaystyle \frac{1}{\sqrt{2}}}\mathbb{I}\right] \!
(-1)^n
U^{-1}(G)
=
\left[a_j,~a^{\dag }_j,{\displaystyle \frac{1}{\sqrt{2}}}\mathbb{I}\right](-1)^n \!
\left[ \!
\begin{array}{ccc}
A^B_{ji}             & \overline{B}^{B}_{ji}              &
{\displaystyle -\frac{\overline{x}^{B}_j}{\sqrt{2}}} \\
B^B_{ji}             & \overline{A}^{B}_{ji}              & 
{\displaystyle \frac{x^B_j}{\sqrt{2}}} \\
{\displaystyle \frac{y^B_i}{\sqrt{2}}} & 
{\displaystyle \frac{\overline{y}^{B}_i}{\sqrt{2}}} & z^B  
\end{array} \!
\right] .
\end{array}
\label{transU(G)aadag(-1)n}
\end{eqnarray}\\[-18pt]

Let us introduce new bosons 
$[b, b^\dag,\!\frac{1}{\sqrt{2}} \mathbb{I}]$
defined as
$
[b, b^\dag,\!\frac{1}{\sqrt{2}} \mathbb{I}]
\equiv\!
U(G)[a, a^\dag,\!
\frac{1}{\sqrt{2}} \mathbb{I}]U^{-1}(G)
$
and a new operator $\Theta_x$
defined as
$\Theta_x
\!\equiv\!
a^{\dag }_i x^B _i - a_i \overline{x}^{B}_i
$
where
$x^B _i$ and $\overline{x}^{B}_i$
are Grassmann variables.
We derive the normalization condition
governing the transformation parameters.
For this aim,
we further introduce new boson operators
$\alpha_i$ and $\alpha^\dag_i$, 
$
\alpha_i
\!=\!
a_j A^{B}_{ji}+a^{\dag }_j B^{B}_{ji} 
$
and
$
\alpha^{\dag }_i
\!=\!
a_j \overline{B}^{B }_{ji}+a^{\dag }_j \overline{A}^{B}_{ji} 
$.
First we compute the commutators
$\alpha_i$
and
$\alpha^\dag_i$
between
$\Theta_x$.
Using the anti-commutators
(\ref{Grassmann})
and the first equation of
(\ref{Yvector}),
the computations are made as follows:\\[-18pt]
\begin{eqnarray}
\begin{array}{l}
[\alpha_i, \Theta_x]
\!=\!
[a_j A^{B}_{ji}+a^{\dag }_j B^{B}_{ji},~
a^{\dag }_k x^B_k -  a_k \overline{x}^B_k] 
\!=\!
2 \alpha_i \Theta_x
-
(x^B_j A^{B}_{ji}+\overline{x}^B_j B^{B}_{ji}) .
\end{array} 
\label{alpha_Theta}
\end{eqnarray}\\[-16pt]
On the other hand, using the relations in
(\ref{new(2N+1)Gmatrix}),
the last term of
(\ref{alpha_Theta})
is rewritten as\\[-18pt]
\begin{eqnarray}
\begin{array}{l}
x^B_j A^{B}_{ji}+\overline{x}^B_j B^{B}_{ji}
=
y^B_i
\!+\!
2 \overline{X}^B_j \! X^B_j
y^B_i 
=
( 2 (Z^B)^2 -1)
y^B_i
=
z^B y^B_i .
\end{array} 
\label{zy}
\end{eqnarray}\\[-18pt]
Then we get the commutation relation
$
[\alpha_i, \Theta_x]
\!=\!
2 \alpha_i \Theta_x
\!-\!
z^B y^B_i  
$.
Using
(\ref{transU(G)aadag(-1)n2}),
$[ \Theta_x, y^B_i ] \!=\! 0$
and the relation
$\alpha_i \Theta_x \!=\! - \Theta_x \alpha_i \!+\! z^B y^B_i$,
we have\\[-14pt]
\begin{eqnarray}
\begin{array}{l}
b_i
\!\!=\!\!
\left( \! 
\alpha_i \!+\! {\displaystyle \frac{1}{2}} y^B_i \!
\right) \!
(z^B \!-\! \Theta_x) 
\!=\!
(z^B \!+\! \Theta_x) \!
\left( \!
\alpha_i \!-\! {\displaystyle \frac{1}{2}} y^B_i \!
\right) \! ,
\end{array} 
\label{b_alpha}
\end{eqnarray}\\[-10pt]
while we have
$
[\alpha^\dag_i, \Theta_x]
\!=\!
2 \alpha^\dag_i \Theta_x 
\!-\!
z^B \overline{y}^B_i 
$,
which is also derived in the similar way as the one
made in 
(\ref{alpha_Theta}).
Using
(\ref{transU(G)aadag(-1)n2}),
$[ \Theta_x, \overline{y}^B_i ] \!=\! 0$
and
the relation
$
\alpha^\dag_i \Theta_x
\!=\!
- \Theta_x \alpha^\dag_i \!+\! z^B \overline{y}^B_i
$,
we also have\\[-12pt]
\begin{eqnarray}
\begin{array}{l}
b^\dag_i
\!\!=\!\!
\left( \!
\alpha^\dag_i \!+\! {\displaystyle \frac{1}{2}} \overline{y}^B_i \!
\right)  \!
(z^B \!-\! \Theta_x) 
\!=\!
(z^B \!+\! \Theta_x) \!
\left( \!
\alpha^\dag_i \!-\! {\displaystyle \frac{1}{2}} \overline{y}^B_i \!
\right) \! .
\end{array} 
\label{b_alpha2}
\end{eqnarray}\\[-10pt] 
To find the normalization conditions
of $A^{B}_{ij}, B^{B}_{ij}, x^B_i$ and $y^B_i$ in
(\ref{transU(G)aadag(-1)n2})
for the commutators $[b_i,b^\dag_j]$ and $[b_i,b_j]$,
we calculate the following commutators
with the aid of
(\ref{b_alpha}) and (\ref{b_alpha2}):\\[-14pt]
\begin{eqnarray}
\!\!\!\!
\begin{array}{ll}
&
[b_i,~b^\dag_j]
\!=\!
\left( \!\!
\alpha_i \!+\! {\displaystyle \frac{1}{2}} y^B_i \!\!
\right) \!\!
\left( \!\!
\alpha^\dag_j 
\!-\! 
{\displaystyle \frac{1}{2}} \overline{y}^B_j \!\!
\right) 
\!-\!
\left( \!\!
\alpha^\dag_j 
\!+\! 
{\displaystyle \frac{1}{2}} \overline{y}^B_j \!\!
\right) \!\!
\left( \!\!
\alpha_i \!-\! {\displaystyle \frac{1}{2}} y^B_i \!\!
\right)  \\
\\[-8pt]
&
\!=\!
[\alpha_i,~\alpha^\dag_j]
\!-\!
{\displaystyle \frac{1}{2}} \{\alpha_i,~ \overline{y}^B_j\}
\!+\!
{\displaystyle \frac{1}{2}} \{y^B_i,~ \alpha^\dag_j\}
\!-\!
{\displaystyle \frac{1}{4}} \!
\left( y^B_i \overline{y}^B_j 
\!-\! \overline{y}^B_j y^B_i \right)   
\!=\!
A^{B}_{ki} \overline{A}^{B}_{kj}
\!-\!
B^{B}_{ki} \overline{B}^{B}_{kj}
\!-\!
{\displaystyle \frac{1}{2}}
y^B_i \overline{y}^B_j 
\!=\!
\delta_{ij} ,
\end{array} 
\label{b_bcommus}
\end{eqnarray}
\vspace{0.1cm} 
\begin{eqnarray}
\!\!\!\!
\begin{array}{ll}
&
[b_i,~b_j]
\!=\!
\left( \!\!
\alpha_i \!+\! {\displaystyle \frac{1}{2}} y^B_i \!\!
\right) \!\!
\left( \!\!
\alpha_j \!-\! {\displaystyle \frac{1}{2}} y^B_j \!\!
\right) 
\!-\!
\left( \!\!
\alpha_j \!+\! {\displaystyle \frac{1}{2}} y^B_j \!\!
\right) \!\!
\left( \!\!
\alpha_i \!-\! {\displaystyle \frac{1}{2}} y^B_i \!\!
\right) \\
\\[-8pt]
&
\!=\!
[\alpha_i,~\alpha_j]
\!-\!
{\displaystyle \frac{1}{2}} \{\alpha_i,~ y^B_j\}
\!+\!
{\displaystyle \frac{1}{2}} \{y^B_i,~ \alpha_j\}
\!-\!
{\displaystyle \frac{1}{4}} \!
\left( y^B_i y^B_j \!-\! y^B_j y^B_i \right)   
\!=\!
A^{B}_{ki} B^{B}_{kj}
\!-\!
B^{B}_{ki} A^{B}_{kj}
\!-\!
{\displaystyle \frac{1}{2}}
y^B_i y^B_j 
\!=\!
0 .
\end{array} 
\label{b_bcommus2}
\end{eqnarray}\\[-6pt]
Here we have used 
$(z^B)^2 - \Theta_x^2 \!=\! 1$.
Then we get the following normalization conditions:\\[-16pt]
\begin{eqnarray}
\!\!\!\!
\left.
\begin{array}{ll}
&
\left(
A^{B\dag} \! A^{B}
\!-\!
B^{B\dag} \! B^{B}
\right)_{ij}
\!=\!
\delta_{ij}
\!-\!
{\displaystyle \frac{1}{2}}
y^{B\dag}_i y^B_j ,
(\mbox{due to (\ref{(2N+1)Amplitudes})}) , 
\left(
A^{B\mbox{\scriptsize T}} \! B^{B}
\!-\!
B^{B\mbox{\scriptsize T}} \! A^{B}
\right)_{ij}
\!=\!
{\displaystyle \frac{1}{2}}
y^{B\mbox{\scriptsize T}}_i y^B_j , \\
\\[-4pt]
&
\left(
x^{B\mbox{\scriptsize T}} \! A^{B} 
\!+\! 
x^{B \dag} \! B^{B}
\right)_i
\!=\!
z^B \! y^B_i,~ 
(z^B)^2 \!-\! x^{B\dag} x^B
\!=\! 1.
\end{array} \!\!
\right\} 
\label{normalization}
\end{eqnarray}\\[-8pt]
By using the new bosons
$\alpha_i$ and $\alpha^\dag_i$
and noticing the third column of the matrix in
(\ref{transU(G)aadag(-1)n}),
$
U(G) \!\!
\left[a_i,a^{\dag }_i,
\frac{1}{\sqrt{2}} \mathbb{I}\right] \!\!
U^{\!-1}\!(G),
(\mbox{at~the~region~near~}z^B 
\!\approx\! 
1~\mbox{and~then~}x^B \!\approx\! 0),
$
is approximated as\\[-10pt]
\begin{eqnarray}
\begin{array}{rl}
&\!\!\!\!\!\!\!\!
U(G) \!\!
\left[a_i,a^{\dag }_i,
{\displaystyle \frac{1}{\sqrt{2}}}\mathbb{I}\right] \!\!
U^{\!-1}\!(G) 
\!=\!
\left[
\left(\!\alpha_i 
\!+\!
{\displaystyle \frac{1}{2}} y^B_i\!\right) \!\!
\left(\!z^B \!-\! \Theta_x \!\right),
\left(\!\alpha^{\dag }_i
\!+\!
{\displaystyle \frac{1}{2}} \overline{y}^{B}_i\!\right) \!\!
\left(\!z^B \!-\! \Theta_x \!\right),
{\displaystyle \frac{1}{\sqrt{2}}}
\left(\!z^B \!+\! \Theta_x \!\right)
\right] \\
\\[-8pt]
&\!\!\!\!\!\!\!\!
\!=\!
\left[
\alpha_i \!+\! {\displaystyle \frac{1}{2}} y^B_i,
\alpha^{\dag }_i
\!+\!
{\displaystyle \frac{1}{2}} \overline{y}^{B}_i,
{\displaystyle \frac{1}{\sqrt{2}}}
\left(\!z^B \!+\! \Theta_x \!\right) \! 
\left(\!z^B \!+\! \Theta_x \!\right)
\right] \!\!
\left(\!z^B \!-\! \Theta_x \!\right) \\
\\[-8pt]
&\!\!\!\!\!\!\!\!
\!=\!
\left[
\alpha_i 
\!+\! 
{\displaystyle \frac{1}{2}} y^B_i,
\alpha^{\dag }_i
\!+\! 
{\displaystyle \frac{1}{2}} \overline{y}^{B}_i,
{\displaystyle \frac{1}{\sqrt{2}}} \!
\left\{\!2 (z^B)^2 \!-\! 1 \!+\! 2z^B \Theta_x\!\right\}\!
\right] \!\!
\left(\!z^B \!-\! \Theta_x \!\right) \\
\\[-8pt]
&\!\!\!\!\!\! 
\!\!\approx\!\!
\left[
\alpha_i 
\!+\! 
{\displaystyle \frac{1}{2}} y^B_i,
\alpha^{\dag }_i
\!+\! 
{\displaystyle \frac{1}{2}} \overline{y}^{B}_i,
{\displaystyle \frac{1}{\sqrt{2}}}z^B
\right] \!\!
\left(\!z^B \!-\! \Theta_x \!\right)  .
\end{array}
\label{ApproxtransU(G)aadag(-1)n}
\end{eqnarray}\\[-6pt]
The above approximation is justified by using
$(z^B)^2 - \Theta_x^2 \!=\! 1$ and
$z^B \!\approx\! 1~\mbox{and~then~}x^B \!\approx\! 0$.
Finally, we obtain the approximate expression
at $z^B \!\approx\! 1$, i.e., $x^B \!\approx\! 0$ 
for equation
(\ref{transU(G)aadag(-1)n}) as\\[-12pt]
\begin{eqnarray}
\begin{array}{l}
\left[b_i,~b^{\dag }_i,
{\displaystyle \frac{1}{\sqrt{2}}} \mathbb{I} \right]
\!=\!
\left[a_j,~a^{\dag }_j,
{\displaystyle \frac{1}{\sqrt{2}}} \mathbb{I} \right] \!\!
\left(z^B \!-\! \Theta_x \right) \!\!\!
\left[ \!\!\!
\begin{array}{ccc}
A^B_{ji}             & \overline{B}^{B}_{ji}              &
{\displaystyle -\frac{\overline{x}^{B}_j}{\sqrt{2}}} \\
B^B_{ji}             & \overline{A}^{B}_{ji}              & 
{\displaystyle \frac{x^B_j}{\sqrt{2}}} \\
{\displaystyle \frac{y^B_i}{\sqrt{2}}} & 
{\displaystyle \frac{\overline{y}^{B}_i}{\sqrt{2}}} & z^B  
\end{array} \!\!
\right] \! .
\end{array}
\label{transU(G)aadag(-1)n2}
\end{eqnarray}


\newpage

\setcounter{equation}{0}
\renewcommand{\theequation}{\arabic{section}.\arabic{equation}}

\section{INVERSE OF BOSON BOGOLIUBOV TRANSFORMATION}

\vspace{-0.5cm}

~~~~To derive an inverse transformation of
(\ref{transU(G)aadag(-1)n2}),
we introduce new boson operators
$\beta^\dag_i$ and $\beta^\dag_i$\\[-16pt]
\begin{eqnarray}
\beta_i
\!=\!
b_j A^{B\dag}_{ji}
\!+\!
b^{\dag }_j B^{B\mbox{\scriptsize T}}_{ji},~~
\beta^\dag_i
\!=\!
b_j B^{B\dag}_{ji}
\!+\!
b^{\dag }_j A^{B\mbox{\scriptsize T}}_{ji} .
\label{beta_b}
\end{eqnarray}\\[-16pt]
The inverse of the canonical transformation
(\ref{transU(G)aadag(-1)n2})
is made in the similar way to
(\ref{b_alpha})
and
(\ref{b_alpha2}).
Using the expression
(\ref{transU(G)aadag(-1)n2}),
$[\Theta_y, x^B_i] 
\!=\! 
[\Theta_y, \overline{x}^B_i] \!=\! 0$
and the relations
$
\beta_i \Theta_y
\!=\!
-\Theta_y \beta_i
\!+\!
z^B x^B_i
$
and
$
\beta^\dag_i \Theta_y
\!=\!
-\Theta_y \beta^\dag_i
\!+\!
z^B \overline{x}^B_i 
$,
we also have \\[-16pt]
\begin{eqnarray}
\left.
\begin{array}{ll}
&
a_i
\!=\!
\left( \!\!
\beta_i \!-\! {\displaystyle \frac{1}{2}} x^B_i \!\!
\right) \!
(z^B \!+\! \Theta_y) 
\!=\!
(z^B \!-\! \Theta_y) \!
\left( \!\!
\beta_i \!+\! {\displaystyle \frac{1}{2}} x^B_i \!\!
\right) \! ,
\\ \\[-10pt]
&
a^\dag_i
\!=\!
\left( \!\!
\beta^\dag_i
\!-\!
{\displaystyle \frac{1}{2}} \overline{x}^B_i \!\!
\right)  \!
(z^B \!+\! \Theta_y)  
\!=\!
(z^B \!-\! \Theta_y) \!
\left( \!\!
\beta^\dag_i
\!+\!
{\displaystyle \frac{1}{2}} \overline{x}^B_i \!\!
\right) \! .
\end{array} \!\!
\right\} 
\label{b_beta2}
\end{eqnarray}\\[-6pt] 
To find also the normalization conditions
of $A^{B}_{ij}, B^{B}_{ij}, x^B_i$ and $y^B_i$ in
(\ref{transU(G)aadag(-1)n2})
for the commutators $[a_i,a^\dag_j]$ and $[a_i,a_j]$,
we calculate the following commutators
with the aid of
(\ref{b_beta2}):\\[-16pt]
\begin{eqnarray}
\!\!\!\!\!\!\!\!
\left.
\begin{array}{ll}
&
[a_i,a^\dag_j]
\!=\!
[\beta_i,\beta^\dag_j]
\!\!+\!\!
{\displaystyle \frac{1}{2}} \{\beta_i,\overline{x}^B_j\}
\!\!-\!\!
{\displaystyle \frac{1}{2}} \{x^B_i,\beta^\dag_j\}
\!\!-\!\!
{\displaystyle \frac{1}{4}} \!
\left( x^B_i \overline{x}^B_j
\!\!-\!\!
\overline{x}^B_j x^B_i \right) 
\!\!=\!\!
A^{B\dag}_{ki} \! A^{B\mbox{\scriptsize T}}_{kj}
\!\!-\!\!
B^{B\mbox{\scriptsize T}}_{ki} \! B^{B\dag}_{kj}
\!\!-\!\!
{\displaystyle \frac{1}{2}}
x^B_i \overline{x}^B_j  
\!=\!
\delta_{ji} ,
\\ \\[-10pt]
&
[a_i,~a_j]
\!=\!
[\beta_i,\beta_j]
\!\!+\!\!
{\displaystyle \frac{1}{2}} \{\beta_i,x^B_j\}
\!\!-\!\!
{\displaystyle \frac{1}{2}} \{x^B_i,\beta_j\}
\!\!-\!\!
{\displaystyle \frac{1}{4}} \!
\left( x^B_i x^B_j
\!\!-\!\!
x^B_j x^B_i \right) 
\!\!=\!\!
A^{B\dag}_{ki} \! B^{B\mbox{\scriptsize T}}_{kj}
\!\!-\!\!
B^{B\mbox{\scriptsize T}}_{ki} \! A^{B\dag}_{kj}
\!\!-\!\!
{\displaystyle \frac{1}{2}}
x^B_i x^B_j 
\!=\!
0 .
\end{array} \!\!\!\!
\right\}  
\label{a_bcommus2}
\end{eqnarray}\\[-4pt] 
Then we get the following normalization conditions:\\[-18pt]
\begin{eqnarray}
\!\!\!\!
\left.
\begin{array}{ll}
&
\left( \!
A^{B} \! A^{B\dag}
\!-\!
\overline{B}^{B} \! B^{B\mbox{\scriptsize T}} \!
\right)_{\!ij}
\!=\!
\delta_{ij}
\!-\!
{\displaystyle \frac{1}{2}}
\overline{x}^B_i x^{B\mbox{\scriptsize T}}_j ,
(\mbox{due to (\ref{(2N+1)Amplitudes})}) , 
\left( \!
\overline{A}^{B} \! B^{B\mbox{\scriptsize T}}
\!-\!
B^{B} \! A^{B\dag} \!
\right)_{\!ij}
\!=\!
{\displaystyle \frac{1}{2}}
x^B_i x^{B\mbox{\scriptsize T}}_j , \\
\\[-6pt]
&
\left(
y^{B} \! A^{B\dag} \!+\! \overline{y}^{B} \!
B^{B\mbox{\scriptsize T}}
\right)_i
\!=\!
z^B \! x^{B\mbox{\scriptsize T}}_i  ,~
(z^B)^2 \!-\!  y^B y^{B\dag}
\!=\! 1.
\end{array} \!\!
\right\} 
\label{normalization2}
\end{eqnarray}\\[-12pt] 
By using another new bosons
$\beta_i$ and $\beta^\dag_i$, 
(\ref{beta_b}),
$
U(G) \!\!
\left[
b_i,~b^{\dag }_i,
\frac{1}{\sqrt{2}} \mathbb{I}
\right] \!\!
U^{\!-1}\!(G)
$
is approximated as\\[-14pt]
\begin{eqnarray}
\begin{array}{rl}
&\!\!\!\!\!\!\!\!
U(G) \!\!
\left[b_i,~b^{\dag }_i,
{\displaystyle \frac{1}{\sqrt{2}}} \mathbb{I}\right] \!\!
U^{\!-1}\!(G) \\
\\[-8pt]
&
\!=\!
\left[
\left( \beta_i \!-\! {\displaystyle \frac{1}{2}} x^B_i \right) \!
\left(z^B \!+\! \Theta_y \right),
\left( \beta^{\dag }_i
\!-\!
{\displaystyle \frac{1}{2}} \overline{x}^{B}_i \right) \!
\left(z^B \!+\! \Theta_y \right),
{\displaystyle \frac{1}{\sqrt{2}}}
\left(b^\dag_j y^B_j 
\!+\! 
b_j \overline{y}^{B}_j \!+\! z^B \right)
\right] \\
\\[-6pt]
&
\!=\!
\left[
\beta_i \!-\! {\displaystyle \frac{1}{2}} x^B_i,~
\beta^{\dag }_i
\!+\!
{\displaystyle \frac{1}{2}} \overline{x}^{B}_i,~
{\displaystyle \frac{1}{\sqrt{2}}} 
\right] \!
\left(z^B \!+\! \Theta_y \right)
\!+\!
\left[
0,~x^B_i \! \left(z^B \!+\! \Theta_y \right),
~\sqrt{2}\!~b_i \overline{y}^{B}_i
\right]\\
\\[-6pt]
&
\!\approx\!
\left[
\beta_i \!-\! {\displaystyle \frac{1}{2}} x^B_i,~
\beta^{\dag }_i
\!+\!
{\displaystyle \frac{1}{2}} \overline{x}^{B}_i,~
{\displaystyle \frac{1}{\sqrt{2}}}z^B
\right] \!
\left(z^B \!+\! \Theta_y \right) ,~
(z^B \!\approx\! 1,~x^B \!\approx\! 0,~y^B \!\approx\! 0) .
\end{array}
\label{ApproxtransU(G)aadag(-1)n2}
\end{eqnarray}\\[-8pt]
Thus we have the inverse transformation for
(\ref{transU(G)aadag(-1)n2})
as\\[-18pt] 
\begin{eqnarray}
\!\!\!\!\!\!\!
\begin{array}{rl}
\left[a_i,~a^{\dag }_i,
{\displaystyle \frac{1}{\sqrt{2}}}\mathbb{I}\right] \!\!
&
\!\!=\!\!
\left[b_j,~b^{\dag }_j,
{\displaystyle \frac{1}{\sqrt{2}}}\mathbb{I}\right] \!
G^\dag \!
\left( \! z^B \!+\! \Theta_y \! \right) \!,~
G_{ji}^\dag
\!=\!\!
\left[ \!\!
\begin{array}{ccc}
A^{B\dag}_{ji}             & B^{B\dag}_{ji}              &
{\displaystyle \frac{y^{B\dag}_j}{\sqrt{2}}} \\
B^{B\mbox{\scriptsize T}}_{ji}         & 
A^{B\mbox{\scriptsize T}}_{ji}& 
{\displaystyle \frac{y^{B\mbox{\scriptsize T}}_j}{\sqrt{2}}} \\
{\displaystyle - \frac{x^{B\mbox{\scriptsize T}}_i}{\sqrt{2}}} & 
{\displaystyle \frac{x^{B\dag}_i}{\sqrt{2}}} & z^B  
\end{array} \!\!
\right] .
\end{array}
\label{transU(G)aadag(-1)n4}
\end{eqnarray}\\[-8pt]
The transformations
(\ref{transU(G)aadag(-1)n2})
and 
(\ref{transU(G)aadag(-1)n4})
hold the orthogonality conditions
but do not exactly:\\[-14pt]
\begin{eqnarray}
G^{\dag }
\widetilde{I}_{2N+1}
G
\!=\!
\widetilde{I}_{2N+1},~
G
\widetilde{I}_{2N+1}
G^{\dag }
\!=\!
\widetilde{I}_{2N+1}.~
\widetilde{I}_{2N+1}
\!\equiv\!
\left[ \!\!
\begin{array}{cc}
\widetilde{I}_{2N}  & \!\! 0  
\\
0   & \!\! 1
\end{array} \!\!
\right] ,~
(z^B \!\approx\! 1,x^B \!\approx\! 0,y^B \!\approx\! 0) ,
\label{Gmat}
\end{eqnarray}\\[-12pt]
though we omit the proof here
because their explicit expressions become too long to write.

\newpage


\setcounter{equation}{0}
\renewcommand{\theequation}{\arabic{section}.\arabic{equation}}

\section{GHB MEAN-FIELD HAMILTONIAN AND ITS DIAGONALIZATION}

\vspace{-0.5cm}

~~~ 
Let us consider the following Hamiltonian
consisting of the generalized Hartree-Bogoliubov (GHB) 
mean-field Hamiltonian (MFH)
for which
we assume a linear MFH
expressed in terms of the generators of the $\mathfrak{Jacobi~hsp}$  algebra,
the $\mathfrak{Jacobi}$  algebra,
(\ref{operatorset})
as follows:\\[-20pt]
\begin{eqnarray}
\begin{array}{rl}
&\!\!\!\!
H_{\mathfrak{Jacobi~hsp}}
=
i F^{B}_{ij} \!
E^{i }_{~j }
{\displaystyle 
+
i
\frac{1}{2}
D^B_{ij}E^{i j}
-
i
\frac{1}{2}
\overline{D}^{B }_{ij}E_{i j}
}
+
i M^B_i \! a^{\dag }_i - i \overline{M}_i^{B} \! a_i \\
\\[-10pt]
\!\!
=
&\!\!\!
{\displaystyle \frac{1}{2}} \!
\left[a,a^{\dag },\!
{\displaystyle \frac{1}{\sqrt{2}}}\mathbb{I}\right] \!
\left[ \!\!
\begin{array}{ccc}
0& 1_N& 0 \\
\\[-10pt]
1_N& 0& 0 \\
\\[-10pt]
0& 0& 1 
\end{array} \!\!
\right] \!
\mathcal{F}^B \!
\left[ \!\!\!
\begin{array}{ccc}
0& 1_N& 0 \\
\\[-10pt]
1_N& 0& 0 \\
\\[-10pt]
0& 0& 1  
\end{array} \!\!\!
\right] \!\!
\left[ \!\!\!
\begin{array}{c}
a^{\dag } \! , \\
\\[-16pt]
a, \\
\\[-16pt]
{\displaystyle \frac{1}{\sqrt{2}}}\mathbb{I}
\end{array} \!\!\!
\right] 
\!=\!
{\displaystyle \frac{1}{2}} \!
\left[a,a^{\dag },\!
{\displaystyle \frac{1}{\sqrt{2}}}\mathbb{I}\right] \!
\widetilde{\mathcal{F}}^B \!
\left[ \!\!\!
\begin{array}{c}
a^{\dag } \! , \\
\\[-16pt]
a, \\
\\[-16pt]
{\displaystyle \frac{1}{\sqrt{2}}}\mathbb{I}
\end{array} \!\!\!
\right] .
\end{array}
\label{mean-field Hamiltonian}
\end{eqnarray}
Here
$F^B, D^B$ and $M^B$ are the SCF parameters
satisfying the properties
\begin{eqnarray}
\begin{array}{l}
F^{B\dag }=-F^B,~~
D^{B\mbox{\scriptsize T}}=D^B, ~~
M^{B\mbox{\scriptsize T}}
=
[M^B_1,\cdots, M^B_N ] , 
\end{array}
\end{eqnarray}\\[-20pt]
and
matrices $\mathcal{F}^B$ and
$\widetilde{\mathcal{F}}^B$
are given by\\[-14pt]
\begin{eqnarray}
\left.
\begin{array}{rl}
&\mathcal{F}^B
\!=\!
\left[ \!\!
\begin{array}{ccc}
i F^B    & i D^B    & i \sqrt{2}M^B \\
\\[-10pt]
- i \overline{D}^{B} & - i \overline{F}^{B} &
- i \sqrt{2}~\! \overline{M}^{B } \\
\\[-10pt]
-i \sqrt{2}M^{B\dag } &
i \sqrt{2}M^{B\mbox{\scriptsize T}} & 0 
\end{array} \!\!
\right] , \\
\\
&\widetilde{\mathcal{F}}^B
=\!\!
\left[ \!\!
\begin{array}{ccc}
0& 1_N& 0 \\
\\[-10pt]
1_N& 0& 0 \\
\\[-10pt]
0& 0& 1 
\end{array} \!
\right] \!
\mathcal{F}^B \!
\left[ \!
\begin{array}{ccc}
0& 1_N& 0 \\
\\[-10pt]
1_N& 0& 0 \\
\\[-10pt]
0& 0& 1  
\end{array} \!
\right]
\!=\!
\left[ \!\!
\begin{array}{ccc}
- i \overline{F}^{B}& -i  \overline{D}^{B}&
- i \sqrt{2} ~\! \overline{M}^{B } \\
\\[-10pt]
i D^B   & i F^B   & i \sqrt{2}M^B \\
\\[-10pt]
i \sqrt{2}M^{B\mbox{\scriptsize T}} &
- i \sqrt{2}M^{B\dag } & 0 
\end{array} \!\!
\right] .
\end{array}
\right\}
\label{tildeF}
\end{eqnarray}\\[-12pt]
The fermion GHB MFH
is studied intensively by one of the present authors (Nishiyama)
\cite{Ni.98}.

Using the generalized Bogoliubov transformation
(\ref{transU(G)aadag(-1)n2}) 
and its inverse transformation
(\ref{transU(G)aadag(-1)n2})
and the operator relation
$(z^B)^2 \!-\! \Theta^2_y \!=\! 1$, 
we can diagonalize the MFH $H_{\mathfrak{Jacobi~hsp}}$
(\ref{mean-field Hamiltonian})
in the following form:\\[-26pt]
\begin{eqnarray}
\begin{array}{rl}
&\!\!\!\!\!\!\!\!\!\!
H_{\mathfrak{Jacobi~hsp}}
\!=\!
{\displaystyle \frac{1}{2}} \!\!
\left[b,b^{\dag },\!
{\displaystyle \frac{1}{\sqrt{2}}}\mathbb{I}\right] \!\!
\left[
G^\dag \!\!
\left(z^B \!+\! \Theta_y \right) \!\!
\widetilde{\mathcal{F}}^B \!\!
\left(z^B \!-\! \Theta_y \right) \!
G
\right] \!\!
\left[ \!\!\!
\begin{array}{c}
b^{\dag }, \! \\
\\[-16pt]
b, \\
\\[-16pt]
{\displaystyle \frac{1}{\sqrt{2}}}\mathbb{I}
\end{array} \!\!\!
\right] 
\!\!=\!\!
{\displaystyle \frac{1}{2}} \!\!
\left[b,b^{\dag },\!
{\displaystyle \frac{1}{\sqrt{2}}}\mathbb{I}\right] \!
G^\dag \! \tilde{F}^B \! G \!\!
\left[ \!\!\!
\begin{array}{c}
b^{\dag } \! , \\
\\[-16pt]
b, \\
\\[-16pt]
{\displaystyle \frac{1}{\sqrt{2}}}\mathbb{I}
\end{array} \!\!\!
\right] \! ,
\end{array} 
\label{diaHm}
\end{eqnarray}
\vspace{-0.8cm}
\begin{eqnarray}
\!\!\!
G^\dag
\widetilde{\mathcal{F}}^B
G
\!\!=\!\!
\left[ \!\!\!
\begin{array}{cc}
E_{2N}  \!\cdot\! I_{2N} &\!\! 0 \\
\\[-10pt]
0  &\!\! \epsilon 
\end{array}  \!\!\!
\right]
\!\equiv\!
\tilde{E}~(\epsilon \mbox{;real number}),
E_{2N}
\!\!=\!\!
\left[ \!E_{\mbox{diag.}}, E_{\mbox{diag.}}  \!\right] ,
E_{\mbox{diag.}}
\!\equiv\!
\left[
E_1,  \!\cdots\!, E_N 
\right] ,
\label{diagonalHm}
\end{eqnarray}\\[-14pt]
where $E_i$ is a quasi-particle energy
and $\epsilon$ is another excitation energy
which is not artificial but possibly exists
mathematically.
Then, we get
$
H_{\mathfrak{Jacobi~hsp}}
\!=\!
\sum_{i=1}^N \! 
\left( \!
E_i b^{\dag }_i b_i 
\!+\!
\frac{1}{2} E_i \!
\right)
\!+\!
\frac{1}{2}\epsilon .
$
Using
(\ref{Gmat}),
(\ref{diagonalHm}) and the second of
(\ref{tildeF}),
we obtain the following eigenvalue equation:\\[-10pt]
\begin{eqnarray}
\begin{array}{l}
\mathcal{F}^B \!
\left[ \!
\begin{array}{ccc}
0& 1_N& 0 \\
\\[-10pt]
1_N& 0& 0 \\
\\[-10pt]
0& 0& 1 
\end{array}
\right] \!
G 
=\!
\left[ \!\!
\begin{array}{ccc}
0& -1_N& 0 \\
\\[-10pt]
1_N& 0& 0 \\
\\[-10pt]
0& 0& 1 
\end{array}
\right] \!
G \!
\left[ \!
\begin{array}{ccc}
1_N & 0& 0 \\
\\[-10pt]
0 & -1_N& 0 \\
\\[-10pt]
0& 0& 1  
\end{array} \!
\right] \!
\tilde{E} ,
\end{array}
\label{FG}
\end{eqnarray}\\[-6pt]
which
is explicitly written by\\[-16pt] 
\begin{eqnarray}
\!\!\!\!
\left[ \!
\begin{array}{ccc}
i F^B    & i D^B    & i \sqrt{2}M^B \\
\\
\\[-16pt]
- i \overline{D}^{B} & - i \overline{F}^{B} &
-i \sqrt{2}~\! \overline{M}^{B} \\
\\
\\[-16pt]
- i \sqrt{2}M^{B\dag } &
i \sqrt{2}M^{B\mbox{\scriptsize T}} &  0 
\end{array} \!
\right] \!
\left[ \!
\begin{array}{ccc}
b^B
\!+\!
{\displaystyle \frac{1}{1\!+\!z^B}}
{\displaystyle \frac{x^B}{\sqrt{2}}}
{\displaystyle \frac{y^B}{\sqrt{2}}} &
\overline{a}^{B}
\!+\!
{\displaystyle \frac{1}{1\!+\!z^B}}
{\displaystyle \frac{x^B}{\sqrt{2}}}
{\displaystyle \frac{\overline{y}^B}{\sqrt{2}}} &
 {\displaystyle \frac{x^B}{\sqrt{2}}} \\
\\[-12pt]
a^B
\!-\!
{\displaystyle \frac{1}{1\!+\!z^B}}
{\displaystyle \frac{\overline{x}^B}{\sqrt{2}}}
 {\displaystyle \frac{y^B}{\sqrt{2}}}&
\overline{b}^{B}
 \!-\!
{\displaystyle \frac{1}{1\!+\!z^B}}
{\displaystyle \frac{\overline{x}^B}{\sqrt{2}}}
{\displaystyle \frac{\overline{y}^B}{\sqrt{2}}} &
 -{\displaystyle \frac{\overline{x}^B}{\sqrt{2}}} \\
\\[-14pt]
{\displaystyle \frac{y^B}{\sqrt{2}}} &
{\displaystyle \frac{\overline{y}^B}{\sqrt{2}}} &
z^B 
\end{array} \!
\right]  \nonumber \\ \nonumber
\\
\!=\!
\left[ \!
\begin{array}{ccc}
-
b^B
\!-\!
{\displaystyle \frac{1}{1\!+\!z^B}}
{\displaystyle \frac{x^B}{\sqrt{2}}}
{\displaystyle \frac{y^B}{\sqrt{2}}} &
- \overline{a}^{B}
\!-\!
{\displaystyle \frac{1}{1\!+\!z^B}}
{\displaystyle \frac{x^B}{\sqrt{2}}}
{\displaystyle \frac{\overline{y}^B}{\sqrt{2}}} &
-{\displaystyle \frac{x^B}{\sqrt{2}}} \\
\\[-12pt]
a^B
\!-\!
{\displaystyle \frac{1}{1\!+\!z^B}}
{\displaystyle \frac{\overline{x}^B}{\sqrt{2}}}
 {\displaystyle \frac{y^B}{\sqrt{2}}} &
\overline{b}^{B}
\!-\!
{\displaystyle \frac{1}{1\!+\!z^B}}
{\displaystyle \frac{\overline{x}^B}{\sqrt{2}}}
{\displaystyle \frac{\overline{y}^B}{\sqrt{2}}} &
-{\displaystyle \frac{\overline{x}^B}{\sqrt{2}}} \\
\\[-14pt]
{\displaystyle \frac{y^B}{\sqrt{2}}} &
{\displaystyle \frac{\overline{y}^B}{\sqrt{2}}} &
z^B 
\end{array} \!
\right] \!
\left[ \!
\begin{array}{ccc}
e + \varepsilon & 0 & 0 \\
\\
\\[-13pt]
0 & - e - \varepsilon & 0 \\
\\
\\[-13pt]
0 & 0 & \epsilon 
\end{array} \!
\right] ,
\label{extendedeigeneq}
\end{eqnarray}\\[-8pt]
where $e$ is the eigenvalue of
the boson Bogoliubov equation
appeared soon later.
From the first column in both sides of equations of 
(\ref{extendedeigeneq}),
we get  the following set of equations:\\[-12pt]
\begin{eqnarray}
\left.
\begin{array}{l}
i F^B \!
\left(
b^B
\!+\!
{\displaystyle \frac{1}{1\!+\!z^B}}
{\displaystyle \frac{x^B}{\sqrt{2}}}
{\displaystyle \frac{y^B}{\sqrt{2}}} 
\right)
\!+\!
i D^B \!
\left(
a^B
\!-\!
{\displaystyle \frac{1}{1\!+\!z^B}}
{\displaystyle \frac{\overline{x}^B}{\sqrt{2}}}
{\displaystyle \frac{y^B}{\sqrt{2}}}
 \right) 
\!+\!
i \sqrt{2} M^{B}
 {\displaystyle \frac{y^B}{\sqrt{2}}} \\
\\[-10pt]
\!=\!
i F^B b^B \!+\! i D^B a^B
\!+\!
{\displaystyle \frac{1}{1\!+\!z^B}}
i \! \left( \!
F^B {\displaystyle \frac{x^B}{\sqrt{2}}}
\!-\!
D^B {\displaystyle \frac{\overline{x}^B}{\sqrt{2}}} \!
\right) \!
{\displaystyle \frac{y^B}{\sqrt{2}}}
\!+\!
i \sqrt{2} M^{B}
 {\displaystyle \frac{y^B}{\sqrt{2}}} \\
\\[-10pt]
\!=\!
-
\left(
b^B
\!+\!
{\displaystyle \frac{1}{1\!+\!z^B}}
{\displaystyle \frac{x^B}{\sqrt{2}}}
{\displaystyle \frac{y^B}{\sqrt{2}}} 
\right)
\left( e + \varepsilon  \right)
\!=\!
-b^B e
\!-\!
b^B \varepsilon
\!-\!
{\displaystyle \frac{1}{1\!+\!z^B}}
{\displaystyle \frac{x^B}{\sqrt{2}}}
{\displaystyle \frac{y^B}{\sqrt{2}}} \!
\left( e + \varepsilon  \right), \\
\\
- i \overline{D}^B \!
\left(
b^B
\!+\!
{\displaystyle \frac{1}{1\!+\!z^B}}
{\displaystyle \frac{x^B}{\sqrt{2}}}
{\displaystyle \frac{y^B}{\sqrt{2}}} 
\right)
\!-\!
i \overline{F}^B \!
\left(
a^B
\!-\!
{\displaystyle \frac{1}{1\!+\!z^B}}
{\displaystyle \frac{\overline{x}^B}{\sqrt{2}}}
 {\displaystyle \frac{y^B}{\sqrt{2}}}
 \right) 
\!-\!
i \sqrt{2} ~\! \overline{M}^{B}
 {\displaystyle \frac{y^B}{\sqrt{2}}}  \\
\\[-12pt]
\!=\!
- i \overline{D}^B \! b^B \!-\! i \overline{F}^B a^B
\!-\!
{\displaystyle \frac{1}{1\!+\!z^B}}
i \! \left( \!
\overline{D}^B {\displaystyle \frac{x^B}{\sqrt{2}}}
\!-\!
\overline{F}^B {\displaystyle \frac{\overline{x}^B}{\sqrt{2}}} \!
\right) \!
{\displaystyle \frac{y^B}{\sqrt{2}}}
\!-\!
i \sqrt{2} ~\! \overline{M}^{B}
 {\displaystyle \frac{y^B}{\sqrt{2}}} \\
\\[-12pt]
\!=\!
\left(
a^B
\!-\!
{\displaystyle \frac{1}{1\!+\!z^B}}
{\displaystyle \frac{\overline{x}^B}{\sqrt{2}}}
 {\displaystyle \frac{y^B}{\sqrt{2}}} 
\right)
\left( e + \varepsilon  \right)
\!=\!
a^B e
+
a^B \varepsilon
\!-\!
{\displaystyle \frac{1}{1\!+\!z^B}}
{\displaystyle \frac{\overline{x}^B}{\sqrt{2}}}
{\displaystyle \frac{y^B}{\sqrt{2}}} \!
\left( e + \varepsilon  \right), \\
\\[-2pt]
- i \sqrt{2}\!M^{B\dag }
\left(
b^B
\!+\!
{\displaystyle \frac{1}{1\!+\!z^B}}
{\displaystyle \frac{x^B}{\sqrt{2}}}
{\displaystyle \frac{y^B}{\sqrt{2}}} 
\right)
\!+\!
i \sqrt{2}\!M^{B\mbox{\scriptsize T}}
\left(
a^B
\!-\!
{\displaystyle \frac{1}{1\!+\!z^B}}
{\displaystyle \frac{\overline{x}^B}{\sqrt{2}}}
{\displaystyle \frac{y^B}{\sqrt{2}}}
 \right) \\
\\[-8pt]
\!=\!
- i \sqrt{2}\!M^{B\dag } b^B
\!+\!
i \sqrt{2}\!M^{B\mbox{\scriptsize T}} a^B
\!-\!
{\displaystyle \frac{1}{1\!+\!z^B}} 
i \! \left( \!
\sqrt{2}\!M^{B\dag } {\displaystyle \frac{x^B}{\sqrt{2}}}
\!+\!
\sqrt{2}\!M^{B\mbox{\scriptsize T}}
{\displaystyle \frac{\overline{x}^B}{\sqrt{2}}} \!
\right) \!
{\displaystyle \frac{y^B}{\sqrt{2}}}
\!=\!
{\displaystyle \frac{y^B}{\sqrt{2}}} e
\!+\!
{\displaystyle \frac{y^B}{\sqrt{2}}} \varepsilon .
\end{array}
\right\}
\label{eigeneq00}
\end{eqnarray}\\[-10pt]
Then we have\\[-14pt]
\begin{eqnarray}
\!\!\!\!
\left.
\begin{array}{c}
i F^B b^B \!+\! i D^B a^B = -b^B e , ~~~~
- i \overline{D}^B \! b^B \!-\! i \overline{F}^B a^B = a^B e ,  \\
\\[-6pt]
{\displaystyle \frac{1}{1\!+\!z^B}}
i \! \left( \!
F^B {\displaystyle \frac{x^B}{\sqrt{2}}}
\!-\!
D^B {\displaystyle \frac{\overline{x}^B}{\sqrt{2}}} \!
\right) \!
{\displaystyle \frac{y^B}{\sqrt{2}}}
\!+\!
i \sqrt{2} M^{B}
 {\displaystyle \frac{y^B}{\sqrt{2}}}
\!=\!
-b^B \varepsilon
\!-\!
{\displaystyle \frac{1}{1\!+\!z^B}}
{\displaystyle \frac{x^B}{\sqrt{2}}}
{\displaystyle \frac{y^B}{\sqrt{2}}} \!
\left( e + \varepsilon  \right) , \\
\\[-6pt]
-
{\displaystyle \frac{1}{1\!+\!z^B}}
i \left( \!
\overline{D}^B {\displaystyle \frac{x^B}{\sqrt{2}}}
\!-\!
\overline{F}^B {\displaystyle \frac{\overline{x}^B}{\sqrt{2}}} \!
\right) \!
{\displaystyle \frac{y^B}{\sqrt{2}}}
\!-\!
i \sqrt{2} \overline{M}^{B}
{\displaystyle \frac{y^B}{\sqrt{2}}}
\!=\! 
a^B \varepsilon
\!-\!
{\displaystyle \frac{1}{1\!+\!z^B}}
{\displaystyle \frac{\overline{x}^B}{\sqrt{2}}}
{\displaystyle \frac{y^B}{\sqrt{2}}} \!
\left( e + \varepsilon  \right) ,  \\
\\[-6pt]
- i \sqrt{2}\!M^{B\dag } b^B
\!+\!
i \sqrt{2}\!M^{B\mbox{\scriptsize T}} a^B 
\!=\! 
{\displaystyle \frac{y^B}{\sqrt{2}}}
e , ~~~~
-
{\displaystyle \frac{1}{1\!+\!z^B}} 
i \! \left( \!\!
\sqrt{2}\!M^{B\dag }{\displaystyle \frac{x^B}{\sqrt{2}}}
\!+\!
\sqrt{2}\!M^{B\mbox{\scriptsize T}}
{\displaystyle \frac{\overline{x}^B}{\sqrt{2}}} \!
\right) 
\!=\! 
\varepsilon ,
\end{array} \!\!
\right\}
\label{eigeneq}
\end{eqnarray}
Through the second column
in both sides of equations of
(\ref{extendedeigeneq}),
we get the set of equations\\[-12pt]
\begin{eqnarray}
\left.
\begin{array}{l}
i F^B \!
\left(
\overline{a}^{B}
\!+\!
{\displaystyle \frac{1}{1\!+\!z^B}}
{\displaystyle \frac{x^B}{\sqrt{2}}}
{\displaystyle \frac{\overline{y}^B}{\sqrt{2}}}
\right)
\!+\!
i D^B \!
\left(
\overline{b}^{B}
 \!-\!
{\displaystyle \frac{1}{1\!+\!z^B}}
{\displaystyle \frac{\overline{x}^B}{\sqrt{2}}}
{\displaystyle \frac{\overline{y}^B}{\sqrt{2}}}
 \right) 
\!+\!
i \sqrt{2} M^{B}
{\displaystyle \frac{\overline{y}^B}{\sqrt{2}}}  \\
\\[-10pt]
\!=\!
i F^B \overline{a}^B \!+\! i D^B \overline{b}^B
\!+\!
{\displaystyle \frac{1}{1\!+\!z^B}} 
i \! \left( \!
F^B {\displaystyle \frac{x^B}{\sqrt{2}}}
\!-\!
D^B {\displaystyle \frac{\overline{x}^B}{\sqrt{2}}} \!
\right) \!
{\displaystyle \frac{\overline{y}^B}{\sqrt{2}}}
\!+\!
i \sqrt{2} M^{B}
 {\displaystyle \frac{\overline{y}^B}{\sqrt{2}}} \\
\\[-12pt]
\!=\!
\left(
\overline{a}^{B}
\!+\!
{\displaystyle \frac{1}{1\!+\!z^B}}
{\displaystyle \frac{x^B}{\sqrt{2}}}
{\displaystyle \frac{\overline{y}^B}{\sqrt{2}}}
\right)
\left( e + \varepsilon  \right)
\!=\!
\overline{a}^{B} e
+
\overline{a}^{B} \varepsilon
\!+\!
{\displaystyle \frac{1}{1\!+\!z^B}}
{\displaystyle \frac{x^B}{\sqrt{2}}}
{\displaystyle \frac{\overline{y}^B}{\sqrt{2}}}
\left( e + \varepsilon  \right) , \\
\\[2pt]
- i \overline{D}^B \!
\left(
\overline{a}^{B}
\!+\!
{\displaystyle \frac{1}{1\!+\!z^B}}
{\displaystyle \frac{x^B}{\sqrt{2}}}
{\displaystyle \frac{\overline{y}^B}{\sqrt{2}}} 
\right)
\!-\!
i \overline{F}^B \!
\left(
\overline{b}^{B}
 \!-\!
{\displaystyle \frac{1}{1\!+\!z^B}}
{\displaystyle \frac{\overline{x}^B}{\sqrt{2}}}
{\displaystyle \frac{\overline{y}^B}{\sqrt{2}}}
 \right) 
\!-\!
i \sqrt{2} ~\! \overline{M}^{B}
{\displaystyle \frac{\overline{y}^B}{\sqrt{2}}} \\
\\[-10pt]
\!=\!
- i \overline{D}^B \overline{a}^B
\!-\!
i \overline{F}^B \overline{b}^B 
\!-\!
{\displaystyle \frac{1}{1\!+\!z^B}}
i \! \left( \!
\overline{D}^B {\displaystyle \frac{x^B}{\sqrt{2}}}
\!-\!
\overline{F}^B {\displaystyle \frac{\overline{x}^B}{\sqrt{2}}} \!
\right) \!
{\displaystyle \frac{\overline{y}^B}{\sqrt{2}}}
\!-\!
i \sqrt{2} ~\! \overline{M}^{B}
 {\displaystyle \frac{\overline{y}^B}{\sqrt{2}}} \\
\\[-12pt]
\!=\!
-
\left(
\overline{b}^{B}
\!-\!
{\displaystyle \frac{1}{1\!+\!z^B}}
{\displaystyle \frac{\overline{x}^B}{\sqrt{2}}}
{\displaystyle \frac{\overline{y}^B}{\sqrt{2}}}
\right)
\left( e + \varepsilon  \right)
\!=\!
-
\overline{b}^{B} e
\!-\!
\overline{b}^{B} \varepsilon
\!+\!
{\displaystyle \frac{1}{1\!+\!z^B}}
{\displaystyle \frac{\overline{x}^B}{\sqrt{2}}}
{\displaystyle \frac{\overline{y}^B}{\sqrt{2}}}
\left( e + \varepsilon  \right) , \\
\\[2pt]
- i \sqrt{2}\!M^{B\dag }
\left(
\overline{a}^{B}
\!+\!
{\displaystyle \frac{1}{1\!+\!z^B}}
{\displaystyle \frac{x^B}{\sqrt{2}}}
{\displaystyle \frac{\overline{y}^B}{\sqrt{2}}} 
\right)
\!+\!
i \sqrt{2}\!M^{B\mbox{\scriptsize T}}
\left(
\overline{b}^{B}
 \!-\!
{\displaystyle \frac{1}{1\!+\!z^B}}
{\displaystyle \frac{\overline{x}^B}{\sqrt{2}}}
{\displaystyle \frac{\overline{y}^B}{\sqrt{2}}}
\right) \\
\\[-10pt]
\!=\!
- i \sqrt{2}\!M^{B\dag } \overline{a}^B
\!+\!
i \sqrt{2}\!M^{B\mbox{\scriptsize T}} \overline{b}^B
\!\!-\!
{\displaystyle \frac{1}{1\!+\!z^B}}
i \! \left( \!\!
\sqrt{2}\!M^{B\dag } {\displaystyle \frac{x^B}{\sqrt{2}}}
\!+\!
\sqrt{2}\!M^{B\mbox{\scriptsize T}}
{\displaystyle \frac{\overline{x}^B}{\sqrt{2}}} \!\!
\right) \!\!
{\displaystyle \frac{\overline{y}^B}{\sqrt{2}}}
\!=\!
- {\displaystyle \frac{\overline{y}^B}{\sqrt{2}}} e
\!-\!
{\displaystyle \frac{\overline{y}^B}{\sqrt{2}}} \varepsilon ,
\end{array} \!\!
\right\}
\label{eigeneq20}
\end{eqnarray}\\[-12pt]
from which we have\\[-16pt]
\begin{eqnarray}
\!\!\!\!
\left.
\begin{array}{c}
i F^B \overline{a}^B \!+\! i D^B \overline{b}^B
=
\overline{a}^B e , ~~~~
- i \overline{D}^B \overline{a}^B
\!-\!
i \overline{F}^B \overline{b}^B = -\overline{b}^B e , \\
\\[-6pt]
{\displaystyle \frac{1}{1\!+\!z^B}} 
i \! \left( \!
F^B {\displaystyle \frac{x^B}{\sqrt{2}}}
\!-\!
D^B {\displaystyle \frac{\overline{x}^B}{\sqrt{2}}} \!
\right) \!
{\displaystyle \frac{\overline{y}^B}{\sqrt{2}}}
\!+\!
i \sqrt{2} M^{B}
 {\displaystyle \frac{\overline{y}^B}{\sqrt{2}}}
\!=\! 
\overline{a}^B \varepsilon
\!+\!
{\displaystyle \frac{1}{1\!+\!z^B}}
{\displaystyle \frac{x^B}{\sqrt{2}}}
{\displaystyle \frac{\overline{y}^B}{\sqrt{2}}}
\left( e + \varepsilon  \right) ,  \\
\\[-6pt]
-
{\displaystyle \frac{1}{1\!+\!z^B}}
i \! \left( \!
\overline{D}^B {\displaystyle \frac{x^B}{\sqrt{2}}}
\!-\!
\overline{F}^B {\displaystyle \frac{\overline{x}^B}{\sqrt{2}}} \!
\right) \!
{\displaystyle \frac{\overline{y}^B}{\sqrt{2}}}
\!-\!
i \sqrt{2} \overline{M}^{B}
{\displaystyle \frac{\overline{y}^B}{\sqrt{2}}}
\!=\!
-\overline{b}^{B} \varepsilon
\!+\!
{\displaystyle \frac{1}{1\!+\!z^B}}
{\displaystyle \frac{\overline{x}^B}{\sqrt{2}}}
{\displaystyle \frac{\overline{y}^B}{\sqrt{2}}}
\left( e + \varepsilon  \right) , \\
\\[-8pt]
- i \sqrt{2}\!M^{B\dag } \overline{a}^B
\!+\!
i \sqrt{2}\!M^{B\mbox{\scriptsize T}} \overline{b}^B
\!=\! 
- {\displaystyle \frac{\overline{y}^B}{\sqrt{2}}}
e , ~~~~
-
{\displaystyle \frac{1}{1\!+\!z^B}} 
i \! \left( \!\!
\sqrt{2}\!M^{B\dag } {\displaystyle \frac{x^B}{\sqrt{2}}}
\!+\!
\sqrt{2}\!M^{B\mbox{\scriptsize T}}
{\displaystyle \frac{\overline{x}^B}{\sqrt{2}}} \!
\right) \!
\!=\!
- \varepsilon .
\end{array} \!\!
\right\}
\label{eigeneq2}
\end{eqnarray}\\[-10pt]
Through the third column
in both sides of equations of
(\ref{extendedeigeneq}),
we get the set of equations\\[-12pt]
\begin{eqnarray}
\!\!\!\!\!\!\!\!
\left.
\begin{array}{c}
i F^B {\displaystyle \frac{x^B}{\sqrt{2}}} 
\!-\!
i D^B {\displaystyle \frac{\overline{x}^B}{\sqrt{2}}}
\!+\!
i \sqrt{2} z^B M^{B} 
\!=\!
-{\displaystyle \frac{x^B}{\sqrt{2}}}
~\! \epsilon ,~~
- i \overline{D}^B {\displaystyle \frac{x^B}{\sqrt{2}}} 
\!+\!
i \overline{F}^B \! {\displaystyle \frac{\overline{x}^B}{\sqrt{2}}}
\!-\!
i \sqrt{2} z^B \overline{M}^{B} 
\!\!=\!
-{\displaystyle \frac{\overline{x}^B}{\sqrt{2}}}
~\! \epsilon ,\\
\\[-10pt]
\left( 
\overline{F^B {\displaystyle \frac{x^B}{\sqrt{2}}}} 
\!=\!
-
\overline{F}^B \! 
{\displaystyle \frac{\overline{x}^B}{\sqrt{2}}},~~
\overline{D^B {\displaystyle \frac{\overline{x}^B}{\sqrt{2}}}}
\!=\!
-
\overline{D}^B \! {\displaystyle \frac{x^B}{\sqrt{2}}} 
\right) \! ,\\
\\[-10pt]
- i \sqrt{2}M^{B\dag }{\displaystyle \frac{x^B}{\sqrt{2}}}  
- i \sqrt{2}M^{B\mbox{\scriptsize T}}
{\displaystyle \frac{\overline{x}^B}{\sqrt{2}}}
\!=\!
z^B \! \epsilon ,~~
\left(
\overline{- i \sqrt{2}M^{B\mbox{\scriptsize T}}
{\displaystyle \frac{\overline{x}^B}{\sqrt{2}}}}
\!=\!
- i \sqrt{2}M^{B\dag }
{\displaystyle \frac{x^B}{\sqrt{2}}} \! 
\right) \! .
\end{array} \!\!\!
\right\}
\label{xthetaepsilon}
\end{eqnarray}\\[-8pt]
The first and second equations of
(\ref{eigeneq}) and (\ref{eigeneq2})
are set of the boson Bogoliubov equation
with eigenvalue $e$
\cite{BR.86}.
Note that inside round brackets 
the SCF parameters $F^{B}, D^{B}$ and $M^{B}$
are regarded as the Grassmann-like numbers,
to ensure the conjugate complexity of the equations
and also to assure the real number of the $\epsilon$.
The additional eigenvalues, $\varepsilon$, however,
become to be $\varepsilon \!=\! 0$ due to
the last equations of
(\ref{eigeneq}) and (\ref{eigeneq2}). 
The division of $E$ into $e$ and $\varepsilon$
is a useful means to treat the MF eigenvalue equation
and to make clear distinct computational steps.

\newpage


\setcounter{equation}{0}
\renewcommand{\theequation}{\arabic{section}.\arabic{equation}}

\section{SCF CONDITION AND MFH LINEAR IN $\mathfrak{Jacobi}$ 
GENERATOR}

\vspace{-0.5cm}

~~~ 
According to suggestion by Ozaki 
\cite{Ozaki.08},
multiplying the first of 
(\ref{eigeneq}) 
by $b^{B\dag}$ 
and 
the first of
(\ref{eigeneq2})
by $a^{B\mbox{\scriptsize T}}$ 
from the right, 
we obtain\\[-20pt] 
\begin{eqnarray}
i F^B b^B b^{B\dag}\!+\! i D^B a^B b^{B\dag}
\!=\!
- b^B e b^{B\dag}, ~~
i F^B \overline{a}^B a^{B\mbox{\scriptsize T}}
\!+\!
i D^B \overline{b}^B a^{B\mbox{\scriptsize T}}
\!=\! 
\overline{a}^B e a^{B\mbox{\scriptsize T}} ,
\label{eigeneq3}
\end{eqnarray}\\[-20pt] 
and multiplying the first of 
(\ref{eigeneq}) 
by $a^{B\dag}$
and 
the first of
(\ref{eigeneq2})
by $b^{B\mbox{\scriptsize T}}$ 
from the right,
we obtain\\[-18pt] 
\begin{eqnarray}
i F^B b^B a^{B\dag} \!+\! i D^B a^B a^{B\dag}
\!=\!
- b^B e a^{B\dag}, ~~
i F^B \overline{a}^B b^{B\mbox{\scriptsize T}}
\!+\!
i D^B \overline{b}^B b^{B\mbox{\scriptsize T}}
\!=\! 
\overline{a}^B e b^{B\mbox{\scriptsize T}} .
\label{eigeneq4}
\end{eqnarray}\\[-22pt] 
Subtracting the second of
(\ref{eigeneq3}) 
from the first
and the second of 
(\ref{eigeneq4}) 
from the first,
we have\\[-20pt] 
\begin{eqnarray}
i F^B
\!=\!
\overline{a}^B e a^{B\mbox{\scriptsize T}} + b^B e b^{B \dag}
\!=\!
- i F^{B\dag}, ~~
i D^B
\!=\!
-
\left(b^B e a^{B\dag}
+
\overline{a}^B e b^{B\mbox{\scriptsize T}}\right)
\!=\!
i D^{B\mbox{\scriptsize T}}.
\label{XY}
\end{eqnarray}\\[-22pt] 
Multiplying the first equation in the last line of 
(\ref{eigeneq})
by $b^{B\dag }$ and the second equation
by $a^{B\mbox{\scriptsize T}}$ from the right
and using
$
z^B \! y^B
\!=\!
x^{B\mbox{\scriptsize T}} \! A^B \!+\! x^{B\dag } \! B^B
\!=\!
x^{B\mbox{\scriptsize T}} \! a^B \!+\! x^{B\dag } b^B
\!-\!
\left(\! 1 \!-\! z^B \!\right) \!
y^B
$,
respectively,
we have\\[-22pt] 
\begin{eqnarray}
\left.
\begin{array}{rl}
&\!\!\!\!\!\!\!\!\!\!
- i \sqrt{2} z^B M^{B\dag } b^B b^{B\dag }
\!+\!
i \sqrt{2} z^B M^{B\mbox{\scriptsize T}} a^B b^{B\dag }
\!=\! 
{\displaystyle \frac{x^{B\mbox{\scriptsize T}}}
{\sqrt{2}}} a^B e b^{B\dag }
\!+\!
{\displaystyle \frac{x^{B\dag }}{\sqrt{2}}} b^B e b^{B\dag }
\!-\!
\left( 1 \!-\! z^B \right) \!
{\displaystyle \frac{y^B }{\sqrt{2}}}
e
b^{B\dag } \! , \\
\\[-12pt]
&\!\!\!\!\!\!\!\!\!\!
- i \sqrt{2} z^B \! M^{B\dag} 
\overline{a}^B \! a^{B\mbox{\scriptsize T}}
\!+\!
i \sqrt{2} z^B \! M^{B\mbox{\scriptsize T} }
\overline{b}^B \!\! a^{B\mbox{\scriptsize T}}
\!=\!
- \!
\left( \!\!  
{\displaystyle \frac{x^{B\dag }}{\sqrt{2}}}
\overline{a}^B \! e a^{B\mbox{\scriptsize T}}
\!+\!
{\displaystyle \frac{x^{B\mbox{\scriptsize T}} }{\sqrt{2}}}
\overline{b}^B \!\! e a^{B\mbox{\scriptsize T}} \!\!
\right)
\!+\!
\left( 1 \!-\! z^B \right) \!\!
{\displaystyle \frac{\overline{y}^B }{\sqrt{2}}}
e
a^{B\mbox{\scriptsize T}} \! .
\end{array} \!\!
\right\} 
\label{zxtheta}
\end{eqnarray}\\[-16pt] 
Subtracting the first equation of
(\ref{zxtheta}) 
from the second one 
and using (\ref{matrixF}) and (\ref{XY}),
we get\\[-20pt] 
\beqa
\!\!\!\!
\BA{ll}
-i \sqrt{2} z^B \! M^{B\dag }
&\!\!\!\!
\!=\!
- 
{\displaystyle \frac{x^{B\dag }}{\sqrt{2}}} \!\!
\left( \!
\overline{a}^B \! e a^{B\mbox{\scriptsize T}} 
\!\!+\! 
b^B \! e b^{B \dag} \!
\right)   
\!-\!
{\displaystyle \frac{x^{B\mbox{\scriptsize T}} }{\sqrt{2}}} \!\!
\left( \! 
 \overline{b}^B \!\! e a^{B\mbox{\scriptsize T}} 
 \!\!+\! 
 a^B \! e b^{B \dag} \!
\right)
\!+\!
\left( 1 \!-\! z^B \right) \!\!
\left\{ \!\!
{\displaystyle \frac{y^B }{\sqrt{2}}}
e
b^{B\dag }
\!+\!
{\displaystyle \frac{\overline{y}^B }{\sqrt{2}}}
e
a^{B\mbox{\scriptsize T}} \!\!
\right\} \\
\\[-12pt]
&\!\!\!\!
\!=\!
- {\displaystyle \frac{x^{B\dag }}{\sqrt{2}}} i F^B
\!-\!
{\displaystyle \frac{x^{B\mbox{\scriptsize T}} }{\sqrt{2}}} 
i \overline{D}^B
\!+\!
\left( 1 \!-\! z^B \right) \!\!
\left\{ \!\!
{\displaystyle \frac{y^B }{\sqrt{2}}}
e
b^{B\dag }
\!+\!
{\displaystyle \frac{\overline{y}^B }{\sqrt{2}}}
e
a^{B\mbox{\scriptsize T}} \!\!
\right\} \! ,
\EA
\label{sqrt2M}
\eeqa\\[-12pt]
whose hermitian adjoint and
complex conjugation become to be\\[-20pt]
\begin{eqnarray}
i \sqrt{2} z^B M^B 
\!=\!
- i F^B
{\displaystyle \frac{x^B}{\sqrt{2}}} 
+
i D^B
{\displaystyle \frac{\overline{x}^B}{\sqrt{2}}}
+
\left( 1 \!-\! z^B \right) \!
\left\{ \!
b^{B}
e
{\displaystyle \frac{y^{B\dag } }{\sqrt{2}}}
+
\overline{a}^B
e
{\displaystyle \frac{y^{B\mbox{\scriptsize T}}}{\sqrt{2}}} \!
\right\} ,
\label{hermitiansqrt2M}
\end{eqnarray}
\vspace{-0.8cm}
\begin{eqnarray}
- i \sqrt{2} z^B \overline{M}^B 
\!=\!
- i \overline{F}^B \!
{\displaystyle \frac{\overline{x}^B}{\sqrt{2}}} 
+ i \overline{D}^B
{\displaystyle \frac{x^B}{\sqrt{2}}}
+
\left( 1 \!-\! z^B \right) \!
\left\{ \!
\overline{b}^{B} \!\!
e
{\displaystyle \frac{y^{B\mbox{\scriptsize T}}}{\sqrt{2}}}
+
a^B \!
e
{\displaystyle \frac{y^{B\dag}}{\sqrt{2}}} \!
\right\} .
\label{complexsqrt2M}
\end{eqnarray}\\[-16pt]
Owing to the relations in the first line of
(\ref{xthetaepsilon})
with
$\epsilon \!=\! 0$,
equations
(\ref{hermitiansqrt2M}) and (\ref{complexsqrt2M})
give the additional conditions\\[-24pt]
\beqa
b^{B} e y^{B\dag }
+
\overline{a}^B e y^{B\mbox{\scriptsize T}}
\!=\!
0,
~~
\overline{b}^{B} \!\! e y^{B\mbox{\scriptsize T}}
+
a^B \! e y^{B\dag}
\!=\!
0 .
\label{additional conditions}
\eeqa\\[-18pt]
Multiplying
(\ref{hermitiansqrt2M})
by $F^B$
and
(\ref{complexsqrt2M})
by $D^B$
and
adding them,
we get\\[-18pt] 
\begin{eqnarray}
\!\!\!\!\!\!\!
\BA{ll}
&
i F^B \!\! \sqrt{2} z^B \! M^B
\!+\!
i D^B \!\! \sqrt{2} z^B \overline{M}^B
\!=\!
- i F^B \! F^B \! {\displaystyle \frac{x^B}{\sqrt{2}}} 
\!+\!
i F^B \! D^B \! {\displaystyle \frac{\overline{x}^B}{\sqrt{2}}}
\!+\!
\left(1 \!-\! z^B \right) \!\!
\left( \!\!
F^B b^B e {\displaystyle \frac{y^{B\dag} }{\sqrt{2}}} 
\!+\!
F^B \overline{a}^B e
{\displaystyle \frac{y^{B\mbox{\scriptsize T}}}{\sqrt{2}}} \!\!
\right) \\
\\[-12pt]
&~\!
\!-\!
i D^B \overline{F}^B
{\displaystyle \frac{\overline{x}^B}{\sqrt{2}}} 
\!+\!
i D^B \overline{D}^B
{\displaystyle \frac{x^B}{\sqrt{2}}}
\!+\!
\left(1 \!-\! z^B \right) \!\!
\left( \!
D^B \overline{b}^B e
{\displaystyle \frac{y^{B\mbox{\scriptsize T}} }{\sqrt{2}}} 
\!+\!
D^B a^B e {\displaystyle \frac{y^{B\dag}}{\sqrt{2}}} \!
\right)  \\
\\[-12pt]
&~\!
\!=\!
- i \!
\left( F^B F^B \!-\! D^B \overline{D}^B \right) \!
{\displaystyle \frac{x^B}{\sqrt{2}}}
\!+\! i \!
\left( F^B D^B \!-\! D^B \overline{F}^B \right) \!
{\displaystyle \frac{\overline{x}^B}{\sqrt{2}}} 
\!+\! i \!
\left(1 \!-\! z^B \right) \!\!
\left( \!
b^B e e {\displaystyle \frac{y^{B\dag} }{\sqrt{2}}} 
\!+\!
\overline{a}^B
e e
{\displaystyle \frac{y^{B\mbox{\scriptsize T}}}{\sqrt{2}}} \!
\right) .
\EA
\label{FzMDzM}
\end{eqnarray}\\[-10pt]
If the conditions
$ F^B \! D^B \!-\! D^B \overline{F}^B \!=\! 0$ and
$
b^B e e y^{B\dag} 
\!+\!
\overline{a}^B e e y^{B\mbox{\scriptsize T}}
\!=\! 0
$
are assumed,
which lead the condition
$e b^{B-1} \overline{a}^B \!=\! b^{B-1} \overline{a}^B e$,
are satisfied,
then we have the expression for
$x^B$  and $x^{B\dag}$ as\\[-16pt]
\beqa
\left.
\BA{ll}
{\displaystyle \frac{x^B}{\sqrt{2}}}
&\!\!
\!=\!
\left( F^B F^{B\dag} \!+\! D^B D^{B\dag} \right)^{\!-1} \!
\left( 
F^B \sqrt{2} z^B M^B
\!+\!
D^B \sqrt{2} z^B \overline{M}^B 
\right) , \\
\\[-10pt]
{\displaystyle \frac{x^{B\dag}}{\sqrt{2}}}
&\!\!
\!=\!
\left( \!
\sqrt{2} z^B M^{B\dag} \! F^{B\dag}
\!+\!
\sqrt{2} z^B M^{B\mbox{\scriptsize T}} \! D^{B\dag}
\right) \!\!
\left( F^B F^{B\dag} \!+\! D^B D^{B\dag} \right)^{\!-1}  \!\! .
\EA
\right\}
\label{solutionx}
\eeqa\\[-20pt]

On the other hand,
combining the equations in the first and the second lines of
(\ref{eigeneq})
and those in the first line of
(\ref{xthetaepsilon})
and using
$\varepsilon \!=\! 0$,
we have the following relations:\\[-12pt]
\begin{eqnarray}
\!\!\!\!\!\!\!\!\!\!
\left.
\begin{array}{ll}
&
i \! \left( \!\!
1 \!-\! {\displaystyle \frac{z^B}{1\!+\!z^B}} \!\!
\right) \!\!
\sqrt{2} M^B 
{\displaystyle \frac{y^B}{\sqrt{2}}} 
 \!=\!
 -
{\displaystyle \frac{1}{1\!+\!z^B}}
{\displaystyle \frac{x^B}{\sqrt{2}}}
{\displaystyle \frac{y^B}{\sqrt{2}}}
e
\!-\!
\left( \!\!
b^B 
\!+\!   
{\displaystyle \frac{1}{1\!+\!z^B}}
{\displaystyle \frac{x^B}{\sqrt{2}}}
{\displaystyle \frac{y^B}{\sqrt{2}}} \!\!
\right) \!
\varepsilon
\!=\!
-
{\displaystyle \frac{1}{1\!+\!z^B}}
{\displaystyle \frac{x^B}{\sqrt{2}}}
{\displaystyle \frac{y^B}{\sqrt{2}}}
e ,  \\
\\[-4pt]
&
i \! \left( \!\!
1 \!-\! {\displaystyle \frac{z^B}{1\!+\!z^B}} \!\!
\right) \!
\sqrt{2} ~\! \overline{M}^B 
{\displaystyle \frac{y^B}{\sqrt{2}}}
\!=\!
~{\displaystyle \frac{1}{1\!+\!z^B}}
{\displaystyle \frac{\overline{x}^B}{\sqrt{2}}}
{\displaystyle \frac{y^B}{\sqrt{2}}}
e , 
\end{array} \!\!
\right\}
\label{eigeneq00xy}
\end{eqnarray}\\[-6pt]
and further
combining the equations in the first and the second lines of
(\ref{eigeneq2})
and those in the first line of
(\ref{eigeneq3})
and using
$\varepsilon \!=\! 0$,
we have the following relations:\\[-14pt]
\begin{eqnarray}
\begin{array}{l}
i \sqrt{2} M^B 
{\displaystyle \frac{\overline{y}^B}{\sqrt{2}}}
=
{\displaystyle \frac{x^B}{\sqrt{2}}}
{\displaystyle \frac{\overline{y}^B}{\sqrt{2}}}
e , ~~
i \sqrt{2} ~\! \overline{M}^B 
{\displaystyle \frac{\overline{y}^B}{\sqrt{2}}}
=
-
{\displaystyle \frac{\overline{x}^B}{\sqrt{2}}}
{\displaystyle \frac{\overline{y}^B}{\sqrt{2}}}
e .
\end{array}
\label{eigeneq00xbary}
\end{eqnarray}\\[-8pt]
The relations
(\ref{additional conditions})
and
$
e b^{B-1} \overline{a}^B \!=\! b^{B-1} \overline{a}^B \! e
$
give the consistent result
with
(\ref{hermitiansqrt2M}) and (\ref{complexsqrt2M})
as\\[-12pt]
\beqa
b^B e e {\displaystyle \frac{y^{B\dag} }{\sqrt{2}}} 
+
\overline{a}^B
e e
{\displaystyle \frac{y^{B\mbox{\scriptsize T}}}{\sqrt{2}}}
=
b^B e b^{B-1} \!
\left( \!
b^B e {\displaystyle \frac{y^{B\dag} }{\sqrt{2}}} 
+
\overline{a}^B e
{\displaystyle \frac{y^{B\mbox{\scriptsize T}}}{\sqrt{2}}} \!
\right)
=
0 ,~~
\overline{b}^{B} \!\!
e
{\displaystyle \frac{y^{B\mbox{\scriptsize T}}}{\sqrt{2}}}
+
a^B \!
e
{\displaystyle \frac{y^{B\dag}}{\sqrt{2}}}
=
0 .
\label{consitentcondition}
\eeqa\\[-10pt]
In
(\ref{solutionx}),
multiplying the equations in the first and second lines by
$\frac{y}{\sqrt{2}}$
and
$\frac{y^\dag}{\sqrt{2}}$
from the right and left, respectively,
we obtain the equations\\[-14pt]
\beqa
\!\!\!\!\!\!\!\!\!\!
\left.
\BA{ll}
&
{\displaystyle \frac{x^B}{\sqrt{2}}}
{\displaystyle \frac{y^B}{\sqrt{2}}}
\!=\!
\left( F^B F^{B\dag} \!+\! D^B D^{B\dag} \right)^{-1} \!
\left( \!
F^B \sqrt{2} z^B M^B
{\displaystyle \frac{y^B}{\sqrt{2}}}
\!+\!
D^B \sqrt{2} z^B \overline{M}^B
{\displaystyle \frac{y^B}{\sqrt{2}}} \!
\right) \\
\\[-10pt]
&
\!=\!
i z^B \!\!
\left( \! 
F^B \! F^{B\dag} \!+\! D^B \! D^{B\dag} \! 
\right)^{-1} \!\!
\left( \!\!
F^B {\displaystyle \frac{x^B}{\sqrt{2}}}
\!-\!
D^B {\displaystyle \frac{\overline{x}^B}{\sqrt{2}}} \!\!
\right) \!\!
{\displaystyle \frac{y^B}{\sqrt{2}}}
e 
\!=\!
i z^B \!
\left( \! 
F^B \! F^{B\dag} \!\!+\!\! D^B \! D^{B\dag} \! 
\right)^{\!-1} \!\!
\sqrt{2}z^B \! M^B
{\displaystyle \frac{y^B}{\sqrt{2}}}
e , \\
\\[-8pt]
&
{\displaystyle \frac{y^{B\dag}}{\sqrt{2}}}
{\displaystyle \frac{x^{B\dag}}{\sqrt{2}}}
\!=\!
\left( \!
{\displaystyle \frac{y^{B\dag}}{\sqrt{2}}} 
\sqrt{2} z^B M^{B\dag} F^{B\dag}
\!+\!
\sqrt{2} z^B M^{B\mbox{\scriptsize T}} D^{B\dag} \!
\right) \!
\left( F^B F^{B\dag} \!+\! D^B D^{B\dag} \right)^{\!-1} \\
\\[-8pt]
&
\!\!=\!\!
-
i z^B \!
e
{\displaystyle \frac{y^{B\dag}}{\sqrt{2}}} \!\!
\left( \!\!
{\displaystyle \frac{x^{B\dag}}{\sqrt{2}}} \! F^{\!B\dag}
\!-\!
{\displaystyle \frac{x^{B\mbox{\scriptsize T}}}{\sqrt{2}}} \!
D^{B\dag} \!\!
\right) \!\!
\left( \! 
F^{\!B} \! F^{\!B\dag} \!+\! D^B \! D^{B\dag} \! 
\right)^{\!\!-1} 
\!\!\!=\!\!
- i z^B \!
e
{\displaystyle \frac{y^{B\dag}}{\sqrt{2}}} \!
\sqrt{2} \! z^B \! M^{\!B\dag} \!\!
\left( \! 
F^{\!B} \! F^{\!B\dag} \!+\! D^B \! D^{B\dag} \! 
\right)^{\!\!-1} \!\! ,
\EA \!\!\!
\right\}
\label{solutionxy}
\eeqa\\[-6pt]
where we have used the relations
(\ref{eigeneq00xy})
and
(\ref{eigeneq00xbary}).
Further multiplying the first and second equations in
(\ref{solutionxy}) by
$\frac{y^{B\dag}}{\sqrt{2}}$
and
$\frac{y^B}{\sqrt{2}}$
from the right and left, respectively,
we have\\[-8pt]
\beqa
\left.
\BA{ll}
&{\displaystyle \frac{x^B}{\sqrt{2}}}
{\displaystyle \frac{y^B}{\sqrt{2}}}
{\displaystyle \frac{y^{B\dag}}{\sqrt{2}}}
=\!
-
{\displaystyle \frac{x^B}{\sqrt{2}}}
{\displaystyle \frac{1 - (z^B)^2 }{2}} 
=\!
i z^B \!
\left( F^B F^{B\dag} + D^B D^{B\dag} \right)^{-1}  
\sqrt{2} z^B M^B
{\displaystyle \frac{y^B}{\sqrt{2}}}
e
{\displaystyle \frac{y^{B\dag}}{\sqrt{2}}} , \\
\\[-4pt]
&{\displaystyle \frac{y^B}{\sqrt{2}}}
{\displaystyle \frac{y^{B\dag}}{\sqrt{2}}}
{\displaystyle \frac{x^{B\dag}}{\sqrt{2}}}
=\!
-
{\displaystyle \frac{1 \!-\! (z^B)^2}{2}}
{\displaystyle \frac{x^{B\dag}}{\sqrt{2}}}
=\!
- i z^B \! 
{\displaystyle \frac{y^B}{\sqrt{2}}}
e
{\displaystyle \frac{y^{B\dag}}{\sqrt{2}}} \!
\sqrt{2} z^B M^{B\dag} \!
\left( F^B F^{B\dag} \!+\! D^B D^{B\dag} \right)^{\!-1} \! .
\EA 
\right\}
\label{solutionxyydag}
\eeqa\\[-4pt]
Then, at last we could reach the following expressions
for the two vectors
$\frac{x^B}{\sqrt{2}}$
and
$\frac{x^{B\dag}}{\sqrt{2}}$:\\[-10pt]
\beqa
\!\!\!\!\!\!
\left.
\BA{ll}
{\displaystyle \frac{x^B}{\sqrt{2}}}
&\!\!\!\!
=
- i
\left( F^B F^{B\dag} + D^B \! D^{B\dag} \right)^{-1} 
{\displaystyle \frac{2(z^B)^2}{1 - (z^B)^2}} 
\sqrt{2} M^B
{\displaystyle \frac{y^B}{\sqrt{2}}} 
e
{\displaystyle \frac{y^{B\dag}}{\sqrt{2}}} \\
\\[-12pt]
&\!\!\!\!
=
- i 
{\displaystyle \frac{2(z^B)^2}{1  - (z^B)^2}} 
<\!e\!> 
\left( F^B F^{B\dag} + D^B D^{B\dag} \right)^{-1} 
\sqrt{2} M^B , \\
\\[-4pt]
{\displaystyle \frac{x^{B\dag}}{\sqrt{2}}}
&\!\!\!\!
=
i
{\displaystyle \frac{2(z^B)^2}{1 - (z^B)^2}} 
{\displaystyle \frac{y^B}{\sqrt{2}}}
e
{\displaystyle \frac{y^{B\dag}}{\sqrt{2}}} 
\sqrt{2} M^{B\dag} 
\left(  F^B F^{B\dag} + D^B D^{B\dag} \right)^{-1} \\
\\[-10pt]
&\!\!\!\!
=
i
{\displaystyle \frac{2(z^B)^2}{1 - (z^B)^2}} 
<\!e\!> 
\sqrt{2} M^{B\dag}
\left( F^B F^{B\dag} + D^B D^{B\dag} \right)^{-1} \! ,
\EA 
\right\}
\label{finalsolutionx}
\eeqa\\[-12pt]
where $<\!e\!>$ is defined as
$
<\!e\!> 
\equiv
\frac{y^B}{\sqrt{2}}
e
\frac{y^{B\dag}}{\sqrt{2}} 
$.
Therefore, the
$<\!e\!>$
stands for the averaged value of all the eigenvalue distribution.
Thus, this is the first time that
the final solutions for
the vectors
$\frac{x^B}{\sqrt{2}}$
and
$\frac{x^{B\dag}}{\sqrt{2}}$
could be derived within the present
framework of the $\mathfrak{Jacobi~hsp}$ MFT.
The inner product of the vectors 
leads to the relation\\[-12pt]
\beqa
\BA{l}
{\displaystyle \frac{x^{B\mbox{\scriptsize T}}}{\sqrt{2}}}
{\displaystyle \frac{\overline{x}^B}{\sqrt{2}}}
=
{\displaystyle \frac{1 \!-\! (z^B)^2 }{2}}
=
{\displaystyle \frac{4(z^B)^4}{(1 \!-\! (z^B)^2)^2}}  
~\! 2
<\!e\!>^2 \! 
M^{B\mbox{\scriptsize T}} \! 
\left( 
\overline{F}^B F^{B\mbox{\scriptsize T}}
\!+\!
\overline{D}^B D^{B\mbox{\scriptsize T}} 
\right)^{\!-2} 
\overline{M}^B \! ,
\EA 
\label{xdagx}
\eeqa\\[-6pt]
which is the remarkable result in the $\mathfrak{Jacobi~hsp}$
MFT
and is simply rewritten as\\[-10pt]
\beqa
{\displaystyle \frac{16(z^B)^4}{(1 - (z^B)^2)^3 }}
<\!e\!>^2 \!
M^{B\mbox{\scriptsize T}} \! 
\left( 
\overline{F}^B F^{B\mbox{\scriptsize T}}
\!+\!
\overline{D}^B D^{B\mbox{\scriptsize T}} 
\right)^{\!-2} 
\overline{M}^B
\!=\!
1 .
\label{MdagM}
\eeqa\\[-8pt]
The above relation shows that the magnitude of
the additional SCF parameter $M^B$
is inevitably restricted by the behavior of
the SCF parameters $F^B$ and $D^B$
which should be governed by the condition
$F^B D^B \!-\! D^B \overline{F}^B \!=\! 0$.
Remember that this condition is one of
the crucial condition in
(\ref{FzMDzM})
to derive
(\ref{solutionx})
for
$\frac{x^B}{\sqrt{2}}$
and
$\frac{x^{B\dag}}{\sqrt{2}}$.
Finally,
substituting the solutions for
$\frac{x^B}{\sqrt{2}}$
and
$\frac{\overline{x}^B}{\sqrt{2}}$
which is derivable from the complex conjugation
of the first equation of 
(\ref{finalsolutionx}),
into the equation in the last line of
(\ref{xthetaepsilon}),
we can determine  $\epsilon$,
saying another excitation energy,
approximately as\\[-12pt] 
\beqa
\!\!\!\!\!\!\!\!\!\!\!\!\!\!\!\!\!\!\!\!\!\!\!\!\!\!\!\!\!\!\!\!
\BA{ll}
\epsilon
&
\!\!\approx\!\!
~
{\displaystyle \frac{4 z^B}{1 \!-\! (z^B)^2}} \!<\!e\!>\! 
M^{B\mbox{\scriptsize T}} \!
\left( 
\overline{F}^B \! F^{B\mbox{\scriptsize T}}
\!+\!
\overline{D}^B \! D^{B\mbox{\scriptsize T}} 
\right)^{\!-1} \!
\overline{M}^B \\
\\[-8pt]
&~~
\!\!\!+\!
{\displaystyle \frac{4 z^B}{1 \!-\! (z^B)^2}} \!<\!e\!>\!
M^{B\dag} \!
\left( 
F^B F^{B\dag}
\!+\!
D^B D^{B\dag} 
\right)^{\!-1} \!
M^B ,
\EA
\label{solutionepsilon}
\eeqa\\[-6pt]
due to the property of the complex conjugation given in
(\ref{xthetaepsilon}),
$
\overline{- i \sqrt{2}M^{B\mbox{\scriptsize T}}
\frac{\overline{x}^B}{\sqrt{2}}}
=\!
- i \sqrt{2}M^{B\dag }
\frac{x^B}{\sqrt{2}}  
$.
$\epsilon$ means
another type of excitation energy
never been seen in the traditional boson MFT
\cite{BR.86}.

$\!\!\!\!\!$In this section,
we have considered the boson GHB $\!$MFH
$\!H_{\mathfrak{Jacobi~hsp}}\!$
(\ref{mean-field Hamiltonian}).$\!$
Half a century ago,
in his famous textbook,
Berezin already gave this type of Hamiltonian
\cite{Berezin.66}, i.e.,
$
H_{\mbox{\scriptsize Berezin}} 
\!=\!
C_{ij} 
a^{\dag }_ia_j 
\!+\!
\frac{1}{2}
Aa^{\dag }_ia^{\dag }_j
\!+\!
\frac{1}{2}
\overline{A}_{ij}a_ia_j
\!+\!
a^{\dag }_i f_i \!+\! \overline{f}_{\! i} ~\! a_i 
$.
As is shown clearly from the structure of the
$H_{\mbox{\scriptsize Berezin}}$,
we should emphasize that
there exists the essential difference between
$H_{\mbox{\scriptsize Berezin}}$ and our Hamiltonian
$H_{\mathfrak{Jacobi~hsp}}$.
Recently, 
Berceanu,
in his papers
\cite{Berceanu.12,Berceanu.6},
also considered a linear Hamiltonian
in terms of the $\mathfrak{Jacobi}$ generators with
matrices of coefficient, $\epsilon^0,~\epsilon^+$
and $\epsilon^-$, \\[-16pt]
\begin{eqnarray}
\begin{array}{l}
\!\!\!\!
H_{\mbox{\scriptsize Berezin}}
\!=\!
\epsilon_i a_i \!+\! \overline{\epsilon}_i a^{\dag }_i
\!+\!
\epsilon^0_{ij} 
K^0_{i j}
\!+\!
\epsilon^+_{ij} 
K^+_{i j}
\!+\!
\epsilon^-_{ij} 
K^-_{i j} ,~
\left( 
[K^0_{i j},K^+_{i j},K^-_{i j}]
\equiv
{\displaystyle \frac{1}{2}}~
[E^{i }_{~j }, E^{i j }, E_{i j}] 
\right) .
\end{array}
\label{BerceanuH}
\end{eqnarray}\\[-12pt]
The Perelomov coherent state (CS)
$e_{z,w}$
\cite{Perelomov.86}
associated with the $\mathfrak{Jacobi}$ algebra
is defined as\\[-14pt]
\beqa
\BA{c}
e_{z,w}
\!=\!
\exp (\mbox{\boldmath $X$}) e_0,~
\mbox{\boldmath $X$}
\!=\!
\sum^N_{i \!=\!1} z_i a^\dag _i
\!+\!
\sum^N_{i,j \!=\!1} w_{ij} K^+_{i j},~ 
(a_i e_0 \!=\! 0) .
\EA
\label{ezw}
\eeqa\\[-14pt]
We denote $\{w_{ij}\}$ as $w$.
Using such the CS
on the Siegel-$\mathfrak{Jacobi}$ domain involving the
Siegel-$\mathfrak{Jacobi}$ ball or 
Siegel-$\mathfrak{Jacobi}$ upper half plane
\cite{Siegel.43,Siegel.64},
a classical equation of motion arisen from the linear Hamiltonian
(\ref{BerceanuH})
is given by a time-dependent differential equation,
saying, the well-known Riccati equation, \\[-14pt]
\beqa
\BA{l}
i \dot{w}
=
\epsilon^-
+
{\displaystyle \frac{1}{2}} \!
\left(
w \epsilon^0
+
\epsilon^{0\mbox{\scriptsize T}}
w^{\mbox{\scriptsize T}}
\right)
+
w \epsilon^+ w , ~~
i \dot{z}
=
\epsilon
+
w \overline{\epsilon}
+
{\displaystyle \frac{1}{2}}
\epsilon^{0\mbox{\scriptsize T}} z
+
w \epsilon^+ z .
\EA
\label{solutionWandz}
\eeqa\\[-12pt]
Berceanu has also found that a classical equation of motion
on both non-compact and compact
symmetric spaces arisen
from the linear Hamiltonian
is described by the matrix-type Riccati equation
\cite{BerceanuM.92,BerceanuG.92}.
On the other hand,
we also have obtained several years ago
the matrix-type Riccati-Hartree-Bogoliubov equation
on the compact symmetric coset space
$\frac{SO(2N)}{U(N)}$
for a fermion system 
accompanying with a general two-body interaction
\cite{SeiyaJoao.15}.

\newpage


\setcounter{equation}{0}
\renewcommand{\theequation}{\arabic{section}.\arabic{equation}}

\section{GEOMETRICAL STRUCTURE OF
$\frac{Sp(2N+2,\mathbb{R})_\mathbb{C}}{U(N+1)}$
COSET MANIFOLD}

\vspace{-0.5cm}
 
~~~
Let us introduce a $(N\!+\!1) \!\times\! (2N\!+\!2)$ isometric matrix
${\cal U}^{\mbox{\scriptsize T}}$
by\\[-10pt]
\beq
{\cal U}^{\mbox{\scriptsize T}}
\!=\!
\left[ \!\!
\BA{cc} 
{\cal B}^{\mbox{\scriptsize T}}, \!~{\cal A}^{\mbox{\scriptsize T}}
\EA \!\!
\right] ,~
( {\cal A}, {\cal B}:\mbox{given~by}~(\ref{calAcalB})) ,
\label{isomat}
\eeq
Along the direction
taken by the present authors $et~al.$
\cite{SJCF.08,NishiProviCord.11},
here we also use the matrix elements of 
${\cal U}$ and ${\cal U}^\dag $
as the coordinates on the manifold 
$Sp(2N\!\!+\!\!2,\mathbb{R})_\mathbb{C}$
instead of the manifold $SO(2\!N\!\!+\!\!2)$.
$\!\!$Even in the case of 
$\!Sp(2\!N\!\!+\!\!2,\mathbb{R})_\mathbb{C}$,
it turns out that
a real line element can be defined by a hermitian metric tensor 
on the manifold.
Under the transformation
${\cal U} \!\rightarrow\! {\cal VU}$
the metric is invariant.
Then, 
the metric tensor defined on the manifold may become singular,
due to the fact that one can use too many coordinates
through the introduction of another matrix ${\cal V}$.
 
According to Zumino
\cite{Zumino.79},
if ${\cal A}^B$ is non-singular,
we have relations governing ${\cal U}^\dag {\cal U}$ as\\[-12pt]
\beqa
\!\!\!\!\!\!\!\!
\left.
\BA{ll}
&
{\cal U}^\dag \!\!
\left(\!-\tilde{I}_{\!2N\!+\!2}\!\right) \!
{\cal U}
\!\!=\!\!
{\cal A}^{B\dag} \! {\cal A}^B
\!\!-\!\!
{\cal B}^{B\dag} \! {\cal B}^B
\!\!=\!\!
{\cal A}^{B\dag} \!\!
\left\{ \!
1_{\!N\!+\!1} 
\!\!-\!\! 
\left( \!
{\cal B}^B \!\! {\cal A}^{B-1} \!
\right)^\dag \!\!
\left(\!
{\cal B}^B \!\! {\cal A}^{B\!-1} \!
\right) \!
\right\} \!
{\cal A}^B
\!\!=\!\!
{\cal A}^{B\dag} \!\!
\left( \!
1_{\!N\!+\!1}  
\!\!-\!\! 
{\cal Q}^{B\dag} \! {\cal Q}^B \!
\right) \!\!\!
{\cal A}^B \! , \\
\\[-8pt]
&
\ln \det {\cal U}^\dag \!
\left(\!-\tilde{I}_{\!2N\!+\!2}\!\right) \!
{\cal U}
\!=\!
\ln \det {\cal U}^\dag {\cal U}
\!=\!
\ln \det 
\left(
1_{\!N\!+\!1} 
\!-\! 
{\cal Q}^{B\dag} \! {\cal Q}^B
\right)
\!+\!
\ln \det {\cal A}^B
\!+\!
\ln \det {\cal A}^{B\dag} ,
\EA \!\!
\right\}
\label{lndetUUdagger}
\eeqa\\[-6pt]
where we have used the 
$\frac{Sp(2N+2,\mathbb{R})_\mathbb{C}}{U(N+1)}$ coset variable 
${\cal Q}^B (\!=\!Q^{B\mbox{\scriptsize T}})$
given by
(\ref{cosetvariable2})
and the $\tilde{I}_{\!2N\!+\!2}$
of
(\ref{calAcalB}).
Due to the symmetric matrix of $\!{\cal Q}^B\!$,
$\!
1_{\!N\!+\!1} 
\!-\! 
{\cal Q}^{B\dag} \! {\cal Q}^B
\!>\! 0
$
can be verified.
If we take the matrix elements of ${\cal Q}^B$ 
and $\overline{\cal Q}^B$ 
as the coordinates 
on the 
$\frac{Sp(2N\!+\!2,\mathbb{R})_\mathbb{C}}{U(N\!+\!1)}$ 
coset manifold,
the real line element can be well defined by a hermitian metric tensor 
on the coset manifold as\\[-8pt]
\beq
ds^2
\!=\!
{\cal G}_{pq}{~}_{\underline{r}\underline{s}}
d{\cal Q}^{Bpq}d
\overline{{\cal Q}}^{B\underline{r}\underline{s}}~
(
{\cal Q}^{Bpq}
\!=\!
{\cal Q}^B_{pq},
\overline{{\cal Q}}^{B\underline{r}\underline{s}} 
\!=\!
\overline{{\cal Q}}^B_{\underline{r}\underline{s}};~
{\cal G}_{pq}{~}_{\underline{r}\underline{s}}
\!=\!
{\cal G}_{\underline{r}\underline{s}}{~}_{pq};~
\underline{r}, \underline{s}
~\mbox{take over}~
0, i, j
) .
\label{metric}
\eeq
The condition that a certain manifold is 
a K\"{a}hler manifold 
is that its complex structure is 
covariant constant relative to the Riemann connection
and that it has vanishing torsions:\\[-10pt]
\beq
{\cal G}_{pq~\underline{r}\underline{s}~,tu}
\stackrel{\mathrm{def}}{=}
\frac{\partial {\cal G}_{pq~\underline{r}\underline{s}}}
{\partial {\cal Q}^{tu}}
\!=\!
{\cal G}_{tu~\underline{r}\underline{s}~,pq} ,~~
{\cal G}_{pq~\underline{r}\underline{s}~,
\underline{t}\underline{u}}
\stackrel{\mathrm{def}}{=}
\frac{\partial {\cal G}_{pq~\underline{r}\underline{s}}}
{\partial \overline{\cal Q}^{\underline{t}\underline{u}}}
\!=\!
{\cal G}_{pq~\underline{t}\underline{u}~,
\underline{r}\underline{s}} ,
\label{Kcondition}
\eeq\\[-8pt]
As in the case of $SO(2N\!+\!2)$
\cite{SJCF.08,NishiProviCord.11},
the hermitian metric tensor 
${\cal G}_{pq}{~}_{\underline{r}\underline{s}}$
can be locally given through a real scalar function,
the K\"{a}hler potential, 
which takes the well-known form\\[-8pt] 
\beq
{\cal K}({\cal Q}^{B\dag},{\cal Q}^B) 
\!=\!
\ln \det 
\left(
1_{\!N\!+\!1} 
\!-\! 
{\cal Q}^{B\dag} \! {\cal Q}^B
\right) ,
\label{Kaehlerpotential}
\eeq
and the explicit expression for
the components of the metric tensor
is given as\\[-12pt] 
\beqa
\!\!
\BA{l}
{\cal G}_{pq~\underline{r}\underline{s}}
\!\!=\!\!
{\displaystyle 
\frac{\partial ^2 \! {\cal K} 
( \! {\cal Q}^{B\dag} \! , \! {\cal Q}^B \! )}
{\partial {\cal Q}^{Bpq} ~\!\!
\partial \overline{{\cal Q}}^{B\underline{r}\underline{s}}}
}
\!\!=\!\!
\left\{ \!
\left( \!
1_{\!N\!+\!1} 
\!\!-\!\!
{\cal Q}^B \! {\cal Q}^{B\dag} 
\right)^{\!\!-1} \!
\right\}_{\!\!sp} \!
\left\{ \!
\left( \!
1_{\!N\!+\!1} 
\!\!-\!\! 
{\cal Q}^{B\dag} \! {\cal Q}^B
\right)^{\!\!-1} \!
\right\}_{\!\!qr} 
\!\!-\!\!
(\!r \!\leftrightarrow\! s\!)
\!\!-\!\!
(\!p\!\leftrightarrow\! q\!) 
\!\!+\!\!
(\!p \!\leftrightarrow\! q, \! r \!\leftrightarrow\! s\!) .
\EA
\label{metricfromKpot}
\eeqa\\[-8pt]
Notice that the above function does not determine
the K\"{a}hler potential 
${\cal K}({\cal Q}^{B\dag} \! ,{\cal Q}^B)$ 
uniquely
since the metric tensor 
${\cal G}_{pq}{}_{\underline{r}\underline{s}}$
is invariant under transformations of the K\"{a}hler potential,\\[-10pt]
\beq
{\cal K}({\cal Q}^{B\dag},{\cal Q}^B)
\rightarrow
{\cal K}^\prime ({\cal Q}^{B\dag},{\cal Q}^B)
=
{\cal K}({\cal Q}^{B\dag},{\cal Q}^B)
+
{\cal F}({\cal Q}^B)
+
\overline{\cal F}(\overline{\cal Q}^B) .
\label{transKpot}
\eeq
${\cal F}({\cal Q}^B)$ and
$\overline{\cal F}(\overline{\cal Q}^B)$
are analytic functions of
${\cal Q}^B$ and $\overline{\cal Q}^B$,
respectively.
In the case of the K\"{a}hler metric tensor,
we have only the components of the metric connections 
with unmixed indices\\[-8pt]
\beq
\Gamma^{~tu}_{pq~rs}
\!=\!
{\cal G}^{\underline{v}\underline{w}~tu}
{\cal G}_{pq~\underline{v}\underline{w}~,rs} ,~~
\overline{\Gamma }^{~\underline{t}\underline{u}}
_{\underline{p}\underline{q}~
\underline{r}\underline{s}}
\!=\!
{\cal G}^{\underline{t}\underline{u}~vw}
{\cal G}_{vw~\underline{r}\underline{s}~,
\underline{p}\underline{q}} ,~~
{\cal G}^{\underline{v}\underline{w}~tu}
\stackrel{\mathrm{def}}{=}
({\cal G}^{-1})_{\underline{v}\underline{w}~tu} ,
\label{metricconect}
\eeq
and also have only the components of the curvatures\\[-14pt]
\beqa
\left.
\BA{ll}
&
\mbox{\boldmath $R$}_{pq~\underline{r}\underline{s}}
{~}_{tu~\underline{v}\underline{w}}
\!=\!
{\cal G}_{vw~\underline{v}\underline{w}}
\Gamma^{~vw}_{pq~tu}{}_{~, \underline{r}\underline{s}}
\!=\!
{\cal G}_{pq~\underline{v}\underline{w}}
{}_{~,tu~\underline{r}\underline{s}}
\!-\!
{\cal G}_{t^\prime u^\prime~
\underline{v^\prime } \underline{w^\prime }}
\Gamma^{~t^\prime u^\prime~}_{pq~tu}
\overline{\Gamma}^{~\underline{v^\prime } \underline{w^\prime }}
_{\underline{r}\underline{s}~\underline{v}\underline{w}} ,\\
\\[-8pt]
&
\mbox{\boldmath $R$}_{\underline{r}\underline{s}~pq}
{~}_{\underline{v}\underline{w}~tu}
\!=\!
{\cal G}_{tu~\underline{t}\underline{u}}
\overline{\Gamma }^{~\underline{t}\underline{u}}
_{\underline{r}\underline{s}~\underline{v}\underline{w}}
{}_{~, pq} 
\!=\!
\mbox{\boldmath $R$}_{pq~\underline{r}\underline{s}} 
{~}_{tu~\underline{v}\underline{w}} .
\EA
\right\}
\label{curvature}
\eeqa

\newpage


\setcounter{equation}{0}
\renewcommand{\theequation}{\arabic{section}.\arabic{equation}}

\section{EXPRESSION FOR
$\frac{Sp(2N+2,\mathbb{R})_\mathbb{C}}{U(N+1)}$ KILLING POTENTIAL
AND GDM}

\vspace{-0.5cm}

~~~Along the direction
taken in Refs.
\cite{SJCF.08,NishiProviCord.11},
we consider an 
$Sp(2N \! \!+ \!\! 2,\mathbb{R})_\mathbb{C}$ 
infinitesimal left transformation
of an 
$Sp(2N \! \!+ \!\! 2,\mathbb{R})_\mathbb{C}$ 
matrix ${\cal G}$ to ${\cal G}^\prime$,
$
{\cal G}^\prime
\!=\!
(1_{\! 2N \!+\! 2} \!+\! \delta {\cal G}) {\cal G}
$,
by using the first equation of
(\ref{infinitesimalop}):\\[-10pt]
\beqa
{\cal G}^\prime
\!\!=\!\! 
\left[ \!\!
\BA{cc}
1_{\!N \!+\! 1} \!+\! \delta {\cal A}^B & \delta \overline{\cal B}^B \\
\\[-6pt]
\delta {\cal B}^B & 1_{N \!+\! 1} \!+\! \delta \overline{\cal A}^B
\EA \!\!
\right] \!
{\cal G}
\!\!=\!\!
\left[ \!\!
\BA{cc}
{\cal A}^B \!+\! \delta {\cal A}^B \! {\cal A}^B
\!+\!
\delta \overline{\cal B}^B \! {\cal B}^B  & 
\overline{\cal B}^B \!+\! \delta {\cal A}^B \! \overline{\cal B}^B
\!+\! 
\delta \overline{\cal B}^B\!  \overline{\cal A}^B \\
\\[-2pt]
{\cal B}^B \!+\! \delta \overline{\cal A}^B \! {\cal B}^B
\!+\!
\delta {\cal B}^B \! {\cal A}^B  & 
\overline{\cal A}^B
\!+\!
\delta \overline{\cal A}^B \! \overline{\cal A}^B
\!+\! 
\delta {\cal B}^B \! \overline{\cal B}^B
\EA \!\!
\right] \! .
\label{calGprime}
\eeqa
Let us define an 
$\!\frac{Sp(2N\!+\!2,\mathbb{R})_\mathbb{C}}{U(N\!+\!1)}\!$ 
coset variable
$\!
{\cal Q}^{B\prime}
(\!=\! {\cal B}^{B\prime} \!\! {\cal A}^{B\prime -1})
\!$
in the ${\cal G}^\prime$ frame.
With the aid of 
(\ref{calGprime}),
the ${\cal Q}^{B\prime}$ is calculated infinitesimally as\\[-16pt]
\beqa
\BA{ll}
{\cal Q}^{B\prime}
\!=\!
{\cal B}^{B\prime} \!\! {\cal A}^{B\prime -1}
\!\!&
\!\!=\!
\left(
{\cal B}^B \!+\! \delta \overline{\cal A}^B \! {\cal B}
\!+\!
\delta {\cal B}^B \! {\cal A}^B
\right) \!
\left(
{\cal A}^B \!+\! \delta {\cal A}^B \! {\cal A}^B
\!+\!
\delta \overline{\cal B}^B \! {\cal B}^B
\right)^{\!-1} \\
\\[-8pt]
&
\!=\!
{\cal Q}^B \!+\! \delta {\cal B}^B 
\!-\!
{\cal Q}^B \delta {\cal A}^B
\!+\!
\delta \overline{\cal A}^B \! {\cal Q}^B
\!-\!
{\cal Q}^B \delta \overline{\cal B}^B \! {\cal Q}^B .
\EA
\label{calQprime}
\eeqa\\[-22pt]

The K\"{a}hler metrics admit a set of holomorphic isometries,
the Killing vectors,
${\cal R}^{i[k]}({\cal Q}^B)$
and
$\overline{\cal R}^{i[\underline{k }]}(\overline{\cal Q}^B)~
(i \!=\! 1, \cdots, \dim {\cal G})$,
which are the solution of the Killing equation\\[-6pt]
\beq
{\cal R}^i _{~[\underline{l }]}({\cal Q}^B)_{,~[k]}
+
\overline{\cal R}^i _{~[k]}({\cal Q}^B)_{,~[\underline{l }]}
\!=\!
0 ,~~
{\cal R}^i _{~[\underline{l}]}({\cal Q}^B)
\!=\!
{\cal G}_{[k][\underline{[l }]}{\cal R}^{i[k]}({\cal Q}^B) .
\label{Killingeq}
\eeq\\[-12pt]
These isometries define infinitesimal symmetry transformations and 
are described geometrically by the above Killing vectors which
are the generators of infinitesimal coordinate transformations
keeping the metric invariant: 
$
\delta {\cal Q}^B
\!=\!
{\cal Q}^{B\prime} \!-\! {\cal Q}^B
\!=\!
{\cal R}({\cal Q}^B)
$
and
$
\delta \overline{\cal Q}^B
\!=\!
\overline{\cal R}(\overline{\cal Q}^B)
$
such that
$
{\cal G}^\prime ({\cal Q}^B, \overline{\cal Q}^B)
\!=\!
{\cal G} ({\cal Q}^B, \overline{\cal Q}^B)
$.
The Killing equation
(\ref{Killingeq})
is the necessary and sufficient condition for 
an infinitesimal coordinate transformation\\[-10pt]
\beq
\delta{\cal Q}^{B[k]}
\!\!=\!\!
\left(
\delta {\cal B}^B 
\!\!-\!\!
\delta {\cal A}^{B\mbox{\scriptsize T}}{\cal Q}^B
\!\!-\!\!
{\cal Q}^B \delta {\cal A}^B
\!\!-\!\!
{\cal Q}^B \delta {\cal B}^{B\dag} {\cal Q}^B
\right)^{[k]}
\!\!=\!\!
\xi_i{\cal R}^{i[k]}({\cal Q}^B) ,~
\delta \overline{\cal Q}^{B[\underline{k }]}
\!\!=\!\!
\xi_i \overline{\cal R}^{i[\underline{k }]}(\overline{\cal Q}^B) .
\label{infinitesimaltrans}
\eeq
The 
$\xi _i$
is infinitesimal and global group parameter.
Due to the Killing equation,
Killing vectors
${\cal R}^{i[k]}({\cal Q}^B)$
and
$\overline{\cal R}^{i[\underline{k }]}(\overline{\cal Q}^B)$
can be written locally as the gradient of some real scalar function,
namely Killing potential
${\cal M}^i ({\cal Q}^B, \overline{\cal Q}^B)$
such that\\[-10pt]
\beq
{\cal R}^i _{~[\underline{k }]}({\cal Q}^B)
\!=\!
-i{\cal M}^i _{~,[\underline{k }]} ,~~
\overline{\cal R}^i _{~[k]}(\overline{{\cal Q}}^B)
\!=\!
i{\cal M}^i _{~,[k]} .
\label{gradKillingpot}
\eeq

According to van Holten
\cite{NNH.01}
and Refs. $\!\!\!\!$
\cite{SJCF.08,NishiProviCord.11}
and using the infinitesimal 
$\!Sp(2N \!\!+\!\! 2,\mathbb{R})_\mathbb{C}\!$ 
matrix $\delta {\cal G}$
given by first of
(\ref{infinitesimalop}),
the Killing potential ${\cal M}$ 
for the coset space
$\frac{Sp(2N\!+\!2,\mathbb{R})_\mathbb{C}}{U(N\!+\!1)}$
is written as\\[-12pt]
\beqa
\left.
\BA{ll}
&
{\cal M} \!
\left(
\delta {\cal A}, \delta {\cal B},\delta {\cal B}^\dag
\right)
\!=\!
\mbox{Tr} \!
\left( \! \delta {\cal G} \widetilde{{\cal M}} \! \right)
\!=\!
\mbox{tr} \!
\left(
\delta {\cal A} {\cal M}_{ \delta {\cal A}}
\!-\!
\delta {\cal B} {\cal M}_{\delta {\cal B}^\dag }
\!+\!
\delta {\cal B}^\dag {\cal M}_{\delta {\cal B}}
\right) ,\\
\\[-8pt]
&
\widetilde{{\cal M}}
\!\equiv\!
\left[ \!\!
\BA{cc} 
\widetilde{{\cal M}}_{\delta {\cal A}} & 
\widetilde{{\cal M}}_{\delta {\cal B}^\dag }\\
\\
\widetilde{{\cal M}}_{\delta {\cal B}} & 
-\widetilde{{\cal M}}_{\delta {\cal A}^{\mbox{\scriptsize T}}} 
\EA \!\!
\right] ,~~
\BA{c}
{\cal M}_{\delta {\cal A}}
\!=\!
\widetilde{{\cal M}}_{\delta {\cal A}}
\!+\!
\left( \!
\widetilde{{\cal M}}_{\delta {\cal A}^{\mbox{\scriptsize T}}} \!
\right)^{\!\mbox{\scriptsize T}} ,\\
\\
{\cal M}_{\delta {\cal B}}
\!=\!
\widetilde{{\cal M}}_{\delta {\cal B}} ,~~
{\cal M}_{\delta {\cal B}^\dag }
\!=\!
\widetilde{{\cal M}}_{\delta {\cal B}^\dag } .
\EA
\EA
\right\}
\label{KillingpotM}
\eeqa
Trace Tr is taken over 
$(2N \!+\! 2) \!\times\! (2N \!+\! 2)$ matrices,
while trace tr is taken over 
$(N \!+\! 1) \times (N \!+\! 1)$ matrices.
Let us introduce $(N \!+\! 1)$-dimensional matrices 
${\cal R}({\cal Q}^B \! ; \delta {\cal G})$, 
${\cal R}_T({\cal Q}^B \! ; \delta {\cal G})$ and ${\cal X}$ by
\beqa
\!\!\!\!
\left.
\BA{ll}
&
{\cal R}({\cal Q}^B \! ; \delta {\cal G})
\!=\!
\delta {\cal B}^B 
\!\!-\!\!
\delta {\cal A}^{B\mbox{\scriptsize T}}{\cal Q}^B
\!\!-\!\!
{\cal Q}^B \delta {\cal A}^B
\!\!-\!\!
{\cal Q}^B \delta {\cal B}^{B\dag} {\cal Q}^B ,~
{\cal R}_T ({\cal Q}^B \! ; \delta {\cal G})
\!=\!
-
\delta {\cal A}^{B\mbox{\scriptsize T}}
\!\!-\!\! 
{\cal Q}^B \delta {\cal B}^{B\dag} \! , \\
\\[-2pt]
&
{\cal X}
\!=\!
(1_{N \!+\! 1} \!-\! {\cal Q}^B{\cal Q}^{B\dag})^{-1}
\!=\! 
{\mathcal X}^\dag .
\EA \!\!
\right\}
\label{RRTChi}
\eeqa
In 
(\ref{infinitesimaltrans}),
putting $\xi_i$ as $\xi_i \!\!=\!\! 1$,
we have
$\delta \! {\cal Q}^B 
\!\!=\!\! 
{\cal R}(\!{\cal Q}^B \! ; \delta {\cal G}\!)
$ 
which is just 
the Killing vector
in the coset space
$\frac{Sp(2N\!+\!2,\mathbb{R})_\mathbb{C}}{U(N\!+\!1)}$
and tr of  
holomorphic matrix-valued function 
${\cal R}_T ({\cal Q}^B \! ; \delta {\cal G})$, i.e.,
$
\mbox{tr}{\cal R}_T ({\cal Q}^B \! ; \delta {\cal G}) 
\!=\! 
{\cal F}({\cal Q}^B)
$
is a holomorphic K\"{a}hler transformation. 
Then, the Killing potential ${\cal M}$ is given as
\beqa
\!\!\!\!\!\!\!\!
\left.
\BA{rl}
&-i{\cal M} \!
\left( \!
{\cal Q}^B, \overline{\cal Q}^B \! ;\delta {\cal G} \!
\right)
\!=\!
-\mbox{tr}
\Delta \!
\left( \!
{\cal Q}^B, \overline{\cal Q}^B \! ;\delta {\cal G} \!
\right) ,\\
\\[-4pt]
&\Delta \!\!
\left( \!
{\cal Q}^B \! , \overline{\cal Q}^B \! ; \! \delta {\cal G} \!
\right)
\!\stackrel{\mathrm{def}}{=}\!
{\cal R}_T (\! {\cal Q}^B \! ; \! \delta {\cal G} \!)
\!-\!
{\cal R}(\! {\cal Q}^B \! ; \! \delta {\cal G} \!) 
{\cal Q}^{B\dag} \! {\cal X} 
\!=\!
\left( 
{\cal Q}^B \! \delta {\cal A}^B \! {\cal Q}^{B\dag} 
\!\!-\!\! 
\delta {\cal A}^{B\mbox{\scriptsize T}}
\!\!-\!\!
\delta {\cal B}^B \! {\cal Q}^{B\dag} 
\!\!-\!\!
{\cal Q}^B \! \delta {\cal B}^{B\dag} 
\right) \!
{\cal X} \! .
\EA  \!\!\!
\right\}
\label{formKillingpotM} 
\eeqa
From
(\ref{KillingpotM}) and (\ref{formKillingpotM}),
we obtain
\beq
-i{\cal M}_{\delta {\cal B}}
\!=\!
-{\cal X} {\cal Q}^B ,~~
-i{\cal M}_{\delta {\cal B}^\dag }
\!=\!
 {\cal Q}^{B\dag} {\cal X} ,~~
-i{\cal M}_{\delta {\cal A}}
\!=\!
1_{\! N \!+\! 1} \!-\! 2 {\cal \overline{X}} .
\label{componentKillingpotM} 
\eeq
Using the expression for $\widetilde{{\cal M}}$,
equation
(\ref{componentKillingpotM}),
their components are written in the form
\beq
-i\widetilde{{\cal M}}_{\delta {\cal B}}
\!=\!
-{\cal X} \! {\cal Q}^B ,~
-i\widetilde{{\cal M}}_{\delta {\cal B}^\dag }
\!=\!
- {\cal Q}^{B\dag} \! {\cal X} ,~
-i\widetilde{{\cal M}}_{\delta {\cal A}}
\!=\!
-{\cal Q}^{B\dag} \! {\cal X} \! {\cal Q}^B ,~
-i\widetilde{{\cal M}}_{\delta {\cal A}^{\mbox{\scriptsize T}}}
\!=\!
-{\cal X} .
\label{tildecomponentKillingpotM} 
\eeq
As already shown for a fermion system
\cite{SJCF.08,NishiProviCord.11},
it is also easily verified that the result of
(\ref{componentKillingpotM})
satisfies the gradient of the real function ${\cal M}$
(\ref{gradKillingpot}).
If setting $r$ as $r \!=\! 0$ in ${\cal Q}^B$
(\ref{cosetvariable2}),
the Killing potential ${\cal M}$
in the 
$\!\frac{Sp(2N\!+\!2,\mathbb{R})_\mathbb{C}}{U(N\!+\!1)}\!$ coset space
leads 
the Killing potential $M$
in the $\!\frac{Sp(2N,\mathbb{R})_\mathbb{C}}{U(N)}\!$ coset space
obtained by van Holten $et~al.$
\cite{NNH.01}
and also by the present authors et. al.
\cite{SJCF.08,NishiProviCord.11}.

To make clear the meaning of the Killing potential,
using the $(2N \!\!+\!\! 2) \!\times\! (N \!\!+\!\! 1)$ 
isometric matrix 
${\cal U}$,
let us introduce the following 
$(2N \!\!+\!\! 2) \!\times\! (2N \!\!+\!\! 2)$ matrix:
\beq
{\cal W}
=
{\cal U}{\cal U}^\dag
\!=\!
\left[ \!
\BA{cc} 
{\cal R} & {\cal K} \\
\\[-4pt]
\overline{\cal K} & 1_{\!N \!+\! 1} \!+\! \overline{\cal R}
\EA \!
\right] ,~~
\BA{c}
{\cal R}
\!=\!
{\cal B}^B {\cal B}^{B\dag} ,\\
\\[-4pt]
{\cal K}
\!=\!
{\cal B}^B \! {\cal A}^{B\dag}  ,
\EA
\label{densitymat}
\eeq
which is hermitian
on the 
$\!Sp(2N \!\!+\!\! 2,\mathbb{R})_\mathbb{C}\!$ 
manifold.
The $\!{\cal W}\!$ is a natural extension of 
the generalized density matrix (GDM) for boson systems
in the $Sp(2N,\mathbb{R})_\mathbb{C}$ CS rep
to the $Sp(2N \!\!+\!\! 2,\mathbb{R})_\mathbb{C}$ CS rep
given in the textbook
\cite{BR.86}.
Since the matrices ${\cal A}^B$ and ${\cal B}^B$
are represented in terms of
${\cal Q}^B \!=\! ({\cal Q}^B_{pq})$ as\\[-4pt]
\beq
{\cal A}^B
\!=\!
(1_{\!N \!+\! 1} \!-\! {\cal Q}^{B\dag}{\cal Q}^B)^{-\frac{1}{2}} \!
\stackrel{\circ}{{\cal U}} ,~~
{\cal B}^B
\!=\!
{\cal Q}^B
(1_{\!N \!+\! 1} \!-\! {\cal Q}^{B\dag}{\cal Q}^B)^{-\frac{1}{2}} \!
\stackrel{\circ}{{\cal U}} ,~~
\stackrel{\circ}{{\cal U}} \in \! U(N \!+\! 1) ,
\label{matAandB}
\eeq
then, we have
\beq
{\cal R}
\!=\!
{\cal Q}^B (1_{N \!+\! 1} 
- 
{\cal Q}^{B\dag}\!{\cal Q}^B)^{-1}\!{\cal Q}^{B\dag}
\!=\!
{\cal Q}^B \overline{\chi }{\cal Q}^{B\dag}
\!=\!
-1_{N \!+\! 1} + \chi ,~~\!
{\cal K}
\!=\!
{\cal Q}^B
(1_{N \!+\! 1} - {\cal Q}^{B\dag}\!{\cal Q}^B)^{-1}
\!=\!
\chi {\cal Q}^B \! .
\label{matRandK}
\eeq
Substituting
(\ref{matRandK})
into
(\ref{tildecomponentKillingpotM}),
the Killing potential $-i\widetilde{{\cal M}}$
is expressed in terms of submatrices ${\cal R}$ and ${\cal K}$ of 
the GDM for boson systems
(\ref{densitymat}) 
as
\beq
-i\widetilde{{\cal M}}
\!=\!
\left[ \!
\BA{cc} 
- \overline{\cal R} & - \overline{\cal K} \\
\\[-4pt]
-{\cal K} & -(1_{N \!+\! 1} \!+\! {\cal R})
\EA \!
\right] ,
\label{tildecomponentKillingpotW}
\eeq
from which
we finally obtain
\beq
-i\overline{\widetilde{{\cal M}}}
\!=\!
\left[ \!
\BA{cc} 
{\cal R} & {\cal K} \\
\\[-4pt]
\overline{\cal K} & 1_{N \!+\! 1} \!+\!  \overline{\cal R}
\EA \!
\right] .
\label{complextildecomponentKillingpotW}
\eeq
To our great surprise,
the expression for the Killing potential
(\ref{complextildecomponentKillingpotW})
just becomes equivalent with
the GDM for boson systems
(\ref{densitymat}). 
The same assertion was also already made
with respect to the compact manifold
$\frac{SO(2N\!+\!2)}{U(N\!+\!1)}$
in a fermion system
\cite{SJCF.08}.


\newpage

\setcounter{equation}{0}
\renewcommand{\theequation}{\arabic{section}.\arabic{equation}}

\section{DISCUSSIONS AND SUMMARY}

~
Basing on the $\mathfrak{Jacobi~hsp}$ algebra,
the Perelomov CS
$e_{z,w}$
\cite{Perelomov.86}
is generated by the action of
$
\exp \!
\left[ 
z a^\dag 
\!\!+\!\!
\mbox{Tr}(\!w K^+\!) 
\right]  
$
to $e_{0}$
with row vector $z$ and column vector $a^\dag$.
Such the CS makes important roles
in the very broad range
of classical and quantum mechanics,
geometric quantization and 
current topics of quantum optics
\cite{AAG.00}.
Right up to now,
Berceanu
has investigated  
how the operators of the $\mathfrak{Jacobi}$ algebra admit
the realization of differential forms in the Siegel-$\mathfrak{Jacobi}$ domain
\cite{Berceanu.12}
on which
the Siegel-$\mathfrak{Jacobi}$ upper half plane
${\cal X}^J_N ~(= {\cal X}_N \!\times\! \mathbb{R}^{2N})$
\cite{Siegel.43}
is expressed as
$
{\cal X}_N:=
\{
Z \!\in\! M(N, \mathbb{C}) |
Z \!=\! X \!+\! i Y, 
X, Y \!\in\!  M(N, \mathbb{R}),
Y \!>\! 0,~
X^{\mbox{\scriptsize T}} \!=\! X, Y^{\mbox{\scriptsize T}} \!=\! Y
\}
$ and 
on which the Siegel-$\mathfrak{Jacobi}$ ball
${\cal D}^J_N ~(= {\cal D}_N \!\times\! \mathbb{C}^{N})$
\cite{Siegel.43}
is given as
$
{\cal D}_N:=
\{
w \!\in\! M(N, \mathbb{C}) :
w \!=\! w^{\mbox{\scriptsize T}},~
\mathbb{I}_N \!-\! w \overline{w} \!>\! 0
\}
$.
The plane and ball are connected each other by analytic
partial Cayley transform
as follows:\\[-10pt]
\begin{eqnarray}
\BA{ll}
\mbox{partial Cayley transform}
&
\Phi :
{\cal X}^J_N \rightarrow {\cal D}^J_N,
\Phi (Z, u)
\!=\!
(w, z) , \\
\\[-4pt]
&
w
\!=\!
(Z - i \mathbb{I}_N) ~\!
{\displaystyle \frac{\mathbb{I}_N}{Z + i \mathbb{I}_N}} , 	~~
z
\!=\!
2 i {\displaystyle \frac{\mathbb{I}_N}{Z + i \mathbb{I}_N}} u ,
\EA
\label{CayleyTr}
\end{eqnarray}
\vspace{-0.5cm}
\begin{eqnarray}
\BA{ll}
\mbox{inverse partial Cayley transform}
&
\Phi^{-1} :
{\cal D}^J_N \rightarrow {\cal X}^J_N,
\Phi^{-1} (w, z)
\!=\!
(Z, u) , \\
\\[-4pt]
&
Z
\!=\!
i {\displaystyle \frac{\mathbb{I}_N}{w - i \mathbb{I}_N}}
(\mathbb{I}_N + w) , ~~
u
\!=\!
{\displaystyle \frac{\mathbb{I}_N}{w - i \mathbb{I}_N}} z . 
\EA
\label{invCayleyTr}
\end{eqnarray}\\[-6pt]
The Siegel-$\mathfrak{Jacobi}$ domain
is related to the $\mathfrak{Jacobi~hsp}$ group
by the Harish-Chandra embedding.
For this embedding,
see Satake's book
\cite{Satake.80}.
Particularly,
the CS based on the Siegel-$\mathfrak{Jacobi}$ ball
${\cal D}^J_N$
is connected with the Gausson (Gaussian pure state)
proposed by
Simon, Sudarshan and Mukunda
\cite{SimonSudarshanMukunda.88}.
While,
Rowe, Rosensteel and Gilmore also showed the simple explicit forms
for the Cayley transform
$
w
\!=\!
(Z - i \mathbb{I}_N) ~\!\!
\frac{\mathbb{I}_N}{Z + i \mathbb{I}_N}
$
and the inverse Cayley transform
$
Z
\!=\!
i \frac{\mathbb{I}_N}{w - i \mathbb{I}_N}
(\mathbb{I}_N + w)
$
between Siegel upper half plane $S_{N}$ and
Siegel unit disk $D_{N}$ 
\cite{RoweRosensteelGilmore.85}.
Rosensteel and Rowe further showed that
the representation spaces of
discrete series are given as Hilbert spaces of
the holomorphic vector-valued functions on
the Siegel upper half plane $S_{N}$
\cite{RosensteelRowe.77,KashiwaraVergne.78}.
The  $Sp(2N, \mathbb{R})_\mathbb{C}$ vector CS is
the holomorphic vector-valued functions on
the Siegel unit disk $D_{N}$
\cite{RoweRosensteelGilmore.85}.

In the present paper,
we have proposed a new boson mean-field theory (MFT)
basing on the $\mathfrak{Jacobi~hsp}$ algebra, 
studied intensively by Berceanu
\cite{Berceanu.12}
and using the anti-commuting Grassmann variables.
Embedding the $\mathfrak{Jacobi~hsp}$ group into 
an $\!Sp(2N \!+\! 2, \mathbb{R})_\mathbb{C}\!$ group and 
using the 
$\frac{Sp(2N \!+\! 2, \mathbb{R})_\mathbb{C}}{U(N \!+\! 1)}$ 
coset variables,
we have build the new boson MFT on
the K\"{a}hler manifold of the non-compact symmetric space 
$\frac{Sp(2N \!+\! 2, \mathbb{R})_\mathbb{C}}{U(N \!+\! 1)}$.
We have given a linear boson MF Hamiltonian (MFH)
expressed in terms of the operators of 
the $\mathfrak{Jacobi~hsp}$ algebra
and diagonalized the linear MFH.
A new aspect of the eigenvalues $e$ and $\varepsilon$ 
of the MFH has been made clear.
A new excitation energy $\epsilon$ arisen from
the additional SCF parameter $M$
has never been seen in the traditional boson MFT
on the K\"{a}hler coset space 
$\frac{Sp(2N, \mathbb{R})_\mathbb{C}}{U(N)}$.
This kind of excitation energy
has been first derived in this work.
We also  have constructed the Killing potential,
which is the extension of 
Killing potential in the 
$\frac{Sp(2N, \mathbb{R})_\mathbb{C}}{U(N)}$ 
coset space 
to that in the 
$\frac{Sp(2N \!+\! 2, \mathbb{R})_\mathbb{C}}{U(N \!+\! 1)}$ 
coset space.
To our great surprise,
it turns out that
the Killing potential is entirely equivalent 
with the generalized density matrix ${\cal W}$
which is a powerful tool to approach 
difficult quantal-problems occurred
in many boson systems
\cite{BR.86,ChoiChernyakMukamel.03}.


\newpage

\begin{center}
{\bf ACKKNOWLEDGEMENTS}
\end{center}
\vspace{-0.1cm}

$\!\!\!\!\!\!\!\!\!\!\!\!$
The author (S.N.) would like to
express his sincere thanks to
Professor Constan\c{c}a Provid\^{e}ncia for kind and
warm hospitality extended to
him at Centro de F\'\i sica,
Universidade de Coimbra, Portugal.
This work is supported by FCT (Portugal) under the project
CERN/FP/83505/2008.
S.N. is indebted to Professor Emeritus M. Ozaki
of Kochi University
for his invaluable discussions and useful comments
in the early stage of this work.
The authors thank the Yukawa Institute for Theoretical Physics
at Kyoto University. Discussions during the YITP workshop
YITP-W-18-05 on ``Progress in Particle Physics 2018'' and
the YITP workshop YITP-S-18-05 41st Shikoku Seminar on
``Sign Problem and its Application"
are useful to complete this work.\\[4pt]

$\!\!\!\!\!\!\!\!\!\!\!\!$
Dedication to the Memory of Hideo Fukutome.


\newpage

\leftline{\large{\bf Appendix}}
\appendix


\vspace{-0.5cm}


\def\thesection{\Alph{section}}
\setcounter{equation}{0}
\renewcommand{\theequation}{\Alph{section}.\arabic{equation}}
\section{$\!\!\!\!\!$EMBEDDING OF $\mathfrak{Jacobi~hsp}$ GROUP 
INTO AN \\$Sp(\!2\!N\!+\!2\!,\!\mathbb{R}\!)_\mathbb{C}$ GROUP}


\vspace{-0.5cm}

~~~~
Referring to the fermion case proposed by Fukutome
\cite{Fuk.81},
we define the projection operators (operators:OPs) 
$\!P_{\!+}\!$ and $\!P_{\!-}\!$ 
onto the sub spaces of
even- and odd-boson numbers, respectively, by\\[-12pt]
\beq
P_\pm
\stackrel{\mathrm{def}}{=}
{\displaystyle \frac{1}{2}}(1 \pm (-1)^n) ,~~
P_\pm ^2
=
P_\pm ,~~
P_+ P_-
=
0 ,
\eeq\\[-16pt]
where $n$ is the boson-number OP
and define the OPs with the spurious index 0:\\[-22pt]
\beqa
\left.
\BA{ll}
&E^0_{~0}
\stackrel{\mathrm{def}}{=}
-
{\displaystyle \frac{1}{2}} (-1)^n
=
{\displaystyle \frac{1}{2}} (P_- - P_+) ,~
E^i_{~0}
\stackrel{\mathrm{def}}{=}
a^\dagger_i P_- 
=
P_+ a^\dagger_i,~~
E^0_{~i }
\stackrel{\mathrm{def}}{=}
a_i P_+ 
=
P_-a_i ,\\
\\[-10pt]
&E^{i 0}
\stackrel{\mathrm{def}}{=}
-a_i^\dagger P_+ 
=
-P_- a_i^\dagger,~~
E^{0 i }
\stackrel{\mathrm{def}}{=}
E^{i 0} ,~
E_{i 0}
\stackrel{\mathrm{def}}{=}
a_i P_- 
=
P_+ a_i ,~~
E_{0 i }
\stackrel{\mathrm{def}}{=}
E_{i 0} .
\EA
\right\}
\label{spuriousoperators}
\eeqa\\[-14pt]
The projection OPs $P_\pm$ are also introduced by
Brandt and Greenberg
\cite{Brandt and Greenberg.69}
The annihilation-creation OPs can be expressed 
in terms of the OPs
(\ref{spuriousoperators})
as\\[-14pt]
\beq
a_i
=
E_{i 0} + E^0_{~i },~~
a^\dag_i
=
-E^{i 0} + E^i_{~0} .
\eeq\\[-20pt]
We introduce the indices $p, q, \cdots$ running over  
$N\!+\!1$ values $0,1,\!\cdots\!,N\!$.
Then the OPs of 
(\ref{operatorset}) and (\ref{spuriousoperators})
can be denoted in a unified manner as
$E^p_{~q},~E_{pq}$ and $E^{pq}$
except
$E_{00}$ and $E^{00}$.
They satisfy\\[-20pt]
\beqa
\left.
\BA{ll}
&
E^{p \dag }_{~q}
=
E^q_{~p} ,~~
E^{p q } 
=
E^{\dag }_{q p } ,~~
E_{p q }
=
E_{q p } ,~~
(p , q \!=\! 0, 1, \cdots, N) \\
\\[-10pt]
&
[E^p_{~q},~E^r_{~s}]
=
\delta_{q r}E^p_{~s} 
- 
\delta_{p s}E^r_{~q},~~
(\mathfrak{u}(N\!+\!1)~\mbox{algebra}) \\
\\[-8pt]
&
\left.
\BA{ll}
&[E^p_{~q},~E_{r s}]
=
-
\delta_{p s}E_{q r} 
- 
\delta_{p r}E_{q s}, \\
\\[-8pt]
&
[E^{p q},~E_{r s}]
=
-
\delta_{p s}E^q_{~r}
- 
\delta_{q r}E^p_{~s}
-
\delta_{p r}E^q_{~s} 
- 
\delta_{q s}E^p_{~r},~
[E_{pq},~E_{rs}]
=
0 .
\EA
\right]
\EA
\right\}
\label{commurelp}
\eeqa\\[-16pt]
The above commutation relations in
(\ref{commurelp}) 
are of the same form as 
(\ref{commurel1}) and (\ref{commurel2}).
The commutation relations concerning
the OPs $E_{00}$ and $E^{00}$
are assumed at present to satisfy
(\ref{commurelp}).
It is a difficult problem how to define
$E_{00}$ and $E^{00}$
but must be solved in a near future.

The $\mathfrak{Jacobi~hsp}$ group is embedded 
into an $\!Sp(2N \!\!+\!\! 2,\mathbb{R})_\mathbb{C}\!$ group.$\!$ 
The embedding leads us to an unified formulation of 
the $Sp(2N \!\!+\!\! 2,\mathbb{R})_\mathbb{C}$
regular representation in which the paired and unpaired modes are
treated in an equal way. 
Define 
($N$\!+\!1)$\!\times\!$($N$\!+\!1) matrices 
${\cal A}^B$ and ${\cal B}^B$ as\\[-8pt]  
\beq
{\cal A}^B
\!\!\equiv\!\! 
\left[ \!\!
\BA{cc}
A^B &\!\! {\displaystyle -\frac{\overline{x}^B}{2}} \\
\\[-12pt]
{\displaystyle \frac{y^B}{2}} &\!\! 
{\displaystyle \frac{1\!\!+\!\! z^B}{2}}
\EA \!\!
\right],~\!
{\cal B}^B
\!\!\equiv\!\! 
\left[ \!\!
\BA{cc}
B^B &\!\! {\displaystyle \frac{x^B}{2}} \\
\\[-12pt]
{\displaystyle \frac{y^B}{2}} &\!\! 
{\displaystyle \frac{1 \!\!-\!\! z^B}{2}}
\EA \!\!
\right] ,\!\!\!\!\!\!
\BA{ll}
&
{\cal A}^{B\dag} {\cal A}^B
\!\!-\!\!
{\cal B}^{B\dag} {\cal B}^B 
\!\!=\!\! 
1_{\!N\!+\!1} ,~
{\cal A}^{B\mbox{\scriptsize T}}{\cal B}^B 
\!\!-\!\! 
{\cal B}^{B\mbox{\scriptsize T}}{\cal A}^B 
\!\!=\!\! 
0 ,\\
\\[-10pt]
&
{\cal A}^B {\cal A}^{B\dag} 
\!\!-\!\! 
\overline{\cal B}^B {\cal B}^{B\mbox{\scriptsize T}} 
\!\!=\!\! 
1_{\!N\!+\!1} ,~
\overline{\cal A}^B {\cal B}^{B\mbox{\scriptsize T}} 
\!\!-\!\! 
{\cal B}^B{\cal A}^{B\dag} 
\!\!=\!\! 
0 , \\
\\[-10pt]
&
y^B
\!=\!
x^{B\mbox{\scriptsize T}}a^B \!\!+\!\! x^{B\dag }b^B .
\EA
\label{calAcalB}
\eeq\\[-8pt]
Imposing the normalization of the matrix $G$
given in$\!$
(\ref{(2N+1)Gmatrix})
and using the $\!\tilde{I}_{\!2N\!+\!2}\!$ defined 
similarly as the second of
(\ref{Fmat}), 
then the $Sp(2N\!+\!2,\mathbb{R})_\mathbb{C}$ 
matrix ${\cal G}$ (not~unitary) is represented as\\[-10pt]
\beq
{\cal G}
\!=\! 
\left[ \!\!
\BA{cc}
{\cal A}^B & \overline{\cal B}^B \\
\\[-12pt]
{\cal B}^B & \overline{\cal A}^B
\EA \!\!
\right] ,~
{\cal G}^{\dag }
\widetilde{I}_{2N\!+\!2}
~\!
{\cal G}
\!=\!
{\cal G}
\widetilde{I}_{2N\!+\!2}
~\!
{\cal G}^{\dag }
\!=\!
\widetilde{I}_{2N\!+\!2} ,~~
\widetilde{I}_{2N\!+\!2}
\!\equiv\!
\left[ \!\!\!
\begin{array}{cc}
1_{\!N\!+\!1}  &\!\!\! 0  \\
\\[-8pt]
0    &\!\!\! -1_{\!N\!+\!1}     
\end{array} \!\!\!
\right]  ,               
\label{calG}
\eeq\\[-14pt]
The matrix ${\cal G}$ satisfies $\det {\cal G} \!=\! 1$ 
as is proved easily below\\[-8pt]
\beq
\det {\cal G}
\!=\!
\det \!
\left( \!
{\cal A}^B \!-\! \overline{\cal B}^B ~\! 
\overline{\cal A}^{B\!-1} \! {\cal B}^B \!
\right) \!
\det \overline{\cal A}^B
\!=\!
\det \!
\left( \!
{\cal A}^B \! {\cal A}^{B\dag} 
\!-\! 
\overline{\cal B}^B ~\! \overline{\cal A}^{B\!-1} \!
{\cal B}^B \! {\cal A}^{B\dag} \!
\right)
\!=\!
1 .
\label{detcalG}
\eeq\\[-16pt]
Using
the explicit expressions for 
${\cal A}^B$ and ${\cal B}^B$
given in
(\ref{calAcalB}),
$
{\cal A}^{B\dag} {\cal A}^B
\!\!-\!\!
{\cal B}^{B\dag} {\cal B}^B
$
is calculated as\\[-20pt]
\beqa
\BA{l}
{\cal A}^{B\dag} {\cal A}^B
\!-\!
{\cal B}^{B\dag} {\cal B}^B
\!=\!
\left[ \!\!\!
\BA{cc}
A^{B\dag} \! A^B \!\!-\!\! B^{B\dag} \! B^B &\!\!\!\!
-
A^{B\dag}{\displaystyle \frac{\overline{x}^B}{2}}
\!-\!
B^{B\dag}{\displaystyle \frac{x^B}{2}}
\!+\!
z^B {\displaystyle \frac{y^{B\dag}}{2}}\\
\\[-10pt]
-
{\displaystyle 
\frac{x^{B\mbox{\scriptsize T}}}{2}} A^B
\!-\!
{\displaystyle \frac{x^{B\dag}}{2}} B^B
\!+\!
z^B {\displaystyle \frac{y^B}{2}} &\!\!\!\!
{\displaystyle
\frac{x^{B\mbox{\scriptsize T}} \overline{x}^{B}}{4}
}
\!-\!
{\displaystyle \frac{x^{B\dag} x^{B}}{4}}
\!+\!
z^B
\EA \!\!\!
\right] ,
\EA
\eeqa\\[-16pt]
where
each block-matrix is computed with aid of the relation
$
y^{B}
\!\!=\!\!
x^{B\mbox{\scriptsize T}} \! a^{B} \!+\! x^{B\dag} b^{B}
$
as follows:\\[-14pt]

\beqa
\BA{ll}
{\cal A}^{B\dag} {\cal A}^B
\!\!-\!\!
{\cal B}^{B\dag} {\cal B}^B \\
&\\[-16pt]
\!=\!
1_{N}
\!-\!
{\displaystyle \frac{1}{2(1\!+\!z^B)}} \!
\left( \!
a^{B\dag}\overline{x}^{B} 
\!+\! 
b^{B\dag}x^{B} \!
\right) 
y^{B}
\!-\!
{\displaystyle \frac{1}{2(1\!+\!z^B)}} \!
y^{B\dag} \!
\left( \!
x^{B\mbox{\scriptsize T}} a^{B} 
\!+\! 
x^{B\dag} b^{B} \!
\right) \\
&\\[-14pt]
\!=\!
1_{N}
\!-\!
{\displaystyle \frac{1}{1\!+\!z^B}} 
y^{B\dag}y^{B}
\!\approx\!
1_{N} , \\
&\\[-8pt]
-{\displaystyle 
\frac{x^{B\mbox{\scriptsize T}}}{2}} A^B
\!-\!
{\displaystyle \frac{x^{B\dag}}{2}} B^B
\!+\!
z^B {\displaystyle \frac{y^{B}}{2}
}
\!=\!
-{\displaystyle \frac{x^{B\mbox{\scriptsize T}}}{2}} \!
\left( \!
a^{B}
\!-\!
{\displaystyle \frac{\overline{x}^{B}y^{B}}
{2(1\!+\!z^B)}} \!
\right)
\!-\!
{\displaystyle \frac{x^{B\dag}}{2}} \!
\left( \!\!
b^{B}
\!+\!
{\displaystyle \frac{x^{B}y^{B}}{2(1\!\!+\!\!z^B)}} \!\!
\right)
\!+\!
z^B {\displaystyle \frac{y^{B}}{2}} \\
&\\[-14pt]
\!=\!
-{\displaystyle \frac{1}{2}} \!\!
\left( \!
x^{B\mbox{\scriptsize T}} a^{B} \!+\! x^{B\dag} b^{B} \!
\right)
\!+\!
{\displaystyle
\frac{x^{B\mbox{\scriptsize T}} \overline{x}^{B} \! y^{B}}
{4(1\!+\!z^B)}
}
\!-\!
{\displaystyle \frac{x^{B\dag} \! x^{B} \! y^{B}}{4(1\!+\!z^B)}}
\!+\!
z^B {\displaystyle \frac{y^{B}}{2}}
\!=\!
-
\left( 1 \!-\! z^B \right) \!
{\displaystyle \frac{y^{B}}{2}}
\!+\!
\left( 1 \!-\! z^B \right) \!
{\displaystyle \frac{y^{B}}{2}}
\!=\!
0 ,  \\
&\\[-8pt]
{\displaystyle
\frac{x^{B\mbox{\scriptsize T}} \overline{x}^{B}}{4}
}
\!-\!
{\displaystyle \frac{x^{B\dag} x^{B}}{4}}
\!+\!
z^B
\!=\!
{\displaystyle \frac{1 \!-\! (z^B)^2 }{2}}
\!+\!
z^B
\!\approx\! 
1 .
\EA
\label{block-mat1}
\eeqa\\[-8pt]
The block-matrix
$
{\cal A}^{B\mbox{\scriptsize T}} \! {\cal B}^B
\!-\!
{\cal B}^{B\mbox{\scriptsize T}} \!\! {\cal A}^B
$
is also calculated as\\[-16pt]
\beqa
\BA{l}
{\cal A}^{B\mbox{\scriptsize T}} \! {\cal B}^B
\!\!-\!\!
{\cal B}^{B\mbox{\scriptsize T}} \! {\cal A}^B
\!=\!
\left[ \!\!\!
\BA{cc}
A^{B\mbox{\scriptsize T}} \! B^B
\!-\!
B^{B\mbox{\scriptsize T}} \! A^B &\!\!\!\!
A^{B\mbox{\scriptsize T}}{\displaystyle \frac{{x}^B}{2}}
\!\!+\!\! 
B^{B\mbox{\scriptsize T}}
{\displaystyle \frac{\overline{x}^B}{2}}
\!\!-\!\!
z^B 
{\displaystyle \frac{y^{B\mbox{\scriptsize T}}}{2}} \\
\\[-10pt]
-{\displaystyle \frac{x^{B\mbox{\scriptsize T}}}{2}} A^B
\!-\!
{\displaystyle \frac{x^{B\dag}}{2}} B^B
\!\!+\!\!
z^B {\displaystyle \frac{y^{B}}{2}} &\!\!\!\!
-
{\displaystyle
\frac{x^{B\mbox{\scriptsize T}} \overline{x}^{B}}{4}
}
\!\!+\!\!
{\displaystyle \frac{x^{B\dag} x^{B}}{4}}
\EA \!\!\!
\right] ,
\EA
\eeqa\\[-30pt]
\beqa
\!\!\!\!\!\!
\BA{ll}
A^{B\mbox{\scriptsize T}} \! B^B
\!-\! 
B^{B\mbox{\scriptsize T}} \! A^B \\
&\\[-14pt]
\!=\!
{\displaystyle \frac{1}{2(1\!+\!z^B)}} \!
\left\{ \!
\left( 
a^{B\mbox{\scriptsize T}} x^{B}
\!+\!
b^{B\mbox{\scriptsize T}} \overline{x}^{B} 
\right) \!
y^{B}
\!-\!
y^{B\mbox{\scriptsize T}} \!\!
\left( 
x^{B\mbox{\scriptsize T}} \! a^{B} 
\!+\! 
x^{B\dag} b^{B} 
\right) \!
\right\}
\!=\!
{\displaystyle \frac{1}{1\!+\!z^B}} \!
\left(
y^{B\mbox{\scriptsize T}} \! y^{B}
\!-\!
y^{B\mbox{\scriptsize T}}\! y^{B}
\right)
\!=\! 
0, \\
&\\[-4pt]
A^{B\mbox{\scriptsize T}}{\displaystyle \frac{{x}^B}{2}}
\!+\! 
B^{B\mbox{\scriptsize T}}
{\displaystyle \frac{\overline{x}^B}{2}} 
\!+\!
z^B {\displaystyle \frac{y^{B\mbox{\scriptsize T}}}{2}}
=
\left( \!
a^{B\mbox{\scriptsize T}}
\!-\!
{\displaystyle
\frac{y^{B\mbox{\scriptsize T}}x^{B\dag}}{2(1\!+\!z^B)}
} \!
\right) \!
{\displaystyle \frac{{x}^B}{2}}
\!+\! 
 \left( \!
b^{B\mbox{\scriptsize T}}
\!+\!
{\displaystyle
\frac{y^{B\mbox{\scriptsize T}}x^{B\mbox{\scriptsize T}}}
{2(1\!+\!z^B)}} \!
\right) \!
{\displaystyle \frac{\overline{x}^B}{2}} 
\!-\!
z^B {\displaystyle \frac{y^{B\mbox{\scriptsize T}}}{2}
} \\
&\\[-14pt]
\!=\!
{\displaystyle \frac{1}{2}} \!
\left( \!
a^{B\mbox{\scriptsize T}} \! x^B
\!+\!
b^{B\mbox{\scriptsize T}} \overline{x}^B \!
\right)
\!-\!
{\displaystyle \frac{y^{B\mbox{\scriptsize T}} \! 
x^{B\dag} \! x^B}
{4(1\!+\!z^B)}}
\!+\!
{\displaystyle
\frac{y^{B\mbox{\scriptsize T}} \! 
x^{B\mbox{\scriptsize T}} \overline{x}^B}
{4(1\!+\!z^B)}
} 
\!-\!
z^B 
{\displaystyle \frac{1}{2}}
y^{B\mbox{\scriptsize T}}\\
&\\[-14pt]
\!=\!
\left( 1 \!-\! z^B \right) \!
{\displaystyle \frac{1}{2}}
y^{B\mbox{\scriptsize T}}
\!-\!
\left( 1 \!-\! z^B \right) \!
{\displaystyle \frac{1}{2}}
y^{B\mbox{\scriptsize T}}
\!=\!
0 , \\
&\\[-8pt]
-
{\displaystyle
\frac{x^{B\mbox{\scriptsize T}} \overline{x}^{B}}{4}
}
\!+\!
{\displaystyle \frac{x^{B\dag} x^{B}}{4}}
\!=\!
-
{\displaystyle \frac{1 \!-\! (z^B)^2 }{2}}
\!\approx\! 
0 .
\EA
\label{block-mat2}
\eeqa\\[-8pt]
From
(\ref{block-mat1})
and
(\ref{block-mat2}), 
we can prove the orthogonal relation
(\ref{calAcalB}).
This means that
$\mathfrak{Jacobi~hsp}$ group is embedded 
into an $Sp(2N \!\!+\!\! 2,\mathbb{R})_\mathbb{C}\!$ group
for $z^B \!\!\approx\! 1$.
As a result,
the embedding of the $\mathfrak{Jacobi~hsp}$ group into
an $Sp(2N \!\!+\!\! 2,\mathbb{R})_\mathbb{C}$ group is realized
in an efficiently good accuracy.
With the use of definitions of $A^B$ and $B^B$ given in
(\ref{new(2N+1)Gmatrix}),
${\cal A^B}$ and ${\cal B}^B$
are decomposed as\\[-6pt]  
\beq
{\cal A}^B
\!=\!\!
\left[ \!\!\!
\BA{cc}
1_{\!N} 
\!\!-\!\! 
{\displaystyle 
\frac{\overline{x}^B r^{B\mbox{\scriptsize T}}}{2}
} &\!\! 
{\displaystyle -\frac{\overline{x}^B}{2}} \\
\\[-10pt]
{\displaystyle 
\frac{(\!1\!\!+\!\!z^B\!)r^{B\mbox{\scriptsize T}}}{2}
} &\!\! 
{\displaystyle \frac{1\!\!+\!\!z^B}{2}}
\EA \!\!\!
\right] \!\!\!
\left[ \!\!\!
\BA{cc}
a^B &\!\! 0 \\
\\ \\[-8pt]
0 &\!\! 1
\EA \!\!\!
\right] \! ,
{\cal B}^B
\!=\!\! 
\left[ \!\!\!
\BA{cc}
1_{\!N} 
\!\!+\!\! 
{\displaystyle \frac{x^B r^{B\mbox{\scriptsize T}} \! q^{B\!-1}}{2}} &\!\! 
{\displaystyle \frac{x^B}{2}} \\
\\[-10pt]
{\displaystyle 
\frac{(\!1\!\!+\!\!z^B\!)r^{B\mbox{\scriptsize T}} \! q^{B\!-1}}{2}
} &\!\! 
{\displaystyle \frac{1\!\!-\!\!z^B}{2}}
\EA \!\!\!
\right] \!\!\!
\left[ \!\!\!
\BA{cc}
b^B &\!\! 0 \\
\\ \\[-8pt]
0 &\!\! 1
\EA \!\!\!
\right] \! ,
r^{B}
\!=\!
{\displaystyle 
\frac{a^{B\mbox{\scriptsize T}\!-1} \! y^{B\mbox{\scriptsize T}}}
{1\!+\!z^B}
} ,
\label{calApcalBp}
\eeq\\[-4pt]
$
r^B
\!\equiv\!
\frac{1}{1+z^B} \!\!
\left( \! x^B \!\!+\!\! q^B \overline{x}^B \!\right)
\!$
and
$\!
q^B
\!\equiv\!
b^B \! a^{B\!-1}(\!q^{B\mbox{\scriptsize T}} \!\!=\!\! q^B\!)
$.
From
(\ref{calApcalBp}),
we get an 
$\!\frac{Sp(2N\!\!+\!\!2,\mathbb{R})_\mathbb{C}}{U(N\!\!+\!\!1)}\!$ 
coset variable as\\[-6pt]
\beq
{\cal Q}^B
\!\!=\!
Q^{B\mbox{\scriptsize T}} 
\!\!= \!
{\cal B}^B \!\! {\cal A}^{B\!-1}
\!\!=\!\! 
\left[ \!\!\!
\BA{cc}
q^B &\!\!\!\!\!\!
r^B
\!+\!
x^B
{\displaystyle
\frac{x^{B \dag} \! q^B \overline{x}^B}
{2(1\!\!+\!\!z^B)^2}
} \\
\\[-8pt]
z^B \! 
r^{B\mbox{\scriptsize T}} &\!\!\!\!\!\! 1 \!\!-\!\! z^B
\EA \!\!\!
\right] 
\!\approx\! 
\left[ \!\!\!
\BA{cc}
q^B &\!\! r^B \\
\\[-6pt]
r^{B\mbox{\scriptsize T}} &\!\!  0
\EA \!\!\!
\right] , 
~
(x^{B \dag} \! q^B \overline{x}^B \!\!=\!\! 0
~\mbox{and}~
z^B \!\approx\! 1) .
\label{cosetvariable2} 
\eeq\\[-8pt]
Then,
the paired and unpaired mode variables 
$(q^B,r^B)$
become independent variables of the
$\frac{Sp(2N\!+\!2,\mathbb{R})_\mathbb{C}}{U(N\!+\!1)}\!$ coset space
and
are unified into a single paired-mode variable in the 
$Sp(2N\!\!+\!\!2,\mathbb{R})_\mathbb{C}$ group. 
It should be remembered that
the $r^B_i$ is a Grassmann number.

\newpage


\def\thesection{\Alph{section}}
\setcounter{equation}{0}
\renewcommand{\theequation}{\Alph{section}.\arabic{equation}}
\section{$\!\!$EXTENDED BOSON REALIZATION OF 
$\mathfrak{sp}(\!2\!N\!+\!2,\!\mathbb{R}\!)_\mathbb{C}$ 
LIE OPERATORS}


\vspace{-0.5cm}

~~~
Referring to the integral representation of a state vector
in fermion space
given by Fukutome
\cite{Fuk.81},
we consider 
a boson-state vector $\ket {\!\Psi\!}$ 
corresponding to a function 
$\Psi (\!{\cal G}\!)$ in ${\cal G} 
\!\in\! 
Sp(2N\!\!+\!\!2,\mathbb{R})_\mathbb{C}$:\\[-20pt]  
\beqa
\BA{l}
{\displaystyle
\ket {\!\Psi\!}
\!=\!\!
\int \! U ({\cal G}) \ket 0 \bra 0 U^\dag ({\cal G}) \ket {\!\Psi\!} d{\cal G}
\!=\!\!
\int \! U ({\cal G}) \ket 0 \Psi ({\cal G}) d{\cal G}
} .
\EA
\label{statePsi}
\eeqa\\[-18pt]
The ${\cal G}$ is given by (\ref{calAcalB}) and (\ref{calG}) and
the $d{\cal G}$ is an invariant group integration.  
When an infinitesimal operator 
$\mathbb{I}_{\cal G} \!+\! \delta \widehat{{\cal G}}$ and
a corresponding infinitesimal unitary OP 
$U (1_{2N+2} \!+\! \delta {\cal G})$
is operated on $\ket {\!\Psi\!}$,
using
$U^{-1}(1_{2N\!+\!2} \!+\! \delta {\cal G}) 
\!=\! 
U (1_{2N\!+\!2} \!-\! \delta {\cal G})$, 
it transforms $\ket {\!\Psi\!}$ as\\[-22pt]
\beqa
\!\!\!\!\!\!
\BA{ll}
&U (1_{2N\!+\!2} \!-\! \delta {\cal G})  \ket {\!\Psi\!}
=
(\mathbb{I}_{\cal G} \!-\! \delta \widehat{{\cal G}}) \ket {\!\Psi\!}
=\!\!
{\displaystyle \int} \! U ({\cal G}) \ket 0 \bra 0 
U^\dag ((1_{2N\!+\!2} \!+\! \delta {\cal G}){\cal G}) \ket {\!\Psi\!} d{\cal G} \\
\\[-12pt]
&
=\!\!
{\displaystyle
\int \! U ({\cal G}) \ket 0 \Psi ((1_{2N\!+\!2} 
\!+\! 
\delta {\cal G}){\cal G})  d{\cal G} 
=\!\!
\int \! U ({\cal G}) \ket 0 (1_{2N\!+\!2} 
\!+\! 
\delta \mbox{\boldmath ${\cal G}$})
}  
\Psi ({\cal G}) d{\cal G},
\EA
\label{infinitrans}
\eeqa\\[-16pt]
\vspace{-0.5cm} 
\beqa
\!\!\!\!\!\!
\left.
\BA{ll}
&
1_{\!2N\!+\!2} \!+\! \delta {\cal G}
\!=\!
\left[ \!\!
\BA{cc} 
1_{\!N\!+\!1} \!+\! \delta {\cal A}^B & \delta \overline{\cal B}^B \\
\\[-8pt]
\delta {\cal B}^B & 1_{\!N\!+\!1} \!+\! \delta \overline{\cal A}^B  
\EA \!\!
\right] \! ,~
\delta {\cal A}^{B\dag} \!=\! - \delta {\cal A}^B ,~
\mbox{tr}\delta {\cal A}^B \!=\! 0 ,~
\delta {\cal B}^B \!=\! \delta {\cal B}^{B\mbox{\scriptsize T}} , \\
\\[-4pt]
&\delta \widehat{{\cal G}}
\!=\!
\delta {\cal A}^{Bp}_{~q} ~\! E^q_{~p}
\!+\!
{\displaystyle \frac{1}{2}} \!
\left(
\delta {\cal B}^B_{pq} ~\! E^{qp} 
\!+\! 
\delta \overline{\cal B}^B_{pq} ~\! E_{qp}
\right) ,~
\delta \mbox{\boldmath ${\cal G}$}
\!=\!
\delta {\cal A}^{Bp}_{~q} ~\! \mbox{\boldmath ${\cal E}^q_{~p}$}
\!+\!
{\displaystyle \frac{1}{2}} \!
\left(
\delta {\cal B}^B_{pq} ~\! \mbox{\boldmath ${\cal E}^{qp}$}
\!+\!
\delta \overline{\cal B}^B_{pq} ~\! \mbox{\boldmath ${\cal E}_{qp}$}
\right) \! . \!\!
\EA
\right\}
\label{infinitesimalop}
\eeqa\\[-14pt]
Equation
(\ref{infinitrans})
shows that
the operation of $\mathbb{I}_{\cal G} \!-\! \delta \widehat{{\cal G}}$ 
on the $\ket {\!\Psi\!}$ in the boson space 
corresponds to the left multiplication by 
$\!1_{\!2N\!+\!2} \!+\! \delta {\cal G}$
to variable ${\cal G}\!$ of function $\Psi ({\cal G})$.
For a small parameter $\epsilon$,
we obtain a representation on the $\Psi ({\cal G})$ as
\beq
\rho (e^{\epsilon \delta {\cal G}}) \Psi ({\cal G})
=
\Psi (e^{\epsilon \delta {\cal G}} {\cal G})
=
\Psi ({\cal G} + \epsilon \delta {\cal G} {\cal G})
=
\Psi ({\cal G} + d{\cal G}) ,
\label{reponPsi}
\eeq
which leads us to a relation 
$d{\cal G} 
= 
\epsilon \delta {\cal G} {\cal G}$.
From this,
we express it explicitly as,
\beqa
\left.
\BA{ll}
&
d{\cal G}
\!=\!
\left[ \!
\BA{cc} 
d{\cal A}^B & d\overline{\cal B}^B \\
\\[-4pt]
d{\cal B}^B & d\overline{\cal A}^B 
\EA \!
\right]
\!=\!
\epsilon \!
\left[ \!
\BA{cc} 
\delta {\cal A}^B {\cal A}^B
\!+\!
\delta \overline{\cal B}^B {\cal B}^B  & 
\delta {\cal A}^B ~\! \overline{\cal B}^B
\!+\!
\delta \overline{\cal B}^B ~\! \overline{\cal A}^B \\
\\[-4pt]
\delta {\cal B}^B {\cal A}^B
\!+\!
\delta\overline{\cal A}^B {\cal B}^B & 
\delta \overline{\cal A}^B ~\! \overline{\cal A}^B
\!+\!
\delta {\cal B}^B ~\! \overline{\cal B}^B \\ 
\EA \!
\right] , \\
\\
&
d{\cal A}^B
=
\epsilon
{\displaystyle \frac{\partial {\cal A}^B}{\partial \epsilon }}
=
\epsilon
(\delta {\cal A}^B {\cal A}^B 
\!+\! 
\delta \overline{\cal B}^B {\cal B}^B) ,~~
d{\cal B}^B
=
\epsilon
{\displaystyle \frac{\partial {\cal B}^B} {\partial \epsilon }}
=
\epsilon
(\delta {\cal B}^B {\cal A}^B 
\!+\! 
\delta \overline{\cal A}^B {\cal B}^B) .
\EA
\right\}
\label{dGdAdB} 
\eeqa
A differential representation of $\rho (\delta {\cal G})$,
$d\rho (\delta {\cal G})$,
is given as
\beq
d\rho (\delta {\cal G}) \Psi ({\cal G})
\!=\!
\left[
\frac{\partial {\cal A}^{Bp}_{~~~q}}{\partial \epsilon }
\frac{\partial }{\partial {\cal A}^{Bp}_{~~~q} }
+
\frac{\partial {\cal B}^B_{pq}}{\partial \epsilon }
\frac{\partial }{\partial {\cal B}^B_{pq} }
+
\frac{\partial \overline{\cal A}^{Bp}_{~~~q}}{\partial \epsilon }
\frac{\partial }{\partial \overline{\cal A}^{Bp}_{~~~q} }
+
\frac{\partial \overline{\cal B}^B_{pq}}{\partial \epsilon }
\frac{\partial }{\partial \overline{\cal B}^B_{pq}}
\right] \!
\Psi ({\cal G}) .
\label{diffrep} 
\eeq
Substituting
(\ref{dGdAdB})
into
(\ref{diffrep}),
we can get explicit forms of the differential representation
\beq
d\rho (\delta {\cal G}) \Psi ({\cal G})
=
\delta \mbox{\boldmath ${\cal G}$}
\Psi ({\cal G}),
\eeq
where each OP in $\delta \mbox{\boldmath ${\cal G}$}$
is expressed in a differential form as\\[-14pt]
\beqa
\left.
\BA{ll}
&
\mbox{\boldmath ${\cal E}^p_{~q}$}
=
\overline{\cal B}^B_{pr}
{\displaystyle \frac{\partial }{\partial \overline{\cal B}^B_{qr}}}
\!-\!
{\cal B}^B_{qr}
{\displaystyle \frac{\partial }{\partial {\cal B}^B_{pr}}}
\!-\!
\overline{\cal A}^{Bq}_{~~~r}
{\displaystyle 
\frac{\partial }{\partial \overline{\cal A}^{Bp}_{~~~r}}
}
\!+\!
{\cal A}^{Bp}_{~~~r}
{\displaystyle \frac{\partial }{\partial {\cal A}^{Bq}_{~~~r}}} 
=
\mbox{\boldmath ${\cal E}^{q \dag }_{~p}$} ,\\
\\[-14pt]
&
\mbox{\boldmath ${\cal E}_{pq}$}
=
\overline{\cal A}^{Bp}_{~~~r}
{\displaystyle \frac{\partial }{\partial \overline{\cal B}^B_{qr}}}
\!+\!
{\cal B}^B_{qr}
{\displaystyle \frac{\partial }{\partial {\cal A}^{Bp}_{~~~r}}}
\!+\!
\overline{\cal A}^{Bq}_{~~~r}
{\displaystyle \frac{\partial }{\partial \overline{\cal B}^B_{pr}}}
\!+\!
{\cal B}^B_{pr}
{\displaystyle \frac{\partial }{\partial {\cal A}^{Bq}_{~~~r}}} 
=
\mbox{\boldmath ${\cal E}^{qp \dag }$} ,\\
\\[-12pt]
&
\mbox{\boldmath ${\cal E}^{p \dag }_{~q}$}
=
\mbox{\boldmath $\overline{{\cal E}}^{p }_{~q}$},~~~~~~~
\mbox{\boldmath ${\cal E}_{pq }^\dag $}
=
\mbox{\boldmath $\overline{{\cal E}}_{pq }$},~~~~~~~
\mbox{\boldmath ${\cal E}_{pq}$}
=
\mbox{\boldmath ${\cal E}_{qp}$} .
\EA
\right\}
\label{diffops}
\eeqa
We define extended boson OPs 
$\mbox{\boldmath ${\cal A}^{Bp}_{~~~q}$}$,
$\!\mbox{\boldmath $\overline{\cal A}^{Bp}_{~~~q}$}$, etc.,
from every variable
${\cal A}^{Bp}_{~~~q}$,
$\!\overline{\cal A}^{Bp}_{~~~q}$, etc.,
as
\beqa
\left.
\BA{ll}
&
\mbox{\boldmath ${\cal A}^B$}
\stackrel{\mathrm{def}}{=}
{\displaystyle
\frac{1}{\sqrt{2}}
\left(
{\cal A}^B \!+\! \frac{\partial }{\partial \overline{\cal A}^B}
\right) 
} ,~~
\mbox{\boldmath ${\cal A}^{B\dag}$}
\stackrel{\mathrm{def}}{=}
{\displaystyle
\frac{1}{\sqrt{2}}
\left(
\overline{\cal A}^B \!-\! \frac{\partial }{\partial {\cal A}^B}
\right) 
} ,\\
\\[-4pt]
&
\mbox{\boldmath $\bar{\cal A}^B$}
\stackrel{\mathrm{def}}{=}
{\displaystyle
\frac{1}{\sqrt{2}}
\left(
\overline{\cal A}^B \!+\! \frac{\partial }{\partial {\cal A}^B}
\right) 
} ,~~
\mbox{\boldmath ${\cal A}^{B\mbox{\scriptsize T}}$}
\stackrel{\mathrm{def}}{=}
{\displaystyle
\frac{1}{\sqrt{2}}
\left(
{\cal A}^B 
\!-\! 
\frac{\partial }{\partial \overline{\cal A}^B}
\right) 
} , \\
\\[-4pt]
&
[\mbox{\boldmath ${\cal A}^B$},
~\mbox{\boldmath ${\cal A}^{B\dag}$}]
=
1 ,~~
[\mbox{\boldmath $\overline{\cal A}^B$},~
\mbox{\boldmath ${\cal A}^{B\mbox{\scriptsize T}}$}]
=1 ,\\
\\[-4pt]
&
[\mbox{\boldmath ${\cal A}^B$},
~\mbox{\boldmath $\overline{\cal A}^B$}]
=
[\mbox{\boldmath ${\cal A}^B$},~
\mbox{\boldmath ${\cal A}^{B\mbox{\scriptsize T}}$}]
=
0 ,~~
[\mbox{\boldmath ${\cal A}$}^{B\dag} ,
~\mbox{\boldmath $\overline{\cal A}^B$}]
=
[\mbox{\boldmath ${\cal A}$}^{B\dag} ,~
\mbox{\boldmath ${\cal A}^{B\mbox{\scriptsize T}}$}]
=
0 ,
\EA
\right\}
\label{boseops}
\eeqa
where ${\cal A}^B$ is a complex variable.
Similar definitions hold for ${\cal B}^B$
in order to define the extended boson OPs 
$\mbox{\boldmath ${\cal B}^B_{pq}$}$,
$\mbox{\boldmath $\bar{\cal B}^B_{pq}$}$, etc.
By noting the relations\\[-4pt]
\beq
{\displaystyle
\overline{\cal A}^B \frac{\partial }{\partial \overline{\cal A}^B}
\!+\!
{\cal A}^B \frac{\partial }{\partial {\cal A}^B}
\!=\!
\mbox{\boldmath ${\cal A}^{B\dag}$}
\mbox{\boldmath ${\cal A}^B$}
\!+\!
\mbox{\boldmath ${\cal A}^{B\mbox{\scriptsize T}}$}
\mbox{\boldmath $\overline{\cal A}^B$} 
},~~
{\displaystyle
\overline{\cal A}^B \frac{\partial }{\partial \overline{\cal B}^B }
\!+\!
{\cal B}^B \frac{\partial }{\partial {\cal A}^B}
\!=\!
\mbox{\boldmath ${\cal A}^{B\dag}$}
\mbox{\boldmath ${\cal B}^B$}
\!+\!
\mbox{\boldmath ${\cal B}^{B\mbox{\scriptsize T}}$}
\mbox{\boldmath $\overline{\cal A}^B$} 
},
\eeq\\[1pt]
the differential OPs
(\ref{diffops})
can be converted into an extended boson OP rep
\beqa
\left.
\BA{ll}
&
\mbox{\boldmath ${\cal E}^p_{~q}$}
\!=\!
\mbox{\boldmath ${\cal B}^{B\dag} _{pr}$}
\mbox{\boldmath ${\cal B}^B_{qr}$}
\!-\!
\mbox{\boldmath ${\cal B}^{B\mbox{\scriptsize T}}_{qr}$}
\mbox{\boldmath $\overline{\cal B}^B_{pr}$}
\!-\!
\mbox{\boldmath ${\cal A}^{Bq \dag }_{~~~r}$}
\mbox{\boldmath ${\cal A}^{Bp}_{~~~r}$}
\!+\!
\mbox{\boldmath ${\cal A}^{Bp 
\mbox{\scriptsize T}}_{~~~r}$}
\mbox{\boldmath $\overline{\cal A}^{Bq}_{~~~r}$} 
\!=\!
\mbox{\boldmath ${\cal B}^{B\dag} _{p \tilde{r}}$}
\mbox{\boldmath ${\cal B}^B_{q \tilde{r}}$}
\!-\!
\mbox{\boldmath ${\cal A}^{Bq \dag }_{~~~\tilde{r}}$}
\mbox{\boldmath ${\cal A}^{Bp}_{~~~\tilde{r}}$} ,\\
\\[-2pt]
&
\mbox{\boldmath ${\cal E}_{pq}$}
\!=\!
\mbox{\boldmath ${\cal A}^{Bp \dag }_{~~~r}$}
\mbox{\boldmath ${\cal B}^B_{qr}$}
\!+\!
\mbox{\boldmath ${\cal B}^{B\mbox{\scriptsize T}}_{qr}$}
\mbox{\boldmath $\overline{\cal A}^{Bp}_{~~~r}$}
\!+\!
\mbox{\boldmath ${\cal A}^{Bq \dag }_{~~~r}$}
\mbox{\boldmath ${\cal B}^B_{pr}$}
\!+\!
\mbox{\boldmath ${\cal B}^{B\mbox{\scriptsize T}}_{pr}$}
\mbox{\boldmath $\overline{\cal A}^{Bq}_{~r}$} 
\!=\!
\mbox{\boldmath ${\cal A}^{Bp \dag }_{~~~\tilde{r}}$}
\mbox{\boldmath ${\cal B}^B_{q \tilde{r}}$}
\!+\!
\mbox{\boldmath ${\cal A}^{Bq \dag }_{~~~\tilde{r}}$}
\mbox{\boldmath ${\cal B}^B_{p \tilde{r}}$} ,
\EA
\right\}
\eeqa
by using the notation
$\!
\mbox{\boldmath ${\cal A}^{\!Bp 
\mbox{\scriptsize T}}_{~~~r\!+\!N}$} 
\!\!\!=\!\!\!
\mbox{\boldmath ${\cal B}^{B\dag} _{pr}$}\!
$
and
$\!
\mbox{\boldmath ${\cal B}^{B\mbox{\scriptsize T}}_{pr\!+\!N}$}
\!\!\!=\!\!\!
\mbox{\boldmath ${\cal A}^{\!Bp \dag }_{~~~r}$}\!
$
and suffix $\!\tilde{r}$ running from 
0 to $N$ and from $N\!+\!1$ to $2N$.
Then we have the extended boson images of 
the boson 
$\mathfrak{sp}(\!2\!N\!+\!2,\!\mathbb{R}\!)_\mathbb{C}$ 
Lie OPs as
\beqa
\left.
\BA{rl}
\mbox{\boldmath $E^i_{~j }$}
\!=\!
&\!\!\!
\mbox{\boldmath ${\cal E}^i_{~j }$}
\!=\!
\mbox{\boldmath ${\cal B}^{B\dag}_{i \tilde{r}}$}
\mbox{\boldmath ${\cal B}^B_{j \tilde{r}}$}
\!-\!
\mbox{\boldmath ${\cal A}^{Bj \dag }_{~~~\tilde{r}}$}
\mbox{\boldmath ${\cal A}^{Bi}_{~~~\tilde{r}}$} ,~~
\mbox{\boldmath ${\cal E}^0_{~0}$}
\!=\!
0 ,\\
\\[-4pt]
\mbox{\boldmath $E_{i j}$}
\!=\!
&\!\!\!
\mbox{\boldmath ${\cal E}_{i j}$}
\!=\!
\mbox{\boldmath ${\cal A}^{Bi \dag }_{~~~\tilde{r}}$}
\mbox{\boldmath ${\cal B}^B_{j \tilde{r}}$}
\!+\!
\mbox{\boldmath ${\cal A}^{Bj \dag }_{~~~\tilde{r}}$}
\mbox{\boldmath ${\cal B}^B_{i \tilde{r}}$} ,~~
\mbox{\boldmath ${\cal E}_{00}$}
\!=\!
0 ,\\
\\[-6pt]
\mbox{\boldmath $a_{i }$}
\!=\!
&\!\!\!
\mbox{\boldmath ${\cal E}_{i 0}$}
\!+\!
\mbox{\boldmath ${\cal E}^0_{~i }$} 
\!=\!
-
\mbox{\boldmath ${\cal A}^{Bi \dag }_{~~~\tilde{r}}$} \!
\left( \!
\mbox{\boldmath ${\cal A}^{B0}_{~~~\tilde{r}}$}
\!-\!
\mbox{\boldmath ${\cal B}^B_{0 \tilde{r}}$} \!
\right)
\!+\!
\left( \!
\mbox{\boldmath ${\cal A}^{B0 \dag }_{~~~\tilde{r}}$}
\!+\!
\mbox{\boldmath ${\cal B}^{B\dag}_{0 \tilde{r}}$} \!
\right) \!
\mbox{\boldmath ${\cal B}^B_{i \tilde{r}}$} \\
\\[-6pt]
\!=\!
&\!\!\!
\sqrt{2} \!
\left( \!
\mbox{\boldmath ${\cal Y}^{B(+)\dag }_{\tilde{r}}$}
\mbox{\boldmath ${\cal B}^B_{i \tilde{r}}$}
\!-\!
\mbox{\boldmath ${\cal A}^{Bi \dag }_{~~~\tilde{r}}$}
\mbox{\boldmath ${\cal Y}^{B(-)}_{\tilde{r}}$} \!
\right) ,
~~
\mbox{\boldmath ${\cal Y}^{B(\pm)}_{\tilde{r}}$}
\!\stackrel{\mathrm{def}}{=}\!
{\displaystyle \frac{1}{\sqrt{2}}}
(\mbox{\boldmath ${\cal A}^{B0}_{~~~\tilde{r}}$}
\!\pm\!
\mbox{\boldmath ${\cal B}^B_{0 \tilde{r}}$}) .
\EA
\right\}
\label{bosonimage2}
\eeqa\\[-6pt]
The representation for
$\mbox{\boldmath $a_{i }$}$
in
(\ref{bosonimage2})
involves,
in addition to the original
$\mbox{\boldmath $A^{Bi }_{~~~j}$}$
and
$\mbox{\boldmath $B^B_{i j }$}$
bosons,
their complex conjugate bosons.
The complex conjugate bosons arise from the use of the matrix ${\cal G}$
as variables of representation. 
The $\mbox{\boldmath ${\cal Y}^B_{\tilde{r}}$}$ 
bosons also arise from extension of algebra from 
$\mathfrak{sp}(\!2N\!,\!\mathbb{R}\!)_\mathbb{C}$ to 
$\mathfrak{Jacobi~hsp}$ algebra
and embedding of the $\mathfrak{Jacobi~hsp}$ into 
$\mathfrak{sp}(\!2N\!\!+\!\!2,\!\mathbb{R}\!)_\mathbb{C}$ algebra.

Using the relations\\[-6pt]
\beq
{\displaystyle \frac{\partial }{\partial {\cal A}^{Bp}_{~~~q}}}
\det {\cal A}^B
\!=\!
({\cal A}^{B-1})^{~q}_p
\det {\cal A}^B ,~~
{\displaystyle \frac{\partial }{\partial {\cal A}^{Bp}_{~~~q}}}
({\cal A}^{B-1})^{~r}_s
\!=\!
-
({\cal A}^{B-1})^{~q}_s
({\cal A}^{B-1})^{~r}_p ,
\eeq
we get the relations which are valid when operated on functions
on the right coset 
$\frac{Sp(2N\!+\!2,\mathbb{R})_\mathbb{C}}{SU(N+1)}$\\[-12pt]
\beqa
\left.
\BA{ll}
&
{\displaystyle \frac{\partial }{\partial {\cal B}^B_{pq}}}
\!=\!
\sum _{r < p}
({\cal A}^{B-1})^{~q}_r 
{\displaystyle \frac{\partial }{\partial {\cal Q}^B_{pr}}},\\
\\[-10pt]
&
{\displaystyle \frac{\partial }{\partial {\cal A}^{Bp}_{~~~q}}}
\!=\!
-
\sum _{s < r < p}
{\cal Q}^B_{rp}
({\cal A}^{B-1})^{~q}_s 
{\displaystyle \frac{\partial }{\partial {\cal Q}^B_{rs}}}
-
{\displaystyle \frac{i}{2}}
({\cal A}^{B-1})^{~q}_p 
{\displaystyle \frac{\partial }{\partial \tau }} ,
\EA
\right\}
\label{differentialformulas}
\eeqa
from which later we can derive the expressions
(\ref{SO2Nplus2LieopQ}).

\newpage


\def\thesection{\Alph{section}}
\setcounter{equation}{0}
\renewcommand{\theequation}{\Alph{section}.\arabic{equation}}
\section{VACUUM FUNCTION FOR EXTENDED BOSONS}


~~
We show here that the function 
$\Phi_{00}({\cal G})$ in 
${\cal G} \!\in\! Sp(2N\!+\!2,\mathbb{R})_\mathbb{C}$
corresponds to the free-boson vacuum function 
in the physical boson space.
Then the 
$\Phi_{00}({\cal G})$
must satisfy the conditions
\beq
\left( \!
\mbox{\boldmath ${\cal E}^p_{~q}$} 
\!+\! 
\frac{1}{2}\mathbb{I}\delta_{pq} \!
\right)
\Phi_{00}({\cal G})
=
\mbox{\boldmath ${\cal E}_{pq}$}\Phi_{00}({\cal G})
=
0 ,~~
\Phi_{00}(1_{2N\!+\!2})
=
1 .
\label{vacuumcondition}
\eeq
The vacuum function $\Phi_{00}({\cal G})$ which satisfy
(\ref{vacuumcondition})
is given by
$
\Phi_{00}({\cal G}) 
\!=\!
\left[\det (\overline{\cal A}^B)\right]^{\frac{1}{2}} ,
$
the proof of which is made easily as follows:\\[-12pt]
\beqa
\!\!\!\!\!\!\!\!\!\!\!\!
\BA{ll}
&
\left( \!
\mbox{\boldmath ${\cal E}^p_{~q}$} 
\!+\! 
{\displaystyle \frac{1}{2}\mathbb{I}\delta_{pq}} \!
\right) \!
\left[\det (\overline{\cal A}^B)\right]^{\frac{1}{2}} \\
\\[-8pt]
&
=
{\displaystyle \frac{1}{2}\delta_{pq}}
\left[\det (\overline{\cal A}^B)\right]^{\frac{1}{2}}
\!+\!
\left( \!
\overline{\cal B}^B_{pr}
{\displaystyle 
\frac{\partial }{\partial \overline{\cal B}^B_{qr}}
}
-
{\cal B}^B_{qr}
{\displaystyle \frac{\partial }{\partial {\cal B}^B_{pr}}}
-
\overline{\cal A}^{Bq}_{~~~r}
{\displaystyle 
\frac{\partial }{\partial \overline{\cal A}^{Bp}_{~~~r}}
}
+
{\cal A}^{Bp}_{~~~r}
{\displaystyle \frac{\partial }
{\partial {\cal A}^{Bq}_{~~~r}}} \!
\right) \!
\left[\det (\overline{\cal A}^B)\right]^{\frac{1}{2}} \\
\\[-8pt]
&
=\!\!
{\displaystyle \frac{1}{2}\delta_{pq}} \!\!
\left[\!\det (\overline{\cal A}^B)\!\right]^{\!\frac{1}{2}} 
\!\!\!-\!\!
\overline{\cal A}^{Bq}_{~~~r}
{\displaystyle 
\frac{\partial }{\partial \overline{\cal A}^{Bp}_{~~~r}}
} \!\!
\left[\!\det (\overline{\cal A}^B)\!\right]^{\!\frac{1}{2}} 
\!\!=\!\!
{\displaystyle \frac{1}{2}\delta_{pq}} \!\!
\left[\!\det (\overline{\cal A}^B)\!\right]^{\!\frac{1}{2}}
\!\!\!-\!
{\displaystyle 
\frac{1}{2}\frac{1}
{\left[\!\det (\overline{\cal A}^B)\!\right]^{\!\frac{1}{2}}}
}
\overline{\cal A}^{Bq}_{~~~r}\!
{\displaystyle 
\frac{\partial }{\partial \overline{\cal A}^{Bp}_{~~~r}}
} \!
\det (\overline{\cal A}^B) \\
\\[-8pt]
&
=
{\displaystyle \frac{1}{2}\delta_{pq}} \!
\left[\det (\overline{\cal A}^B)\right]^{\frac{1}{2}}
\!\!-
{\displaystyle 
\frac{1}{2}\frac{1}
{\left[\det (\overline{\cal A}^B)\right]^{\frac{1}{2}}}
}
(\overline{\cal A}^B \overline{\cal A}^{B-1})_{qp}
\det (\overline{\cal A}^B) 
= 0 ,
\EA
\label{vacuum1}
\eeqa
\vspace{-0.1cm}
\beqa
\mbox{\boldmath ${\cal E}_{pq}$}
\left[\det (\overline{\cal A}^B)\right]^{\frac{1}{2}} 
\!=\!
\left( \!
\overline{\cal A}^{Bp}_{~~~r}
{\displaystyle \frac{\partial }{\partial \overline{\cal B}^B_{qr}}}
\!+\!
{\cal B}^B_{qr}
{\displaystyle \frac{\partial }{\partial {\cal A}^{Bp}_{~~~r}}}
\!+\!
\overline{\cal A}^{Bq}_{~~~r}
{\displaystyle \frac{\partial }{\partial \overline{\cal B}^B_{pr}}}
\!+\!
{\cal B}^B_{pr}
{\displaystyle \frac{\partial }{\partial {\cal A}^{Bq}_{~~~r}}} \!
\right) \!
\left[\det (\overline{\cal A}^B)\right]^{\frac{1}{2}} 
\!= 0 .
\label{vacuum2}
\eeqa
The vacuum functions 
$\Phi_{00}(\!G\!)$ in $G \!\in\! \mathfrak{Jacobi~hsp}$ group and 
$\Phi_{00}(\!g\!)$ in $g \!\in\! Sp(\!2N,\mathbb{R}\!)_\mathbb{C}$ group
satisfy
\beq
\mbox{\boldmath $a_{i }$}\Phi_{00}(G)
=
\left( \!
\mbox{\boldmath $E^i_{~j }$} \!+\! \frac{1}{2}\mathbb{I}\delta_{i j } \!
\right) \!
\Phi_{00}(G)
=
\mbox{\boldmath $E_{i j}$}\Phi_{00}(G)
=
0 ,~~
\Phi_{00}(1_{2N\!+\!2})
=
1 ,
\label{vacuumcondition2}
\eeq
\beq
\left( \!
\mbox{\boldmath $e^i_{~j }$} \!+\! \frac{1}{2}\mathbb{I}\delta_{i j } \!
\right) \!
\Phi_{00}(g)
=
\mbox{\boldmath $e_{i j }$}\Phi_{00}(g)
=
0 ,~~
\Phi_{00}(1_{2N})
=
1 .
\label{vacuumcondition3}
\eeq

By using the 
$\mathfrak{sp}(2N\!+\!2,\mathbb{R})_\mathbb{C}$ 
Lie OPs $E^{pq}$, 
expression
for the $Sp(2N\!+\!2,\mathbb{R})_\mathbb{C}$ WF $\!\ket {\!G\!}$ 
is given in a form quite
similar to the $Sp(2N,\mathbb{R})_\mathbb{C}$ WF $\!\ket {\!g\!}$ as
\beq
\ket G
\!=\!
\bra 0 U(G) \ket 0
\exp \!
\left(\! \frac{1}{2}\! \cdot\! {\cal Q}^B_{pq}E^{pq} \!\right) \!
\ket 0 .
\label{SO2N+1wf}
\eeq
Equation
(\ref{SO2N+1wf}) 
has the property 
$U(G) \ket 0 \!\!=\!\! U({\cal G}) \ket 0$.
On the other hand, from 
(\ref{calApcalBp})
we get
\beq
\det {\cal A}^B
\!=\!
\frac{1\!+\!z^B}{2} \det a^B \! ,~
\det {\cal B}^B
\!=\!
\left\{ \!
\frac{1\!-\!z^B}{2} 
\!+\! 
\frac{1}{2(1\!+\!z^B)} \!
\left(
x^{B\mbox{\scriptsize T}}q^{B-1}x^B \!-\! x^{B\dag} x^B
\right) \!
\right\} \!
\det b^B
\!=\!
0 .
\label{detAdeta}
\eeq
Then we obtain the vacuum function $\Phi_{00}({\cal G})$ 
expressed in terms of the K\"{a}hler potential as
\beq
\overline{\bra 0 U({\cal G}) \ket 0}
=
\Phi_{00}({\cal G})
=
\left[\det(\overline{\cal A}^B)\right]^{\frac{1}{2}}
=
e^{-\frac{1}{4}{\cal K}({\cal Q}^B,~{\cal Q}^{B\dag} )}
e^{-i\frac{\tau }{2}} , 
\label{Bogowf2}
\eeq
\beq
\Phi_{00}({\cal G}) 
= 
\Phi_{00}(G)
= 
\sqrt{\frac{1\!+\! z^B}{2}}
\left[\det(\overline{a}^B)\right]^{\frac{1}{2}} 
= 
\sqrt{\frac{1\!+\! z^B}{2}}
e^{-\frac{1}{4}{\cal K}(q^B,~q^{B\dag} )}
e^{-i\frac{\tau }{2}} .
\label{SO2Nplus1vacuumf2}
\eeq

\newpage

 
\def\thesection{\Alph{section}}
\setcounter{equation}{0}
\renewcommand{\theequation}{\Alph{section}.\arabic{equation}}
\section{DIFFERENTIAL FORMS FOR EXTENDED BOSONS OVER
$\frac{Sp(2N\!+\!2,\mathbb{R})_\mathbb{C}}{U(N\!+\!1)}$ COSET SPACE}


\vspace{-0.5cm}

~~~Using the differential 
formulas
(\ref{differentialformulas}),
the boson $\mathfrak{sp}(2N\!+\!2,\mathbb{R})_\mathbb{C}$ 
Lie OPs are mapped into
the regular representation consisting of functions 
on the coset space
$\frac{Sp(2N\!+\!2,\mathbb{R})_\mathbb{C}}{U(N\!+\!1)}$. 
The extended boson images of the boson 
$\mathfrak{sp}(2N\!+\!2,\mathbb{R})_\mathbb{C}$ Lie OPs
\mbox{\boldmath ${\cal E}^p_{~q}$} etc. are
represented by the closed first order differential forms
over the 
$\frac{Sp(2N\!\!+\!\!2,\mathbb{R})_\mathbb{C}}{U(N\!\!+\!\!1)}$ 
coset space
in terms of the 
$\frac{Sp(2N\!\!+\!\!2,\mathbb{R})_\mathbb{C}}{U(N\!\!+\!\!1)}$ 
coset variables 
${\cal Q}^B_{pq}$ 
and 
the phase variable
$
\tau \!\!
\left(
\!=\! 
\frac{i}{2} 
\ln \!\!
\left[ 
\frac{\det({\overline{A}^B})}{\det({A^B})}
\right] \!
\right)
$  
of $U(N\!+\!1)$ 
identical with the one,
$
\tau \!\!
\left(
\!=\! 
\frac{i}{2} 
\ln \!\!
\left[ 
\frac{\det({\overline{a}^B})}{\det({a^B})}
\right] \!
\right)
$ 
of $U(N)$
due to the first equation of 
(\ref{detAdeta}), 
as\\[-10pt]
\beqa
\BA{l}
\mbox{\boldmath ${\cal E}^p_{~q}$} 
=
{\displaystyle
\overline{{\cal Q}}^B_{pr}\frac{\partial }
{\partial \overline{{\cal Q}}^B_{qr}}
+
{\cal Q}^B_{qr}\frac{\partial }{\partial {\cal Q}^B_{pr}}
+
i\delta_{pq}\frac{\partial }{\partial \tau }
} ,~~
\mbox{\boldmath ${\cal E}_{pq}$} 
=
{\displaystyle
{\cal Q}^B_{pr}{\cal Q}^B_{sq}\frac{\partial }
{\partial {\cal Q}^B_{rs}}
+
\frac{\partial }{\partial \overline{{\cal Q}}^B_{pq}}
+
i{\cal Q}^B_{pq}\frac{\partial }{\partial \tau }
} ,
\EA
\label{SO2Nplus2LieopQ}
\eeqa\\[-6pt]
derivation of which is made in quite the same way 
as the case of the
$\mathfrak{sp}(2N,\mathbb{R})_\mathbb{C}$.
The extended boson images of the 
$\mathfrak{Jacobi~hsp}$ OPs 
are given with the aid of those of the 
$\mathfrak{sp}(2N\!+\!2,\mathbb{R})_\mathbb{C}$ OPs.
From 
(\ref{SO2Nplus2LieopQ}),
we get representation of the 
$\mathfrak{sp}(2N\!+\!2,\mathbb{R})_\mathbb{C}$ Lie OPs 
in terms of
the variables $q^B_{i j }$ and $r^B_i$.
They are very similar to those of the fermion
$\mathfrak{so}(2N\!+\!1)$ Lie OPs
except that the hermitian adjoint of the OP image
is obtained by the complex conjugate of the OP image
with minus sign 
\cite{Fuk.77}:\\[-10pt]
\beqa
\!\!\!\!\!\!\!\!
\left.
\BA{ll}
&\E^i_{~j }
\!=\!
\mbox{\boldmath ${\cal E}^i_{~j }$}
\!=\!
{\displaystyle
\mbox{\boldmath $e^i_{~j }$}
\!+\!
\overline{r}^B_i \frac{\partial }{\partial \overline{r}^B_j }
\!+\!
r^B_j \frac{\partial }{\partial r^B_i } ,~
\mbox{\boldmath $e^i_{~j }$}
\!=\!
\overline{q}^B_{i l }\frac{\partial }
{\partial \overline{q}^B_{j l }}
\!+\!
q^B_{j l }\frac{\partial }{\partial q^B_{i l }}
\!+\!
i \delta_{i j }\frac{\partial }{\partial \tau }
},~
\E^j_{~i }
\!=\!
\E j_{~i }^\dag
\!=\!
\overline{\E}^j_{~i } ,\\
\\[-10pt]
&\E_{i j }
\!=\!
\mbox{\boldmath ${\cal E}_{i j }$}
\!=\!
{\displaystyle
\mbox{\boldmath $e_{i j }$}
\!+\!
(r^B_i q^B_{j l} \!+\! r^B_j q^B_{i l })
\frac{\partial }{\partial r_l} ,~
\mbox{\boldmath $e_{i j }$}
\!=\!
q^B_{i l }q^B_{k j }\frac{\partial }
{\partial q^B_{l k }}
\!+\!
\frac{\partial }{\partial \overline{q}^B_{i j }}
\!+\!
iq^B_{i j }\frac{\partial }{\partial \tau }
},~
\E^{i j }
\!=\!
\E^\dag_{ j i }
\!=\!
\overline{\E}_{i j } ,
\EA \!\!
\right\}
\label{SO2Nplus1Lieopa}
\eeqa
\vspace{-0.3cm}
\beqa
\mbox{\boldmath $a_{i }$}
\!=\!
\mbox{\boldmath ${\cal E}_{0 i}$}
\!+\!
\mbox{\boldmath ${\cal E}^0_{~i}$}
\!=\!
{\displaystyle
\frac{\partial }{\partial \overline{r}^B_i }
\!+\!
\overline{r}^B_l \frac{\partial }{\partial \overline{q}^B_{i l }}
\!+\!
(r^B_{i}r^B_l  \!+\! q^B_{i l}) \frac{\partial }{\partial r^B_l}
\!+\!
q^B_{i l } r^B_k \frac{\partial }{\partial q^B_{l k }}
\!+\!
ir^B_{i}\frac{\partial }{\partial \tau }
} ,~
\mbox{\boldmath $a^\dagger_{i }$}
\!=\!
\mbox{\boldmath $\overline{a}_{i}$} .
\label{SO2Nplus1Lieopb}
\eeqa

Contrastively,
Berceanu,
in
\cite{Berceanu.12},
has given the differential action of the generators of 
the $\mathfrak{Jacobi~hsp}$ group
on the coherent state
$e_{z,w}$
given by
(\ref{ezw})
in the following forms:\\[-6pt]
\beqa
\!\!\!\!\!\!\!\!
\left.
\BA{ll}
&
\K^0_{ij }e_{z,w}
=
{\displaystyle
\left( \!
\frac{k}{4}
\delta_{i j }
+
\frac{z_j}{2}
\frac{\partial }{\partial z_i }
+
w_{j l } \frac{\partial }{\partial w_{l i}} \!
\right) \!
e_{z,w}
} \! ,~
(k: \mbox{extremal weight}) , \\
\\[-8pt]
&
\K^+_{i j }e_{z,w}
=
{\displaystyle
\frac{\partial }{\partial w_{i j}}
e_{z,w}
} ,\\
\\[-8pt]
&
\K^-_{i j }e_{z,w}
=
{\displaystyle
\left( \!
\frac{k}{2}
w_{i j}
\!+\!
\frac{z_i z_j}{2}
\!+\!
\frac{1}{2}
(z_{i} w_{l j}  \!+\! z_{j} w_{l i}) \frac{\partial }{\partial z_l}
\!+\!
w_{k j } w_{i l } \frac{\partial }{\partial w_{l k }} \!
\right) \!
e_{z,w}
} ,\\
\\[-6pt]
&
\mbox{\boldmath $a^\dagger_{i }$}e_{z,w}
\!=\!
{\displaystyle \frac{\partial }{\partial z_i}}
e_{z,w} ,~
\mbox{\boldmath $a_{i }$}e_{z,w}
\!=\!
{\displaystyle
\left( \!
z_i 
\!+\!
w_{i l} \frac{\partial }{\partial z_l} \!
\right) \!
e_{z,W}
} .
\EA \!\!
\right\}
\label{Sp2Nplus1Lieop}
\eeqa\\[-2pt]
The explicit forms of the expressions 
(\ref{Sp2Nplus1Lieop})
have just a few similarity to those of the ones
(\ref{SO2Nplus1Lieopa})
and
(\ref{SO2Nplus1Lieopb})
obtained exactly using the coset space
$\frac{SO(2N\!+\!2)}{U(N\!+\!1)}$.
Contrary to the similarity,
they have no exact property with respect to
complex conjugacy
in the operator-realization,
as shown from the forms of
the third and last Eqs in
(\ref{Sp2Nplus1Lieop}).
Berceanu has
further investigated
how the $\mathfrak{Jacobi~hsp}$ algebra admits what kind of
realization of the differential OPs
in the Siegel-$\mathfrak{Jacobi}$ space on 
both the Siegel-$\mathfrak{Jacobi}$ ball ${\cal D}^J_N$
and
the Siegel-$\mathfrak{Jacobi}$ upper half plane ${\cal X}^J_N$
\cite{Siegel.43}.

\newpage


\end{document}